\documentclass[usenatbib,onecolumn,fleqn]{mnras}

\usepackage{amsmath} % AMS Math Package
\usepackage{amssymb}	% Math symbols such as \mathbb

\usepackage{ifpdf}
\ifpdf
	\usepackage{graphicx}
	\usepackage{epstopdf}
\else
	\usepackage[dvips]{graphicx}
\fi

\usepackage[section]{placeins}	% Prevents floats from moving to another \section

\renewcommand{\v}[1]{\ensuremath{\boldsymbol{#1}}} % for vectors

\renewcommand{\d}[2]{\frac{{\rm d} #1}{{\rm d} #2}} % for derivatives
\newcommand{\pd}[2]{\frac{\partial #1}{\partial #2}} 
\newcommand{\covd}[2]{{#1}\phantom{}_{;#2}} % for covariant derivatives

\title{Oscillations of superfluid hyperon stars: decoupling scheme and g-modes}

\author[V.~A.~Dommes, M.~E.~Gusakov]
{V.~A.~Dommes$^{1}$,
 M.~E.~Gusakov$^{1,2}$
 \\
$^1$Ioffe Physical-Technical Institute of the Russian Academy of
Sciences,
Polytekhnicheskaya 26, 194021 St.~Petersburg, Russia
\\
$^2$Peter the Great St.~Petersburg Polytechnic University,
Polytekhnicheskaya 29, 195251 St.~Petersburg, Russia
}
\date{\today}

\begin{document}
%\label{firstpage}
\maketitle

\begin{abstract}
We analyse the oscillations of general relativistic superfluid hyperon stars,
	following the approach suggested by Gusakov \& Kantor and Gusakov et al.
	and generalizing it to the nucleon-hyperon matter.
We show that the equations governing the oscillations
	can be split into two weakly coupled systems
	with the coupling parameters $s_{\rm e}$, $s_{\rm \mu}$, and $s_{\rm str}$.
The approximation $s_{\rm e} = s_{\rm \mu} = s_{\rm str} = 0$ (decoupling approximation)
	allows one to drastically simplify the calculations of stellar oscillation spectra.
An efficiency of the presented scheme is illustrated
	by the calculation of sound speeds in the nucleon-hyperon matter
	composed of neutrons (n), protons (p), electrons (e), muons ($\mu$),
	as well as $\rm \Lambda$, ${\rm \Xi}^-$, and ${\rm \Xi}^0$-hyperons.
However, the gravity oscillation modes (g-modes) cannot be treated within this approach,
	and we discuss them separately.
For the first time	we study the composition g-modes
	in superfluid hyperon stars with the $\rm npe\mu\Lambda$ core
	and show that there are two types of g-modes (`muonic' and `$\Lambda$--hyperonic') in such stars.
We also calculate the g-mode spectrum and find out that the eigenfrequencies $\nu$ of the superfluid g-modes
	can be exceptionally large (up to $\nu \approx 742~{\rm Hz}$ for a considered stellar model).
\end{abstract}

\begin{keywords}
stars: interiors -- stars: neutron -- stars: oscillations
\end{keywords}

%%%%%%%%%%%%%%%%%%%%%%%%%%%%%%%%%%%%%%%%%%%%%%%%%%%%%%%%%%%%%%%%%%%%%%%%%%%%%%
\section{Introduction}
%%%%%%%%%%%%%%%%%%%%%%%%%%%%%%%%%%%%%%%%%%%%%%%%%%%%%%%%%%%%%%%%%%%%%%%%%%%%%%

It is interesting to study the oscillations of compact stars\footnote{
		By `compact' we mean neutron, hyperon, or quark stars.
		}
	because of two reasons. 
First, these oscillations can 
	%(and, probably, have already ) 
	be directly observed by analysing electromagnetic 
	radiation from the stellar surface
	(\citealt{2005ApJ...628L..53I,
		2006ApJ...653..593S,
		2007AdSpR..40.1446W,
		2007Ap&SS.308..625W,
		2014ApJ...784...72S,
		2014ApJ...793L..38S})
	and, in the future, gravitational radiation from the oscillating stars
	(\citealt{2003CQGra..20R.105A,
		2004PhRvD..70l4015B,
		2011GReGr..43..409A,
		2012CQGra..29l4013S,
		2013CQGra..30s3002A}).
Secondly, some classes of oscillations of rotating compact stars 
	(the most important are r- and f-modes)
	are generically unstable with respect to excitation of gravitational waves.
Such oscillations can be spontaneously excited in a rotating star 
	and can strongly affect its observational properties
	(\citealt{2007PhRvD..76f4019B,
			2014MNRAS.442.1786A,
			2014MNRAS.442.3037L}),
	even if the oscillations themselves are not directly detected. 

Unfortunately, a realistic modelling of compact star dynamics 
	is a difficult task.
The main difficulties are:
	(i) accounting for the effects of general relativity;
	(ii) an equation of state (EOS) and an actual composition
	of the internal layers of compact stars are not reliably known 
	(nucleon matter? nucleon-hyperon matter? quarks? some other exotica?);
	(iii) possible superfluidity of baryons substantially complicates
	stellar dynamics by increasing the number of independent degrees of freedom
	(velocity fields) involved into the problem.

Because of the general complexity of the problem,
	here we concentrate on its particular piece. 
Namely, in this paper we discuss in detail
	the equations governing the oscillations 
	of general relativistic superfluid {\it hyperon} stars (HSs),
	which are the compact objects hosting hyperons 
	(e.g., ${\rm \Lambda}$, ${\rm \Xi}^{-}$, ${\rm \Xi}^{0}$, ${\rm \Sigma}^{-}$) 
	in their cores.
According to most of the microscopic theories they 
	appear	at densities around $\rho \sim (2 \div 3) \rho_0$, 
	where $\rho_0 \approx 2.8 \times 10^{14}$~g~cm$^{-3}$ 
	is the density in atomic nuclei
	(see e.g.,
	\citealt{2012A&A...543A.157B,
		2012NuPhA.881...62W,
		2012PhRvC..85f5802W,
		2014MNRAS.439..318G}).
Thus, they should exist in the majority of (not too light) neutron stars. 
Meanwhile, up until now, 
	most of the studies of stellar oscillation spectra ignored a possible presence of hyperons
	even when modelling non-superfluid  (`normal') compact stars 
	(but see e.g.,
		\citealt{2002PhRvD..65f3006L,
				2004PhRvD..70l4015B,
				2006PhRvD..73h4001N,
				2014PhRvD..89d4006B,
				2015PhRvD..91d4034C}
	and references therein).
Concerning the superfluid HSs, 
	even the equations driving the oscillations
	of such stars were not established until recently.
The problem was addressed in a series of papers by Gusakov \& Kantor,
	where a dissipative relativistic superfluid hydrodynamics was formulated (\citealt{2008PhRvD..78h3006G})
	and applied to study the sound waves in superfluid nucleon-hyperon mixture (\citealt{2009PhRvD..79d3004K});
	its main ingredients (entrainment matrix and bulk viscosity coefficients) 
	have been calculated by
	\citet{2008PhRvD..78h3006G}
	and
	\citet{2009PhRvC..79e5806G,
		2009PhRvC..80a5803G}.
Subsequently, a multifluid Newtonian hydrodynamics, capable of describing 
	superfluid nucleon-hyperon mixtures, has been formulated by \citet{2012PhRvD..86f3002H};
	prior to that, its simplified version was used 
	by \citet{2010MNRAS.408.1897H}
	to study
	the hyperon bulk viscosity and the resulting r-mode damping 
	in superfluid HSs.

This work is built on the existing research described above
	and is aimed at presenting an approximate scheme
	allowing to decouple the superfluid and normal degrees of freedom and, 
	hence, to substantially simplify modelling of oscillations of HSs.
The presented method is a generalization of a similar method
	suggested and applied
	by
	\citet{2011PhRvD..83h1304G,
		2011MNRAS.418L..54C,
		2012ASPC..466..211K,
		2013MNRAS.428.1518G,
		2014PhRvD..90b4010G}
	in application to superfluid neutron stars
	with neutron-proton-electron cores ($\rm npe$ cores).
We argue
	that this method can be used to study the oscillation modes 
	which survive in barotropic (non-stratified) HSs 
	(such as e.g., f-, p-, and r-modes)
	but is inapplicable to gravity modes (g-modes), 
	whose frequencies are determined by the degree 
	of stratification of the matter in the stellar cores 
	and vanish for purely barotropic stars.
That is why the g-modes in superfluid HSs should be treated separately.
Here we calculate their spectrum for the first time,
	adopting a modern hyperonic EOS from \citet{2014MNRAS.439..318G}
	and following the approach of \citet{2014MNRAS.442L..90K},
	who studied g-modes
	in neutron stars
	with superfluid $\rm npe$ cores with admixture of muons.
	
The paper is organized as follows.
In Section~\ref{sec:hydro} we discuss the main processes 
	of particle transformations in
	a nucleon-hyperon matter and review the relativistic hydrodynamics
	of superfluid nucleon-hyperon mixtures.
In Section~\ref{sec:decoupling} we present an approximate method allowing one to decouple
	the equations describing superfluid and normal degrees of freedom
	and generalize it to allow for stellar rotation which
	substantially complicates the dynamics leading to the formation 
	of arrays of Feynman--Onsager vortices.
In Section~\ref{sec:sound}
	we test our decoupling scheme by the calculation 
	of the sound speeds in superfluid nucleon-hyperon matter
	and comparing them with the exact result.
In Section~\ref{sec:g-modes} we argue that this scheme cannot be used for the analysis of g-modes
	and calculate their spectrum for a one particular model of a HS.
Finally, we sum up in Section~\ref{sec:summary}.

%%%%%%%%%%%%%%%%%%%%%%%%%%%%%%%%%%%%%%%%%%%%%%%%%%%%%%%%%%%%%%%%%%%%%%%%%%%%%%
\section{Relativistic superfluid hydrodynamics of nucleon-hyperon mixture}
\label{sec:hydro}
%%%%%%%%%%%%%%%%%%%%%%%%%%%%%%%%%%%%%%%%%%%%%%%%%%%%%%%%%%%%%%%%%%%%%%%%%%%%%%

%%%%%%%%%%%%%%%%%%%%%%%%%%%%%%%%%%%%%%%%%%%%%%%%%%%%%%%%%%%%%%%%%%%%%%%%%%%%%%
\subsection{Definitions}
%%%%%%%%%%%%%%%%%%%%%%%%%%%%%%%%%%%%%%%%%%%%%%%%%%%%%%%%%%%%%%%%%%%%%%%%%%%%%%

In what follows we use the geometric system of units, 
	in which the gravitational constant $G$ 
	and the speed of light $c$ 
	are equal to unity, $G = c =1$.
A brief glossary of symbols 
	and the main definitions used in the paper
	are collected in Table 1.
	% \ref{tab:glossary}.

%%%%%%%%%%%%%%%%%%%%%%%%%%%%%%%%%%%%%%%%%%%%%%%%%%%%%%%%%%%%%%%%%%%%%%%%%%%%%%
\begin{table}
\label{tab:glossary}
\caption{A brief glossary of symbols}
	\begin{tabular}{ll}
	\hline
		$i,k = {\rm n, p, \Lambda, {\Xi^{-}}, {\Xi^{0}}, {\Sigma^{-}} }$ 	&	indices for baryons
		\\
		$l = {\rm e, \mu }$ 	&	indices for leptons
		\\
		$q_i, q_l$ 	&	electric charge of a given particle
		\\
		$\alpha, \beta, \gamma = 0,1,2,3$		&	spacetime indices %for four-vector and four-tensor components
		\\
		$g^{\alpha\beta}$	&	metric tensor
		\\
		$u^\alpha$ 	&	 four-velocity of normal fluid
		\\
		$v^\alpha_{{\rm sfl}(i)}$		&	 four-velocity of superfluid baryon species $i$
		\\
		$n_i, n_l$	&	number density for particles $i$, $l$
		\\
		$n_{\rm b} = \sum_{i} n_i $	&	baryon number density
		\\
		$\mu_i, \mu_l$	&	relativistic chemical potential for particles $i$, $l$
		\\
		$w^\alpha_{(i)} = \mu_i (v^\alpha_{{\rm sfl}(i)}- u^\alpha)$ 	&	 superfluid four-vector, convenient to use
		 instead of 	$v^\alpha_{{\rm sfl}(i)}$
		\\
		$Y_{ik}$		&  symmetric relativistic entrainment matrix
		\\
		$j_{(i)}^\alpha = n_{i} u^\alpha + Y_{ik} w^\alpha_{(k)}$ 	&	 four-current for baryon species $i$
		\\
		$j_{({\rm e})}^\alpha = n_{\rm e} u^\alpha$, $j_{({\rm \mu})}^\alpha = n_{\rm \mu} u^\alpha$ 	&	 four-currents for electrons and muons
		\\
		$j_{({\rm b})}^\alpha = \sum\limits_{i} j_{(i)}^\alpha \equiv n_{\rm b} U_{({\rm b})}^\alpha$	&	baryon four-current
		\\
		$U_{({\rm b})}^\alpha = j_{({\rm b})}^\alpha / n_{\rm b} $	&	baryon four-velocity
		\\
		$W^\alpha = U_{({\rm b})}^\alpha - u^\alpha$	& difference between baryon and normal four-velocities 
		\\
		$S_i	$		&	strangeness of particle $i$
		\\
		$n_{{{\rm str}}} = - \sum_{i} S_i n_i 
						= n_{\rm \Lambda} + 2 n_{{\rm \Xi}^{-}} + 2 n_{{\rm \Xi}^{0}} + n_{{\rm \Sigma}^{-}}
						$		
						&	(minus) strangeness number density
		\\
		$j_{({{\rm str}})}^\alpha = - \sum\limits_{i} S_i j_{(i)}^\alpha$	&	`strange' four-current
		\\
		$U_{({\rm str})}^\alpha = j_{({{\rm str}})}^\alpha / n_{{\rm str}}$
			& strangeness four-velocity
		\\
		$T^{\alpha\beta}$		&		energy-momentum tensor
		\\
		$\partial_\alpha X = \pd{}{x^\alpha}X$	&	partial derivative of a quantity $X$ (scalar, vector, or tensor)
		\\
		$X_{;\alpha}$		&	covariant derivative of a quantity $X$ (scalar, vector, or tensor)
		\\
	\hline
	\end{tabular}
\end{table}
%%%%%%%%%%%%%%%%%%%%%%%%%%%%%%%%%%%%%%%%%%%%%%%%%%%%%%%%%%%%%%%%%%%%%%%%%%%%%%

%%%%%%%%%%%%%%%%%%%%%%%%%%%%%%%%%%%%%%%%%%%%%%%%%%%%%%%%%%%%%%%%%%%%%%%%%%%%%%%%%%%%%%%
\subsection{Main processes of particle transformations in nucleon-hyperon matter}
%%%%%%%%%%%%%%%%%%%%%%%%%%%%%%%%%%%%%%%%%%%%%%%%%%%%%%%%%%%%%%%%%%%%%%%%%%%%%%%%%%%%%%%
\label{sec:reactions}

We consider a HS matter consisting of
	neutrons ($\rm n$), protons ($\rm p$),
	electrons ($\rm e$), muons ($\rm \mu$),
	as well as 
	$\rm \Lambda$, ${\rm \Xi}^{-}$, ${\rm \Xi}^0$ and ${\rm \Sigma}^-$--hyperons.
The most effective reactions in such a matter 
	are the following fast
	processes due to strong interaction of particles
	(see e.g., \citealt{2014MNRAS.439..318G}):
\begin{gather}
\label{eq:LL-nX0}
	{\rm \Lambda} + {\rm \Lambda} \leftrightarrow {\rm n} + {\rm \Xi} ^{0}
	,\\
\label{eq:LL-pXm}
	{\rm \Lambda} + {\rm \Lambda} \leftrightarrow {\rm p} + {\rm \Xi} ^{-}
	,\\
\label{eq:nX0-pXm}
	{\rm n} + {\rm \Xi} ^{0} \leftrightarrow {\rm p} + {\rm \Xi} ^{-}
	,\\
\label{eq:nL-pSm}
	{\rm n} + {\rm \Lambda} \leftrightarrow {\rm p} + {\rm \Sigma} ^{-}
	,\\
\label{eq:nXm-LSm}
	{\rm n} + {\rm \Xi} ^{-} \leftrightarrow {\rm \Lambda} + {\rm \Sigma} ^{-}
	,\\
\label{eq:LXm-X0Sm}
	{\rm \Lambda} + {\rm \Xi} ^{-} \leftrightarrow {\rm \Xi} ^{0} + {\rm \Sigma} ^{-}
.
\end{gather}
We assume that the perturbed matter is always 
	in equilibrium with respect to these reactions,
	which means
\begin{gather}
\label{eq:dmu-LLnX0}
2 \mu_\Lambda = \mu_{\rm n} + \mu_{{\rm \Xi}^{0}}
	,\\
\label{eq:dmu-LLpXm}
	2 \mu_\Lambda = \mu_{\rm p} + \mu_{{\rm \Xi}^{-}}
	,\\
\label{eq:dmu-nLpSm}
	\mu_{\rm n} + \mu_\Lambda = \mu_{\rm p} + \mu_{{\rm \Sigma}^{-}}
,
\end{gather}
where $\mu_i$ is the relativistic chemical potential for a particle species $i$.

The unperturbed matter is also in equilibrium
	with respect to a number of reactions
	due to weak interaction.
The latter include various Urca processes and weak nonleptonic reactions such as, 
	e.g.,~$n+n \rightarrow n + \Lambda$.
The corresponding conditions of chemical equilibrium
	(for the unperturbed matter only!)
	are 
\begin{gather}
\label{eq:dmu-Urca1}
	\mu_{{\rm n}} = \mu_{{\rm p}} + \mu_{{\rm e}}
,\\
\label{eq:dmu-Urca2}
	\mu_{{\rm n}} = \mu_{{\rm p}} + \mu_{{\rm\mu}}
,\\
\label{eq:dmu-nL}
	\mu_{{\rm n}} = \mu_{{\rm \Lambda}}
.
\end{gather}

%%%%%%%%%%%%%%%%%%%%%%%%%%%%%%%%%%%%%%%%%%%%%%%%%%%%%%%%%%%%%%%%%%%%%%%%%%%%%%
\subsection{Hydrodynamic equations}
%%%%%%%%%%%%%%%%%%%%%%%%%%%%%%%%%%%%%%%%%%%%%%%%%%%%%%%%%%%%%%%%%%%%%%%%%%%%%%

In this section we give a brief overview
	of the superfluid relativistic hydrodynamics 
	(see e.g., \citealt{2006MNRAS.372.1776G, 2013MNRAS.428.1518G} for details).
For definiteness, all baryons are assumed to be in superfluid state.
In what follows, the indices $i$ and $k$ are reserved for baryons,
$i$, $k = n$, $\rm p$, $\rm \Lambda$, ${\rm \Xi}^{-}$, ${\rm \Xi}^0$, ${\rm \Sigma}^-$, 
while the index $l$ is for leptons, $l = e$, $\rm \mu$.
Unless otherwise stated, 
a summation is assumed over the repeated space--time indices 
(Greek letters $\alpha, \beta, \gamma$, $\ldots$)
and particle indices (Latin letters).

In a superfluid matter a motion with few independent velocities is possible. 
These are the `superfluid' four-velocities $v^\alpha_{{\rm sfl}(i)}$
	describing the motion of baryon condensates%
%%%%%%%%%%
\footnote{To avoid any confusion, 
here by superfluid 
velocity we mean the quantity 
$v^\alpha_{{\rm sfl}(i)} \equiv \hbar \, \partial^{\alpha}\Phi_i/(2\mu_i)$,
where $\Phi_i$ is the phase of the Cooper-pair condensate wavefunction for particle species $i$,
and $\hbar$ is Planck's constant.}
%%%%%%%%%%
  (each can flow with its own velocity),
	as well as the `normal' four-velocity $u^\alpha$
	with which the `normal' (non-superfluid) baryon fraction
	and leptons move.
The latter velocity is normalized by the standard condition,
$u^{\alpha} u_{\alpha}=-1$.
Instead of $v^\alpha_{{\rm sfl}(i)}$
	it is often more convenient to use the four-vectors
	$w^\alpha_{(i)} \equiv \mu_i (v^\alpha_{{\rm sfl}(i)} - u^\alpha)$.
In terms of the quantities $u^\alpha$ and $w^\alpha_{(i)}$
	the particle density currents $j_{(i)}^\alpha$ can be represented as
\begin{gather}
\label{eq:j_i}
	j_{(i)}^\alpha = n_i u^\alpha + Y_{ik} w^\alpha_{(k)}
,\\
\label{eq:j_l}
	j_{(l)}^\alpha = n_l u^\alpha
,
\end{gather}
where $Y_{ik}$($=Y_{ki}$) is the relativistic entrainment matrix,
	which is a generalization of the concept of superfluid density
	to strongly interacting superfluid mixtures
	(see \citealt{2006MNRAS.372.1776G,
		2007PhRvD..76h3001G,
		2009PhRvC..79e5806G,
		2009PhRvC..80a5803G,
		2014MNRAS.439..318G}).
Generally, it is a function of the particle number densities $n_i$ and 
	ratios $T/T_{{\rm c}i}$, where $T$ is the temperature and $T_{{\rm c}i}$ is the critical
	temperature for transition of a particle species $i$ to superfluid state.

The system of hydrodynamic equations describing
	non-magnetized superfluid mixtures 
	is formulated below and includes the following.

(i) The continuity equation for baryons,
\begin{gather}
	\covd{j_{({\rm b})}^\alpha}{\alpha} = 0, \quad 
	j_{({\rm b})}^\alpha
	\equiv n_{\rm b} U_{(\rm b)}^\alpha
	= n_{\rm b} u^\alpha + \sum_i Y_{ik} w_{(k)}^\alpha,
\label{eq:continuity-b}
\end{gather}
where we introduce the baryon number density
	$n_{\rm b} = \sum_{i} n_i $
	and the baryon four-velocity 
	$U^\alpha_{(\rm b)} = u^\alpha+1/n_{\rm b} \, \sum_i Y_{ik} w_{(k)}^\alpha$.

(ii) The continuity equations for electrons, muons, and strangeness.
We assume that the weak processes of particle transformations are slow 
	on a typical hydrodynamic time-scale
	(see e.g., \citealt{2002A&A...394..213H} and references therein).
Hence, the corresponding continuity equations can be written as 
\begin{gather}
\label{eq:continuity}
  \covd{j_{({\rm e})}^\alpha}{\alpha} 
	= \covd{j_{({\rm \mu})}^\alpha}{\alpha}
	= \covd{j_{({\rm str})}^\alpha}{\alpha}
	= 0
,
\end{gather}
where $j_{({\rm str})}^\alpha
		\equiv n_{\rm str} U_{(\rm str)}^\alpha
		\equiv - \sum S_i j^{\alpha}_{(i)}$
	is the `strange' four-current
	and $S_i$ is the strangeness of particle species $i$.
Here we also introduced
	the (minus) strangeness number density 
	$n_{\rm str}
		= - \sum_{i} S_i n_i 
		= n_{\rm \Lambda} + 2 n_{{\rm \Xi}^{-}} + 2 n_{{\rm \Xi}^{0}} + n_{{\rm \Sigma}^{-}}$
	and the strangeness four-velocity
	$U_{({\rm str})}^\alpha
		= j_{({{\rm str}})}^\alpha / n_{{\rm str}}$.

(iii) Quasineutrality condition,
\begin{gather}
\label{eq:quasineutrality}
	q_i j_{(i)}^\alpha
		+ q_l j_{(l)}^\alpha
	= 0
,
\end{gather}
which implies the following two relations ($q_j$ is the electric charge of particle species $j$),
\begin{gather}
\label{eq:qn=0}
	q_i n_{i} + q_l n_{l} = 0
,\\
\label{eq:qYw=0}
	q_i Y_{ik} w_{(k)}^\alpha= 0
.
\end{gather}

(iv) Einstein equations
\begin{gather}
\label{eq:Einstein}
	R^{\alpha\beta} - \frac{1}{2} g^{\alpha\beta} R = 8\pi T^{\alpha\beta}
,
\end{gather}
where $R^{\alpha \beta}$, $R$, and $g_{\alpha \beta}$ are the Ricci tensor, the scalar curvature,
and the metric tensor, respectively;
$T^{\alpha \beta}$ is the energy-momentum tensor 
of superfluid matter,
\begin{gather}
\label{eq:Tab0}
	T^{\alpha\beta}=(P+\varepsilon) u^\alpha u^\beta + P g^{\alpha\beta}
		+ Y_{ik} \left[  w^\alpha_{(i)} w^\beta_{(k)} +\mu_i w^\alpha_{(k)} u^\beta + \mu_k w^\beta_{(i)} u^\alpha  \right]
\end{gather}
which satisfies energy-momentum conservation 
(compatible with equation~\ref{eq:Einstein}),
\begin{gather}
\label{eq:Tabb}
	{T^{\alpha\beta}}_{;\beta}=0
.
\end{gather}
In equation \eqref{eq:Tab0} $P$ is the pressure and $\varepsilon$ is the energy density.
For future purposes it is convenient to 
	rewrite the expression for $T^{\alpha\beta}$ 
	by making use of the chemical equilibrium conditions 
\eqref{eq:dmu-LLnX0}--\eqref{eq:dmu-nLpSm} and the quasineutrality condition \eqref{eq:qYw=0}:
\begin{multline}
\label{eq:Tab}
	T^{\alpha\beta}
		= (P+\varepsilon) u^\alpha u^\beta + P g^{\alpha\beta}
		+ \mu_{\rm n} n_{\rm b} \left( W^\alpha u^\beta + W^\beta u^\alpha \right)
		%+ \text{(quadratically small terms)}
		\\- \underline{
			\left(
				\Delta\mu_{\rm \Lambda} Y_{{\rm \Lambda} k}
				+ 2 \Delta\mu_{\rm \Lambda} Y_{{\rm \Xi}^- k}
				+ 2 \Delta\mu_{\rm \Lambda} Y_{{\rm \Xi}^0 k}
				+ \Delta\mu_{\rm \Lambda} Y_{{\rm \Sigma}^- k}
			\right)
			\left(
				w^\alpha_{(k)} u^\beta + w^\beta_{(k)} u^\alpha
			\right)
			}
		+ \underline{ Y_{ik} w^\alpha_{(i)} w^\beta_{(k)} }
,
\end{multline}
where
	$W^\alpha \equiv 1 / n_{\rm b} \sum_i Y_{ik} w_{(k)}^\alpha$
	and $\Delta \mu_{\Lambda} \equiv \mu_{\rm n}-\mu_\Lambda$.

(v) The equation stating that the motion
	of superfluid species $i$ is purely potential
	(a more general equation describing rotating superfluids, containing 
	Onsager-Feynman vortices, is discussed in Section~\ref{sec:decoupling:rotation}):
\begin{gather}
\label{eq:potentiality}
	\covd{\left(
			w_{(i)\alpha} + \mu_i u_\alpha + q_i A_\alpha
		\right)}{\beta}
	- \covd{\left( 
			w_{(i)\beta} + \mu_i u_\beta + q_i A_\beta
		\right)}{\alpha}
	= 0
,
\end{gather}
where $A_\alpha$ is the four-potential of the electromagnetic field.

The hydrodynamic equations given above should be supplemented
by the definition of the comoving frame in which we measure (define)
such thermodynamic quantities as $n_i$, $P$, $\varepsilon$, etc.
Below we define the comoving frame as the frame in which 
$u^\alpha = \left(1,0,0,0\right)$. 
This imposes a number of conditions
on 
$j_{(i)}^\alpha$, $T^{\alpha\beta}$, and $w^\alpha_{(i)}$,
\begin{gather}
	u_\alpha j_{(i)}^\alpha = - n_i	
	,\\
	u_\alpha u_\beta T^{\alpha\beta} = \varepsilon
	,\\
	u_\alpha w^\alpha_{(i)} = 0
.
\label{eq:comoving}
\end{gather}

The thermodynamic quantities in equations 
	\eqref{eq:continuity-b}--\eqref{eq:comoving} 
	are related by the following well-known conditions
	(see e.g., \citealt{1980stph.book.....L};
		the last term in equations~\ref{eq:dE} and \ref{eq:dP}
		arises due to superfluidity,
		see e.g.,
		\citealt{2006MNRAS.372.1776G} for details):
\begin{gather}
\label{eq:P-E}
	P + \varepsilon = \mu_i n_i + \mu_l n_l + TS
,\\
\label{eq:dE}
	{\rm d} \varepsilon = \mu_i {\rm d} n_i + \mu_l {\rm d} n_l 
		+ T {\rm d} S + \frac{Y_{ik}}{2} {\rm d} \left[ w_{(i)}^\alpha w_{(k)\alpha}  \right]
,\\
\label{eq:dP}
		{\rm d} P = n_i {\rm d} \mu_i + n_l {\rm d} \mu_l
			+ S {\rm d} T - \frac{Y_{ik}}{2} {\rm d} \left[ w_{(i)}^\alpha w_{(k)\alpha}  \right]
.
\end{gather}

These equations can be conveniently presented in the form
\begin{gather}
\label{eq:thermodyn1}
	P + \varepsilon
		= \mu_{\rm n} n_{\rm b}
			- \Delta\mu_{\rm e} n_{\rm e}
			- \Delta\mu_{\rm \mu} n_{\rm \mu} 
			- \Delta\mu_{\rm \Lambda} n_{\rm str} 
			+ \underline{TS}
,\\
\label{eq:thermodyn2}
	{\rm d} \varepsilon =  \mu_{\rm n} {\rm d} n_{\rm b}
		- \Delta\mu_{\rm e} {\rm d} n_{\rm e} - \Delta\mu_{\rm \mu} {\rm d} n_{\rm \mu} - \Delta\mu_{\rm \Lambda} {\rm d} n_{\rm str} 
		+ \underline{T {\rm d} S}
		+ \underline{\frac{Y_{ik}}{2} {\rm d} \left[ w_{(i)}^\alpha w_{(k)\alpha}  \right]}
,\\
\label{eq:thermodyn3}
	{\rm d} P =  n_{\rm b} {\rm d} \mu_{\rm n}
		- n_{\rm e} {\rm d} \Delta\mu_{\rm e} - n_{\rm \mu} {\rm d} \Delta\mu_{\rm \mu} - n_{\rm str} {\rm d} \Delta\mu_{\rm \Lambda}
		+ \underline{S {\rm d} T}
		- \underline{\frac{Y_{ik}}{2} {\rm d} \left[ w_{(i)}^\alpha w_{(k)\alpha}  \right]}
,
\end{gather}
where $\Delta\mu_{\rm e} \equiv \mu_{\rm n} - \mu_{\rm p} - \mu_{\rm e}$ and 
$\Delta\mu_{\rm \mu} \equiv \mu_{\rm n} - \mu_{\rm p} - \mu_\mu$.

%%%%%%%%%%%%%%%%%%%%%%%%%%%%%%%%%%%%%%%%%%%%%%%%%%%%%%%%%%%%%%%%%%%%%%%%%%%%%%%%%%%%%%%%%
\subsection{Superfluid degrees of freedom}
\label{sec:sfl-degrees-of-freedom}
%%%%%%%%%%%%%%%%%%%%%%%%%%%%%%%%%%%%%%%%%%%%%%%%%%%%%%%%%%%%%%%%%%%%%%%%%%%%%%%%%%%%%%%%%

Let us inspect a number of independent superfluid degrees of freedom in our problem.
The potentiality equations
	\eqref{eq:potentiality} (with $\beta = 0$)
	along with the chemical equilibrium conditions \eqref{eq:dmu-LLnX0}--\eqref{eq:dmu-nLpSm}
	result in the three equations connecting six superfluid four-velocities $w_{(i)\alpha}$:
\begin{gather}
\label{eq:w-LLpXm-nonlin}
	\pd{}{t} \left[
		2 w_{({\rm \Lambda})\alpha} - w_{({\rm p})\alpha} - w_{({\rm \Xi}^{-})\alpha}
	\right]
	=
	\pd{}{x^\alpha} \left[
		2 w_{({\rm \Lambda})0} - w_{({\rm p})0} - w_{({\rm \Xi}^{-})0}
	\right]
,\\
\label{eq:w-LLnX0-nonlin}
	\pd{}{t} \left[
		2 w_{({\rm \Lambda})\alpha} - w_{({\rm n})\alpha} - w_{({\rm \Xi}^{0})\alpha}
	\right]
	=
	\pd{}{x^\alpha} \left[
		2 w_{({\rm \Lambda})0} - w_{({\rm n})0} - w_{({\rm \Xi}^{0})0}
	\right]
,\\
\label{eq:w-nLpSm-nonlin}
	\pd{}{t} \left[
		w_{({\rm n})\alpha} + w_{({\rm \Lambda})\alpha} - w_{({\rm p})\alpha} - w_{({\rm \Sigma}^{-})\alpha}
	\right]
	=
	\pd{}{x^\alpha} \left[
		w_{({\rm n})0} + w_{({\rm \Lambda})0} - w_{({\rm p})0} - w_{({\rm \Sigma}^{-})0}
	\right]
,
\end{gather}
	which, in the case of small harmonic perturbations
	(when $w_{(i)\alpha} \propto {\rm e}^{{\rm i}\omega t}$
	and $w_{(i)0} = 0$,
	see Section~\ref{sec:decoupling:1} below)
	reduce to a set of simple algebraic relations:
\begin{gather}
\label{eq:w-LLpXm}
	2 w_{({\rm \Lambda})\alpha} = w_{({\rm p})\alpha} + w_{({\rm \Xi}^{-})\alpha}
	,\\
\label{eq:w-LLnX0}
	2 w_{({\rm \Lambda})\alpha} = w_{({\rm n})\alpha} + w_{({\rm \Xi}^{0})\alpha}
	,\\
\label{eq:w-nLpSm}
	w_{({\rm n})\alpha} + w_{({\rm \Lambda})\alpha} = w_{({\rm p})\alpha} + w_{({\rm \Sigma}^{-})\alpha}
.
\end{gather}
The quasineutrality condition \eqref{eq:qYw=0}
	provides one more relation.
Consequently, only $6-4=2$ superfluid four-vectors 
	(e.g., $w_{({\rm n})\alpha}$ and $w_{({\rm \Lambda})\alpha}$) are independent. 
Thus, there are only two superfluid degrees of freedom in the problem.

The same analysis can be performed for other cases, 
	when some particles are absent or non-superfluid.
Namely, it can be shown that, if the thresholds for the appearance of hyperons $n_{({\rm b})}^{(i)}$ 
	satisfy the inequality
	$n_{({\rm b})}^{({\rm \Lambda})} < n_{({\rm b})}^{({\rm \Xi}^{-})} < n_{({\rm b})}^{({\rm \Xi}^{0})} < n_{({\rm b})}^{({\rm \Sigma}^{-})}$
	(which is true for all the EOSs GM1A, GM1`B and TM1C studied here),
	then in each case there are no more than two superfluid degrees of freedom.
Three degrees of freedom arise only
	in the (nonrealistic) situation, when
	at some density ${\rm \Sigma}^-$--hyperons
	as well as ${\rm \Xi}^-$-- and/or ${\rm \Xi}^0$--hyperons
	are present while $\rm \Lambda$--hyperons are absent.

%%%%%%%%%%%%%%%%%%%%%%%%%%%%%%%%%%%%%%%%%%%%%%%%%%%%%%%%%%%%%%%%%%%%%%%%
\section{Decoupling of superfluid and normal equations}
\label{sec:decoupling}
%%%%%%%%%%%%%%%%%%%%%%%%%%%%%%%%%%%%%%%%%%%%%%%%%%%%%%%%%%%%%%%%%%%%%%%%
%%%%%%%%%%%%%%%%%%%%%%%%%%%%%%%%%%%%%%%%%%%%%%%%%%%%%%%%%%%%%%%%%%%%%%%%
\subsection{Equilibrium four-vectors $u^{\alpha}$ and $w^{\alpha}_{(i)}$ and
	small deviations from equilibrium}
\label{sec:decoupling:1}
%%%%%%%%%%%%%%%%%%%%%%%%%%%%%%%%%%%%%%%%%%%%%%%%%%%%%%%%%%%%%%%%%%%%%%%%

We assume that deviations from the equilibrium are small,
	so that one can use linearized hydrodynamic equations
	to study a perturbed nucleon-hyperon matter
	of HSs.
We further assume that in equilibrium 
	the superfluid components comove with the normal (non-superfluid) liquid component, i.e.,
	$w^{1}_{(i)} = w^{2}_{(i)} = w^{3}_{(i)} = 0$ (\citealt{2006MNRAS.372.1776G}).
Finally, everywhere except in Section~\ref{sec:decoupling:rotation} 
	we assume that the normal component of the star is at rest, 
	$u^{\alpha} = (u^0,0,0,0)$.
	[In Section~\ref{sec:decoupling:rotation} we briefly discuss the case of a rotating HS,
	for which $u^{\alpha} = (u^0,0,0,u^{\phi})$.]
From the condition \eqref{eq:comoving} it then follows that $w^0_{(i)}=0$ 
for both rotating and non-rotating stellar configurations, so that 
all the components of the four-vectors $w^{\alpha}_{(i)}$ vanish in equilibrium, 
$w^{\alpha}_{(i)}=0$.
A perturbation of an arbitrary quantity $A$ 
	from its equilibrium value 
	will be denoted as $\delta A$.
Note that this notation will not be used for the four-vectors $w^{\alpha}_{(i)}$
	and scalars $\Delta \mu_{\rm e}$, $\Delta \mu_{\rm \mu}$, and $\Delta \mu_{\rm \Lambda}$
	since $\delta w^{\alpha}_{(i)}=w^{\alpha}_{(i)}$, $\delta \Delta \mu_{\rm e}=\Delta \mu_{\rm e}$, etc.\
	(remember that 
		$\Delta\mu_{\rm e} = \Delta\mu_{\rm \mu} = \Delta\mu_{\rm \Lambda} = 0$
		in equilibrium,
		see equations~\ref{eq:dmu-Urca1}--\ref{eq:dmu-nL}).

We will further use a simplified version of 
	equations \eqref{eq:thermodyn1}--\eqref{eq:thermodyn3}
	by noticing that, in a strongly degenerate matter, 
	one can neglect small temperature-dependent terms
	$TS$, $T {\rm d} S$, and $S {\rm d} T$ there. 
We shall also neglect the quadratically small terms 
	in equations \eqref{eq:Tab}, \eqref{eq:thermodyn2}, and \eqref{eq:thermodyn3}
	which depend on the superfluid four-vectors $w^{\alpha}_{(i)}$.
Overall, all the underlined terms 
	in equations \eqref{eq:Tab} and \eqref{eq:thermodyn1}--\eqref{eq:thermodyn3}
	will be neglected.

%%%%%%%%%%%%%%%%%%%%%%%%%%%%%%%%%%%%%%%%%%%%%%%%%%%%%%%%%%%%%%%%%%%%%%%%
\subsection{Normal equations and coupling parameters}
%%%%%%%%%%%%%%%%%%%%%%%%%%%%%%%%%%%%%%%%%%%%%%%%%%%%%%%%%%%%%%%%%%%%%%%%

In the linear approximation a perturbation $\delta T^{\alpha\beta}$ 
	of the energy-momentum tensor \eqref{eq:Tab}
	can be rewritten as
\begin{gather}
\label{eq:dTab}
	\delta T^{\alpha\beta}
	=
	(\delta P + \delta\varepsilon) U_{({\rm b})}^\alpha U_{({\rm b})}^\beta
	+ (P + \varepsilon)
		\left[
			U_{({\rm b})}^\alpha \delta U_{({\rm b})}^\beta + U_{({\rm b})}^\beta \delta U_{({\rm b})}^\alpha
		\right]
	+ \delta P g^{\alpha\beta} + P \delta g^{\alpha\beta}
,
\end{gather}
	where the quantities $U_{({\rm b})}^\alpha$, $P$, $\varepsilon$, and $g^{\alpha\beta}$ 
	are taken in equilibrium
	(note that in equilibrium $U_{({\rm b})}^\alpha = u^\alpha$).

If $\delta T^{\alpha\beta}$ does not depend on superfluid degrees of freedom,
	then the system of hydrodynamic equations contains a subsystem that coincides
	with the equations of ordinary (non-superfluid) hydrodynamics.
Let us find an approximation which leads to this case.
One can describe perturbations in superfluid
	$\rm npe\mu\Lambda\Xi^{-}\Xi^{0}\Sigma^{-}$ matter 
	with the following independent `normal' variables
	$\delta g^{\alpha\beta}$, $\delta U_{({\rm b})}^\alpha$
	and `superfluid' variables $w_{(i)}^\alpha$
	(e.g., $w_{({\rm n})}^\alpha$ and $w_{({\rm \Lambda})}^\alpha$).

Using the continuity equations
	\eqref{eq:continuity-b} and \eqref{eq:continuity},
	one can schematically write
\begin{gather}
\label{eq:dnb}
	\delta n_{\rm b}
		= \delta n_{\rm b} (\delta U_{({\rm b})}^\alpha, \delta g^{\alpha\beta})
	,\\
	\delta n_{\rm e}
		= \delta n_{\rm e} (\delta u^\alpha, \delta g^{\alpha\beta})
		= \delta n_{\rm e} (\delta U_{({\rm b})}^\alpha,\delta g^{\alpha\beta}, w_{(i)}^\alpha)
	,\\
	\delta n_{\rm \mu}
		= \delta n_{\rm \mu} (\delta u^\alpha, \delta g^{\alpha\beta})
		= \delta n_{\rm \mu} (\delta U_{({\rm b})}^\alpha, \delta g^{\alpha\beta}, w_{(i)}^\alpha)
	,\\
	\delta n_{\rm str}
		= \delta n_{\rm str} (\delta U_{({\rm str})}^\alpha, \delta g^{\alpha\beta})
		= \delta n_{\rm str} (\delta U_{({\rm b})}^\alpha, \delta g^{\alpha\beta}, w_{(i)}^\alpha)
	,
\end{gather}
	where, for example, the first equation means that 
	the perturbation $\delta n_{\rm b}$ of baryon number density $n_{\rm b}$ 
	can be expressed through (depends on) 
	the perturbations $\delta U_{({\rm b})}^\alpha$ and $\delta g^{\alpha\beta}$.
Now let us split $\delta n_{\rm e}$
	into the sum of two terms,
	$\delta n_{{\rm e}({\rm norm})}$
	and $\delta n_{{\rm e}({\rm SFL})}$,
	which depend on normal and superfluid degrees of freedom,
	respectively,
\begin{gather}
	\delta n_{\rm e} (\delta U_{({\rm b})}^\alpha,\delta g^{\alpha\beta}, w_{(i)}^\alpha)
		= \delta n_{{\rm e}({\rm norm})}
			(\delta U_{({\rm b})}^\alpha,\delta g^{\alpha\beta})
		  + \delta n_{{\rm e}({\rm SFL})}
				(w_{(i)}^\alpha)
	,
\end{gather}
and do the same for $\delta n_{\rm \mu}$ and $\delta n_{\rm str}$,
\begin{gather}
	\delta n_{\rm \mu} (\delta U_{({\rm b})}^\alpha,\delta g^{\alpha\beta}, w_{(i)}^\alpha)
		= \delta n_{{\rm \mu}({\rm norm})}
			(\delta U_{({\rm b})}^\alpha,\delta g^{\alpha\beta})
		  + \delta n_{{\rm \mu}({\rm SFL})}
				(w_{(i)}^\alpha)
,\\
	\delta n_{\rm str} (\delta U_{({\rm b})}^\alpha,\delta g^{\alpha\beta}, w_{(i)}^\alpha)
		= \delta n_{{\rm str}({\rm norm})}
			(\delta U_{({\rm b})}^\alpha,\delta g^{\alpha\beta})
		  + \delta n_{{\rm str}({\rm SFL})}
				(w_{(i)}^\alpha)
	.
\end{gather}
Any thermodynamic quantity (e.g. $\varepsilon$ or $P$)
	in a degenerate matter can be presented as a function of
	$(n_{\rm b}, n_{\rm e}, n_{\rm \mu}, n_{\rm str})$,
	hence its perturbation is known function of
	$(\delta n_{\rm b}, \delta n_{\rm e}, \delta n_{\rm \mu}, \delta n_{\rm str})$
	or $(\delta U_{({\rm b})}^\alpha, \delta g^{\alpha\beta}, w_{(i)}^\alpha)$.

Guided by this observation,
	let us express $\delta \varepsilon$ and $\delta P$
	through the perturbations of number densities,
\begin{gather}
\label{eq:coupling:de}
	\delta \varepsilon = \mu_{\rm n} \delta n_{\rm b}
,\\
\label{eq:coupling:dP}
\begin{split}
	\frac{\delta P}{P}
		= & \pd{\ln P(n_{\rm b},n_{\rm e},n_{\rm \mu},n_{\rm str})}{\ln n_{\rm b}}
		\left(
			\frac{\delta n_{\rm b}}{n_{\rm b}}
		\right.
			+ \frac{\tilde{s}_{\rm e} \delta n_{\rm e (norm)} + s_{\rm e} \delta n_{\rm e (SFL)}}{n_{\rm e}} 
			+ \frac{\tilde{s}_{\rm \mu} \delta n_{\rm \mu (norm)} + s_{\rm \mu} \delta n_{\rm \mu (SFL)}}{n_{\rm \mu}} 
			\\
			& \left.
			+ \frac{\tilde{s}_{\rm str} \delta n_{\rm str (norm)} + s_{\rm str} \delta n_{\rm str (SFL)}}{n_{\rm str}} 
		\right)
,
\end{split}
\end{gather}
where
\begin{gather}
\label{eq:ss}
	\tilde{s}_{\rm e} = s_{\rm e} = \frac{\partial \ln P / \partial \ln n_{\rm e}}{\partial \ln P / \partial \ln n_{\rm b}}
	,\quad
	\tilde{s}_{\rm\mu} = s_{\rm\mu} = \frac{\partial \ln P / \partial \ln n_{\rm \mu}}{\partial \ln P / \partial \ln n_{\rm b}}
	,\quad
	\tilde{s}_{\rm str} = s_{\rm str} = \frac{\partial \ln P / \partial \ln n_{\rm str}}{\partial \ln P / \partial \ln n_{\rm b}}
.
\end{gather}
To obtain equation~\eqref{eq:coupling:de}
	we used equation~\eqref{eq:thermodyn2}
	and neglected quadratically small terms
	$\Delta\mu_{\rm e} \delta n_{\rm e}$,
	$\Delta\mu_{\rm \mu} \delta n_{\rm \mu}$,
	and $\Delta\mu_{\rm \Lambda} \delta n_{\rm str}$.
In equations \eqref{eq:coupling:dP} and \eqref{eq:ss} 
	we introduced the `electron', `muon' and `strange' coupling parameters
	$s_{\rm e}$, $s_{\rm \mu}$, and $s_{\rm str}$, respectively,
	and the quantities $\tilde{s}_{\rm e}$, $\tilde{s}_{\rm \mu}$, and $\tilde{s}_{\rm str}$.
We discriminate between the parameters
	$s_{\rm e}$ and $\tilde{s}_{\rm e}$,
	$s_{\rm \mu}$ and $\tilde{s}_{\rm \mu}$,
	or $s_{\rm str}$ and $\tilde{s}_{\rm str}$
	due to purely technical reasons:
	it turns out to be convenient
	to develop a perturbation theory in parameters
	$s_{\rm e}$, $s_{\rm \mu}$, and $s_{\rm str}$
	while treating the terms depending on
	$\tilde{s}_{\rm e}$, $\tilde{s}_{\rm \mu}$, and $\tilde{s}_{\rm str}$
	in a non-perturbative way
	(see a discussion in the sections 5 and 6
	in \citealt{2013MNRAS.428.1518G}).
Let us assume for a moment that all the coupling parameters vanish,
	$s_{\rm e} = s_{\rm \mu} = s_{\rm str} = 0$ 
	(hereafter such an approximation will be called `decoupling approximation'). 
In that case
	$\delta T^{\alpha\beta} = \delta T^{\alpha\beta} (\delta U_{({\rm b})}^\alpha, \delta g^{\alpha\beta})$
	does not depend on the superfluid degrees of freedom $w_{(i)}^\alpha$ 
	and has exactly the same form as in the absence of superfluidity.
This means that the perturbed Einstein equation \eqref{eq:Einstein},
	$\delta \left( R^{\alpha\beta} - \frac{1}{2} g^{\alpha\beta} R \right)= 8\pi \delta T^{\alpha\beta}$,
	also does not depend on $w_{(i)}^\alpha$ and hence coincides with the corresponding equations for normal matter.
Solving these equations one can obtain `normal' oscillation modes of a non-superfluid star. 
However, if a star oscillates on a frequency
	which does not coincide with any of the `normal' eigenfrequencies,
	then the eigenfunctions
	$\delta U_{({\rm b})}^\alpha$
	and $\delta g^{\alpha\beta}$
	must vanish,
	$\delta U_{({\rm b})}^\alpha = \delta g^{\alpha\beta} = 0$
	(this also implies $\delta P = \delta n_{\rm b} = 0$, 
		see equations \ref{eq:dnb} and \ref{eq:coupling:dP}
		with $s_{\rm e} = s_{\rm \mu} = s_{\rm str} = 0$),
	which means that perturbations are described
	with superfluid variables $w_{(i)}^\alpha$ only. 
Solving `superfluid' equations (see Section~\ref{sec:decoupling-sfl} below) 
	one can obtain eigenfrequencies and eigenfunctions for superfluid modes.

\begin{figure}
\centering
\begin{minipage}{.32\textwidth}
  \centering
  \includegraphics[width=1.\linewidth]{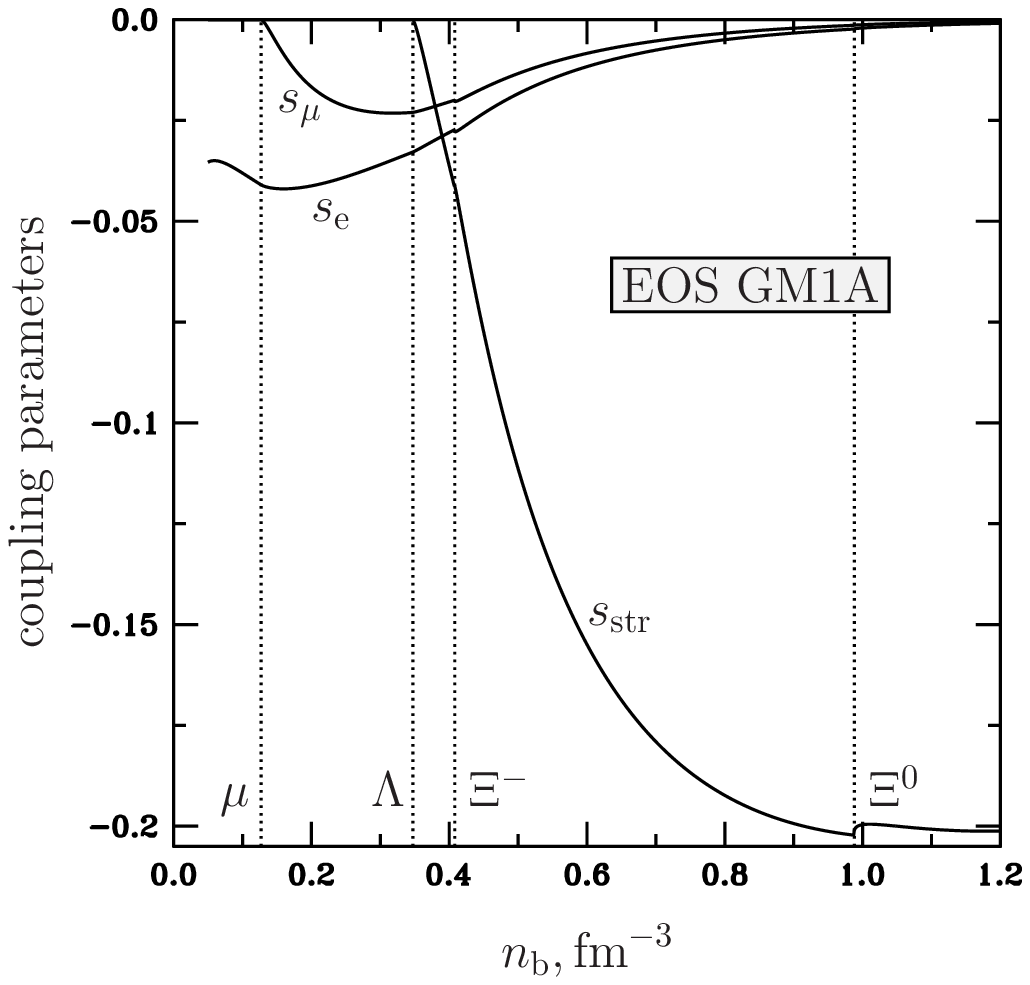}
  \label{fig:coupling:GM1A}
\end{minipage}
\begin{minipage}{.32\textwidth}
  \centering
  \includegraphics[width=1.\linewidth]{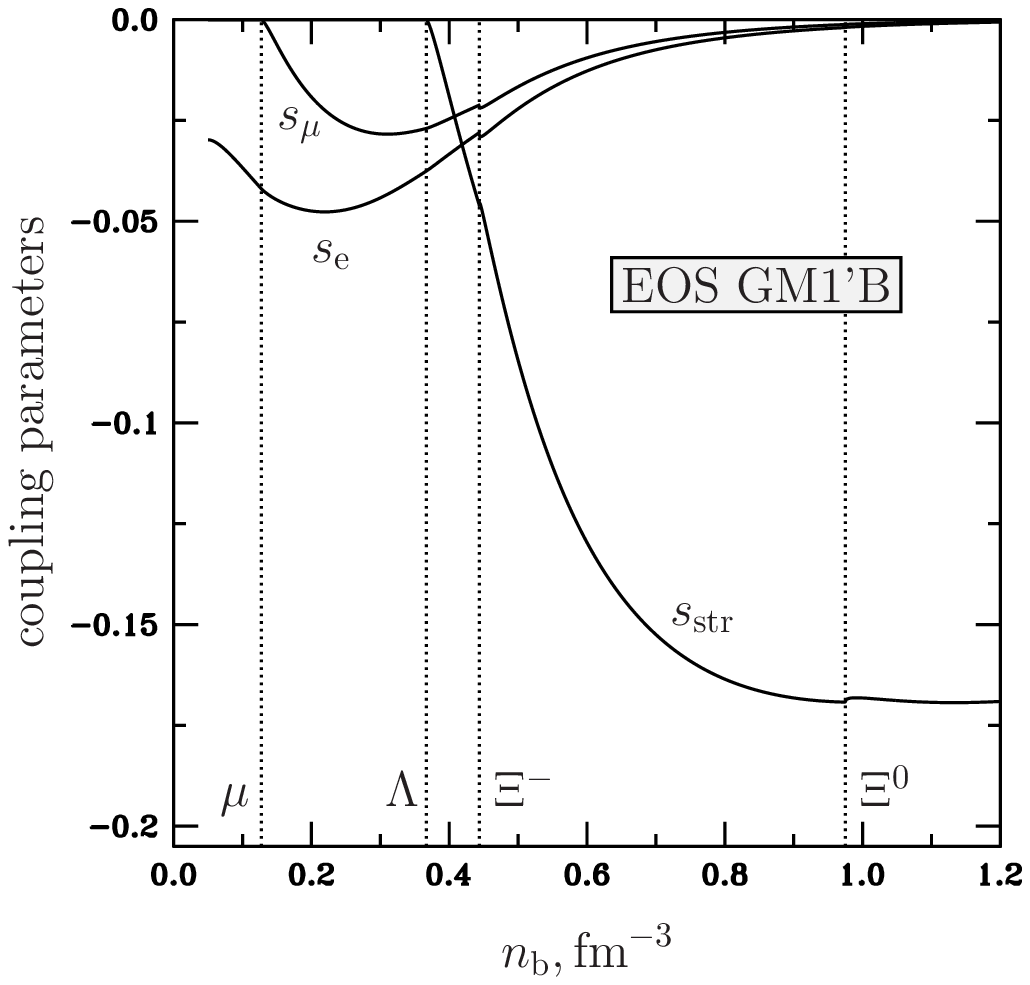}
	\label{fig:coupling:GM1B}
\end{minipage}
\begin{minipage}{.32\textwidth}
  \centering
  \includegraphics[width=1.\linewidth]{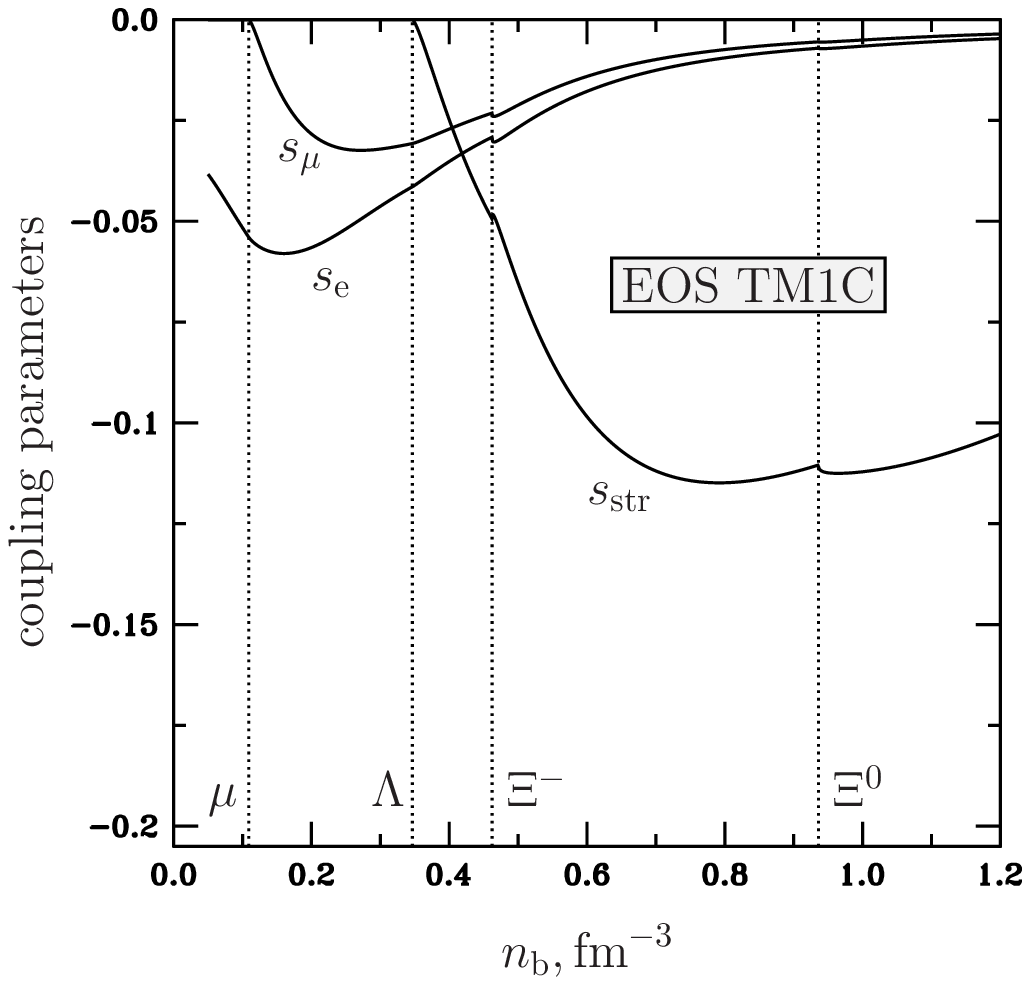}
	\label{fig:coupling:TM1C}
\end{minipage}
\caption{
	Coupling parameters
		$s_{\rm e},~s_{\rm \mu},~s_{\rm str}$
		versus the baryon number density $n_{\rm b}$
		for the EOSs GM1A, GM1'B, TM1C from \citet{2014MNRAS.439..318G}.
	Vertical lines are the thresholds for the appearance
		of muons and $\rm \Lambda$--, ${\rm \Xi}^{-}$--, ${\rm \Xi}^0$--hyperons.
}
\label{fig:coupling}
\end{figure}

If the coupling parameters
	$s_{\rm e},~s_{\rm \mu}$, and $s_{\rm str}$
	are small but finite, then superfluid and normal modes remain approximately decoupled.
These parameters
	are plotted in Fig.~\ref{fig:coupling}
	for the realistic hyperonic EOSs GM1A, GM1'B, TM1C
	from \citet{2014MNRAS.439..318G}.
One sees that the absolute value
	of the largest coupling parameter, $s_{\rm str}$,
	generally does not exceed $0.2$.
Since $s_{\rm str}$ is smaller at low densities,
	one can conclude that decoupling approximation
	works better for low-mass stars.
%%%%%%%%%%%%%%%%%%%%%%%%%%%%%%%%%%%%%%%%%%%%%%%%%%%%%%%%%%%%%%%%%%%
%%%%%%%%%%%%%%%%%%%%%%%%%%%%%%%%%%%%%%%%%%%%%%%%%%%%%%%%%%%%%%%%%%%

%%%%%%%%%%%%%%%%%%%%%%%%%%%%%%%%%%%%%%%%%%%%%%%%%%%%%%%%%%%%%%%%%%%
\subsection{Superfluid equations}
\label{sec:decoupling-sfl}
%%%%%%%%%%%%%%%%%%%%%%%%%%%%%%%%%%%%%%%%%%%%%%%%%%%%%%%%%%%%%%%%%%%
Assuming that all the coupling parameters vanish,
	one can, in principle, 
	study superfluid oscillation modes using 
	the potentiality conditions
	for the motion of superfluid components
	\eqref{eq:potentiality}
	together with the continuity equations \eqref{eq:continuity} 
	and the condition $\delta U_{({\rm b})}^\alpha = \delta g^{\alpha\beta} = 0$ 
	(as it is discussed in the previous section).
However, if the coupling parameters are small but finite 
	(which is the case for realistic EOSs),
	such an approach will lead to significant errors
	(see details in Appendix \ref{sec:appendix-decoupling}).
In this section we derive a set of equations which are more suitable for our decoupling scheme.
These equations are generalization 
	of the superfluid equation discussed by \citet{2011PhRvD..83h1304G}.
To obtain them,	we follow the derivation of that paper.

Using the energy-momentum conservation \eqref{eq:Tabb},
	one can compose a vanishing combination
	$\covd{{T_\alpha}^\beta}{\beta} + u_\alpha u_\gamma \covd{T^{\gamma\beta}}{\beta} = 0$.
Subtracting from it the potentiality condition \eqref{eq:potentiality} 
	for neutrons multiplied by $n_{\rm b} u^\beta$,
	one obtains
\begin{multline}
\label{eq:sfl-osc-general}
	(P+\varepsilon - \mu_{\rm n} n_{\rm b}) u^\beta \covd{u_\alpha}{\beta}
		+ \left(
				\partial_\beta{P} - n_{\rm b} \partial_\beta{\mu_{\rm n}}
			\right) u_\alpha u^\beta
		+ \left(
				\partial_\alpha{P} - n_{\rm b} \partial_\alpha{\mu_{\rm n} }
			\right)
		\\
		+ (g_{\alpha\gamma} + u_\alpha u_\gamma ) u^\beta \covd{( \mu_{\rm n} n_{\rm b} W^\gamma )}{\beta}
		+\mu_{\rm n} n_{\rm b}
		\left(
			\covd{{u^\beta}}{\beta}  W_\alpha
			+ \covd{u_{\alpha}}{\beta} W^\beta
		\right)
	- n_{\rm b} u^\beta
	\left[
		\covd{w_{({\rm n})\alpha}}{\beta}
		-\covd{w_{({\rm n})\beta}}{\alpha}
	\right]
	= 0
,
\end{multline}
or, using the thermodynamic relations \eqref{eq:thermodyn1} and \eqref{eq:thermodyn3},
\begin{multline}
\label{eq:sfl-osc-general2}
		\left(
			- \Delta\mu_{\rm e} n_{\rm e}  - \Delta\mu_{\rm \mu} n_{\rm \mu} - \Delta\mu_{\rm \Lambda} n_{\rm str} 
			\right) u^\beta \covd{u_\alpha}{\beta}
		+ \left(
			- n_{\rm e} \partial_\beta{\Delta\mu_{\rm e}} - n_{\rm \mu} \partial_\beta{\Delta\mu_{\rm \mu}} - n_{\rm str} \partial_\beta{\Delta\mu_{\rm \Lambda}}
			\right) u_\alpha u^\beta
		\\+ \left(
				- n_{\rm e} \partial_\alpha{\Delta\mu_{\rm e}}- n_{\rm \mu} \partial_\alpha{\Delta\mu_{\rm \mu}} - n_{\rm str} \partial_\alpha{\Delta\mu_{\rm \Lambda}}
			\right)
		%+\\
		+ (g_{\alpha\gamma} + u_\alpha u_\gamma ) u^\beta \covd{( \mu_{\rm n} n_{\rm b} W^\gamma )}{\beta}
		+\mu_{\rm n} n_{\rm b}
		\left(
			\covd{{u^\beta}}{\beta}  W_\alpha
			+ \covd{u_{\alpha}}{\beta} W^\beta
		\right)
	\\- n_{\rm b} u^\beta
	\left[
		\covd{w_{({\rm n})\alpha}}{\beta}
		-\covd{w_{({\rm n})\beta}}{\alpha}
	\right]
	= 0
.
\end{multline}

Each term in equation \eqref{eq:sfl-osc-general2}
	depends on one of the small quantities
	$\Delta\mu_{\rm e}$, $\Delta\mu_{\rm \mu}$,
	$\Delta\mu_{\rm \Lambda}$,
	$w_{({\rm n})}^\alpha$ or $W^\alpha$. 
Thus, since we are working in the linear approximation, one can replace all other quantities 
in this equation with their equilibrium values.

Now let us consider a non-rotating equilibrated star with the Schwarzschild metric,
\begin{gather}
\label{eq:Schwarzschild}
	{\rm d}s^2 = - {\rm e}^\nu {\rm d}t^2 + {\rm e}^\lambda {\rm d}r^2 + r^2 ({\rm d}\theta^2 + \sin^2 \theta {\rm d}\phi^2),
\end{gather}
and assume that all its perturbations depend on time as ${\rm e}^{{\rm i}\omega t}$ 
($\omega$ is the perturbation frequency).
In this case the spatial components ($\alpha = 1,\, 2,\, 3$) 
	of the superfluid equation take a very simple form
\begin{gather}
\label{eq:sfl-osc-n}
	{\rm i} \omega n_{\rm b} ( \mu_{\rm n} W_\alpha - w_{({\rm n})\alpha})
	= n_{\rm e}     \pd{}{x^\alpha}\left(\Delta\mu_{\rm e} {\rm e}^{\nu/2} \right)
	+ n_{\rm \mu} \pd{}{x^\alpha}\left( \Delta\mu_{\rm \mu} {\rm e}^{\nu/2} \right)
	+ n_{\rm str} \pd{}{x^\alpha}\left( \Delta\mu_{\rm \Lambda} {\rm e}^{\nu/2} \right)
	,\quad \alpha = 1,2,3
.
\end{gather}
In a similar way
	(using the potentiality condition for $\rm \Lambda$--hyperons instead of neutrons)
	one can derive an equation for $\rm \Lambda$--hyperons,
\begin{gather}
\label{eq:sfl-osc-L}
	{\rm i} \omega n_{\rm b} ( \mu_{\rm n} W_\alpha - w_{({\rm \Lambda})\alpha})
	= n_{\rm e}     \pd{}{x^\alpha}\left(\Delta\mu_{\rm e} {\rm e}^{\nu/2} \right)
	+ n_{\rm \mu} \pd{}{x^\alpha}\left( \Delta\mu_{\rm \mu} {\rm e}^{\nu/2} \right)
	+ (n_{\rm str}-n_{\rm b}) \pd{}{x^\alpha}\left( \Delta\mu_{\rm \Lambda} {\rm e}^{\nu/2} \right)
	,\quad \alpha = 1,2,3
.
\end{gather}
Subtracting equation \eqref{eq:sfl-osc-L}
	from \eqref{eq:sfl-osc-n},
	one can obtain the following simple equation:
\begin{gather}
\label{eq:sfl-osc-nL}
	{\rm i} \omega n_{\rm b}
		(w_{({\rm \Lambda})\alpha} - w_{({\rm n})\alpha})
	= n_{\rm b} \pd{}{x^\alpha}
		\left( \Delta\mu_{\rm \Lambda} {\rm e}^{\nu/2} \right)
	,\quad \alpha = 1,2,3
.
\end{gather}
This equation could also be derived by subtracting
	the potentiality condition for neutrons
	from the potentiality condition for $\rm \Lambda$--hyperons
	(see equation \ref{eq:potentiality}).

As a result, superfluid oscillation modes
	in the decoupling regime 
	can be calculated by using the two equations,
	\eqref{eq:sfl-osc-n} and \eqref{eq:sfl-osc-nL},
	%(or \ref{eq:sfl-osc-n} and \ref{eq:sfl-osc-L}),
	along with the continuity equations \eqref{eq:continuity}
	and the conditions
	$\delta U_{({\rm b})}^\alpha = \delta g^{\alpha\beta} = 0$.
If neutrons or $\rm \Lambda$--hyperons are non-superfluid, 
	one can write similar equations for other particle species 
	(see Appendix~\ref{sec:appendix-generalized-sfl} for more details).

%%%%%%%%%%%%%%%%%%%%%%%%%%%%%%%%%%%%%%%%%%%%%%%%%%%%%%%%%%%%%%%%%%%%

%%%%%%%%%%%%%%%%%%%%%%%%%%%%%%%%%%%%%%%%%%%%%%%%%%%%%%%%%%%%%%%%%
\subsection{Effects of rotation}
%%%%%%%%%%%%%%%%%%%%%%%%%%%%%%%%%%%%%%%%%%%%%%%%%%%%%%%%%%%%%%%%%
\label{sec:decoupling:rotation}

Rotation leads to the formation of Feynman-Onsager vortices inside HSs
	with the interspacing distance $\sim 10^{-2} - 10^{-4} {\rm cm}$.
Neglecting the vortex energy,
	the hydrodynamic equations averaged over the volume
	containing large amount of vortices
	have the same form as the corresponding equations for non-rotating matter
	(\citealt{1961JETP..40..920K,
		1991AnPhy.205..110M}).
The only exception is the potentiality condition \eqref{eq:potentiality},
	which should be replaced (for neutral particles) by
\begin{gather}
\label{eq:rotation:potentiality}
	u^\beta \left[
		\covd{\left( 
			w_{(i)\beta} + \mu_i u_\beta
		\right)}{\alpha}
		- \covd{\left(
			w_{(i)\alpha} + \mu_i u_\alpha
		\right)}{\beta}
	\right]
	= \mu_{\rm n} n_{\rm b} f_{\alpha (i)}
.
\end{gather}
This equation is a generalization of equation (8)
	from \citet{2012ASPC..466..211K}
	to the case of a few neutral superfluids.
The vector $f_{\alpha (i)}$ here is defined as
\begin{gather}
\label{eq:rotation:f}
	f_{(i)}^\alpha
		= \alpha_i (g^{\alpha\beta} + u^\alpha u^\beta) F_{\beta\gamma(i)} W_{(i)}^\gamma
		  + \frac{\beta_i - \gamma_i}{N_i}
			\left[
				F_{\gamma\beta(i)} F_{(i)}^{\beta\alpha}
				+ u^\alpha u_\delta F_{\gamma\beta(i)} F_{(i)}^{\beta\delta} 
				+ u^\beta u_\delta F_{\gamma\beta(i)} F_{(i)}^{\delta\alpha} 
			\right]
			W_{(i)}^\gamma
		  + \gamma_i N_i W_{(i)}^\alpha
,
\end{gather}
	where no summation over index $i$ is assumed, and
\begin{gather}
\label{eq:rotation-W}
	W_{(i)}^\alpha \equiv \frac{Y_{ik} w_{(k)}^\alpha }{n_{\rm b}}
,\\
\label{eq:rotation-Fab}
	F_{\alpha\beta(i)}
		= \covd{\left( 
				w_{(i)\beta} + \mu_i u_\beta
			\right)}{\alpha} 
		  -\covd{\left(
				w_{(i)\alpha} + \mu_i u_\alpha
			\right)}{\beta}
,\\
\label{eq:rotation:N}
	N_i = \left(
			-\frac{1}{2} F_{\alpha\beta(i)} F_{(i)}^{\alpha\beta} 
			- u^\alpha u_\gamma F_{\beta\alpha(i)} F_{(i)}^{\beta\gamma}
		  \right)^{1/2}
.
\end{gather}
Here $\alpha_i$, $\beta_i$, and $\gamma_i$
	are some scalars (kinetic coefficients),
	which, in the non-relativistic limit,
	are equal to the corresponding coefficients
	of non-relativistic hydrodynamics
	describing a rotating superfluid
	(see \citealt{1961JETP..40..920K}).

Let us now inspect how rotation affects the oscillation equations.
The right-hand side of equation \eqref{eq:rotation:potentiality}
	can schematically be presented in the form
	$\mu_{\rm n} n_{\rm b} f_{\alpha(i)} \equiv O_{\alpha\beta(i)} W_{(i)}^\beta$,
	where the tensor $O^{\alpha\beta}_{(i)}$ 
	is defined by the expression (\ref{eq:rotation:f}) for $f^{\alpha}_{(i)}$.
Repeating now the derivation of equation 
	\eqref{eq:sfl-osc-general2}
	and making use of Eq.~\eqref{eq:rotation:potentiality}
	instead of the potentiality condition \eqref{eq:potentiality}, one derives 
	the same equation 
	\eqref{eq:sfl-osc-general2}
	but with the term $n_{\rm b} O_{\alpha\beta({\rm n})} W_{({\rm n})}^\beta$
	in its right-hand side.
This term depends on the small quantity $W_{({\rm n})}^\beta$,
	vanishing in equilibrium,
	so that our reasoning about the mode decoupling remains valid even for the rotating HSs.
 Note that, allowing for rotation, 
	one should 
	use the metric of a rotating star
	instead of the Schwarzschild metric.

Superfluid oscillation modes
	(e.g., superfluid r-modes)
	in a rotating HS are described
	by	the following equations
	for the superfluid velocities
	$w_{({\rm n})}^\alpha$ and $w_{({\rm \Lambda})}^\alpha$
	(both neutrons and $\Lambda$--hyperons
		are assumed to be superfluid):
\begin{multline}
\label{eq:rotation:sfl-osc-n}
		\left(
			- \Delta\mu_{\rm e} n_{\rm e}  - \Delta\mu_{\rm \mu} n_{\rm \mu} - \Delta\mu_{\rm \Lambda} n_{\rm str} 
			\right) u^\beta \covd{u_\alpha}{\beta}
		+ \left(
			- n_{\rm e} \partial_\beta{\Delta\mu_{\rm e}} - n_{\rm \mu} \partial_\beta{\Delta\mu_{\rm \mu}} - n_{\rm str} \partial_\beta{\Delta\mu_{\rm \Lambda}}
			\right) u_\alpha u^\beta
		\\+ \left(
				- n_{\rm e} \partial_\alpha{\Delta\mu_{\rm e}}- n_{\rm \mu} \partial_\alpha{\Delta\mu_{\rm \mu}} - n_{\rm str} \partial_\alpha{\Delta\mu_{\rm \Lambda}}
			\right)
		%+\\
		+ (g_{\alpha\gamma} + u_\alpha u_\gamma ) u^\beta \covd{( \mu_{\rm n} n_{\rm b} W^\gamma )}{\beta}
		+\mu_{\rm n} n_{\rm b}
		\left(
			\covd{{u^\beta}}{\beta}  W_\alpha
			+ \covd{u_{\alpha}}{\beta} W^\beta
		\right)
	\\- n_{\rm b} u^\beta
	\left[
		\covd{w_{({\rm n})\alpha}}{\beta}
		-\covd{w_{({\rm n})\beta}}{\alpha}
	\right]
	=
	n_{\rm b} O_{\alpha\beta({\rm n})} W_{({\rm n})}^\beta
,
\end{multline}
\begin{multline}
\label{eq:rotation:sfl-osc-L}
	\left(
		- \Delta\mu_{\rm e} n_{\rm e}  - \Delta\mu_{\rm \mu} n_{\rm \mu} - \Delta\mu_{\rm \Lambda} n_{\rm str} 
		+ \Delta\mu_{\rm \Lambda} n_{\rm b}
	\right) u^\beta \covd{u_\alpha}{\beta}
	+ \left(
		- n_{\rm e} \partial_\beta{\Delta\mu_{\rm e}} - n_{\rm \mu} \partial_\beta{\Delta\mu_{\rm \mu}} - n_{\rm str} \partial_\beta{\Delta\mu_{\rm \Lambda}}
		+ n_{\rm b} \partial_\beta{\Delta\mu_{\rm \Lambda}}
	\right) u_\alpha u^\beta
	\\+
	\left(
		- n_{\rm e} \partial_\alpha{\Delta\mu_{\rm e}}- n_{\rm \mu} \partial_\alpha{\Delta\mu_{\rm \mu}} - n_{\rm str} \partial_\alpha{\Delta\mu_{\rm \Lambda}}
		+ n_{\rm b} \partial_\alpha{\Delta\mu_{\rm \Lambda}}
	\right)
	+ (g_{\alpha\gamma} + u_\alpha u_\gamma ) u^\beta \covd{( \mu_{\rm n} n_{\rm b} W^\gamma )}{\beta}
	+\mu_{\rm n} n_{\rm b}
	\left(
		\covd{{u^\beta}}{\beta}  W_\alpha
		+ \covd{u_{\alpha}}{\beta} W^\beta
	\right)
	\\- n_{\rm b} u^\beta
	\left[
		\covd{w_{({\rm \Lambda})\alpha}}{\beta}
		-\covd{w_{({\rm \Lambda})\beta}}{\alpha}
	\right]
	= 
	n_{\rm b} O_{\alpha\beta({\rm \Lambda})} W_{({\rm \Lambda})}^\beta
.
\end{multline}
As in the previous section, all the quantities in equations
	\eqref{eq:rotation:sfl-osc-n} and \eqref{eq:rotation:sfl-osc-L}
	except for $\Delta\mu_{\rm e}$, $\Delta\mu_{\rm \mu}$, $\Delta\mu_{\rm \Lambda}$,
	$w_{(i)}^\alpha$, and $W_{(i)}^\alpha$ 
	should be replaced with their equilibrium values.

%%%%%%%%%%%%%%%%%%%%%%%%%%%%%%%%%%%%%%%%%%%%%%%%%%%%%%%%%%%%%%%%%%%%
\section{Example: sound waves in nucleon-hyperon matter}
\label{sec:sound}
%%%%%%%%%%%%%%%%%%%%%%%%%%%%%%%%%%%%%%%%%%%%%%%%%%%%%%%%%%%%%%%%%%%%

In this section we illustrate the decoupling scheme
	developed in Section~\ref{sec:decoupling}
	by the calculation of the speed of sound in a homogeneous nucleon--hyperon matter.
Since this problem can be solved exactly,
	we can use it as a test for our approximate method.
We consider small harmonic perturbations
	($ \sim {\rm e}^{{\rm i}\omega t - {\rm i} \v{kr}} = {\rm e}^{- {\rm i} k_\alpha x^\alpha} $)
	in homogeneous superfluid matter in Minkowski spacetime with the metric 
	$g^{\alpha\beta} = \operatorname{diag}(-1,1,1,1)$.
We assume that all baryons
	($\rm n$, $\rm p$, $\rm \Lambda$, ${\rm \Xi}^{-}$, ${\rm \Xi}^0$, ${\rm \Sigma}^-$)
	can be superfluid.

Perturbations are described by
	the energy-momentum conservation law \eqref{eq:Tabb} 
	and superfluid equations \eqref{eq:sfl-osc-n} and \eqref{eq:sfl-osc-L} 
	for neutrons and $\rm \Lambda$--hyperons\footnote{
		If neutrons or $\rm \Lambda$--hyperons
			are non-superfluid,
			one has to employ
			similar superfluid equations
			\eqref{eq:sfl-osc-A}
			for other particle species.
	}.
In our case these equations 
	take the following simple form:
\begin{gather}
\label{eq:sound-1}
	\omega (P + \varepsilon) \v{\delta U}_{({\rm b})}
		= \v{k} \delta P
	,\\
\label{eq:sound-2}
	\omega n_{\rm b}
		\left(	\mu_{\rm n} \v{W} - \v{w}_{({\rm n})}	\right)
	= - \v{k}
		\left(
			n_{\rm e}	\Delta\mu_{\rm e}
			+ n_{\rm \mu} \Delta\mu_{\rm \mu}
			+ n_{{\rm str}} \Delta\mu_{\rm \Lambda} 
		\right)
	,\\
\label{eq:sound-3}
	\omega n_{\rm b}
		\left(	\mu_{\rm n} \v{W} - \v{w}_{({\rm \Lambda})}	\right)
	= - \v{k}
		\left(
			n_{\rm e}	\Delta\mu_{\rm e}
			+ n_{\rm \mu} \Delta\mu_{\rm \mu}
			+ n_{\rm str} \Delta\mu_{\rm \Lambda} 
			- n_{\rm b} \Delta\mu_{\rm \Lambda} 
		\right)
.
\end{gather}
Here $\v{\delta U}_{({\rm b})}$, $\v{W}$, $\v{w}_{(i)}$, and $\v{k}$ 
are three-vectors composed of spatial components of the corresponding four-vectors.

Now we have to write $\delta P$ and $\Delta\mu_j$
	in terms of
	$\v{\delta U}_{({\rm b})}$, $\v{w}_{({\rm n})}$, and $\v{w}_{({\rm \Lambda})}$.
As a first step, we present them
	as functions of the number density perturbations,
\begin{gather}
\label{eq:sound:delta}
	\delta = \delta n_{\rm b} \pd{}{n_{\rm b}}
			+ \delta n_{\rm e} \pd{}{n_{\rm e}}
			+ \delta n_{\rm \mu} \pd{}{n_{\rm \mu}}
			+ \delta n_{{\rm str}} \pd{}{n_{{\rm str}}}
,
\end{gather}
	and then, with the help of the continuity equations
	\eqref{eq:continuity-b} and \eqref{eq:continuity},
	express the number density perturbations 
	through the velocities $\v{\delta U}_{({\rm b})}$ and $\v{w}_{(i)}$:
\begin{gather}
\label{eq:sound:dnb}
	\delta n_{\rm b} = n_{\rm b} \frac{\v{k}}{\omega} \v{\delta U}_{({\rm b})}
,\\
\label{eq:sound:dne}
	\delta n_{\rm e}	= n_{\rm e} \frac{\v{k}}{\omega} \v{\delta u}
				= n_{\rm e} \frac{\v{k}}{\omega} \left( \v{\delta U}_{({\rm b})} - \frac{1}{n_{\rm b}} \sum_i Y_{ik}\v{w}_{(k)}  \right)
,\\
\label{eq:sound:dnmu}
	\delta n_{\rm \mu}= n_{\rm \mu} \frac{\v{k}}{\omega} \v{\delta u}
				= n_{\rm \mu} \frac{\v{k}}{\omega} \left( \v{\delta U}_{({\rm b})} - \frac{1}{n_{\rm b}} \sum_i Y_{ik}\v{w}_{(k)} \right)
,\\
\label{eq:sound:dnstr}
	\delta n_{{\rm str}}  = n_{{\rm str}} \frac{\v{k}}{\omega} \v{\delta U}_{({\rm str})}
				= \frac{\v{k}}{\omega}
					\left[
						n_{{\rm str}} \v{\delta U}_{({\rm b})} - \frac{n_{{\rm str}}}{n_{\rm b}}\sum_i Y_{ik} \v{w}_{(k)}
						- S_i Y_{ik} \v{w}_{(k)}
					\right]
.					
\end{gather}
Also we should express all the superfluid velocities $\v{w}_{(k)}$
	through $\v{w}_{({\rm n})}$ and 
	$\v{w}_{({\rm \Lambda})}$ using Eqs.
\eqref{eq:qYw=0} and \eqref{eq:w-LLpXm}--\eqref{eq:w-nLpSm}.

After substituting all these relations
	into the system of equations \eqref{eq:sound-1}--\eqref{eq:sound-3}
	one arrives at the linear equation of the form
\begin{gather}
\label{eq:sound:Ax=0}
	\mathsf{\mathbf{A}} \cdot \v{x} = 0
,
\end{gather}
where $\v{x}$ is a vector, $\v{x} = (\delta U_{({\rm b})}, w_{({\rm n})}, w_{({\rm \Lambda})})$, with
	$\delta U_{({\rm b})} = \v{\delta U}_{({\rm b})} \v{k}/k$,
	$w_{({\rm n})} = \v{w}_{({\rm n})} \v{k}/k$, and
	$w_{({\rm \Lambda})} = \v{w}_{({\rm \Lambda})}\v{k}/k$
	(it is clear that the vectors
		$\v{k}$, $\v{\delta U}_{({\rm b})}$, $\v{w}_{({\rm n})}$ and $\v{w}_{({\rm \Lambda})}$
		must be collinear);
$\mathsf{\mathbf{A}}$ is a $3 \times 3$ matrix,
 whose elements depend on thermodynamic quantities,
	entrainment matrix $Y_{ik}$,
	as well as on the frequency $\omega$ and the wavenumber $k$.
The system \eqref{eq:sound:Ax=0} has a nontrivial solution
	only if $\det \mathsf{\mathbf{A}} = 0$.
This condition results in a cubic equation
	for the squared speed of sound,
	$c_{\rm S}^2 \equiv \omega^2 / k^2$.
Three roots of this cubic equation correspond to
	three sound modes in the nucleon--hyperon matter.

Note that in the decoupling approximation
	$\delta P$ does not depend on
	$\delta n_{\rm e (SFL)}$, $\delta n_{\rm \mu (SFL)}$, and $\delta n_{\rm str (SFL)}$
	(see equation~\ref{eq:coupling:dP}),
	so that Eq.~\eqref{eq:sound-1} coincides with 
	the corresponding equation for the normal (non-superfluid) matter 
	and does not contain the superfluid variables $\v{w}_{(i)}$.
This equation describes `normal' sound modes 
	and can be solved separately from equations \eqref{eq:sound-2} and \eqref{eq:sound-3}.
The latter equations describe `superfluid' sound modes.

We calculated sound speeds for the EOSs GM1A, GM1'B, and TM1C
	studied by \citet{2014MNRAS.439..318G}. 
In our calculations we need to specify baryon critical temperatures $T_{{\rm c}i}$,
	which are generally functions of baryon number density $n_{\rm b}$.
These temperatures are poorly known, especially for hyperons
	(see e.g., \citealt{2013arXiv1302.6626P}).
In view of large uncertainties,
	we (somewhat arbitrary)
	adopt the following values for $T_{{\rm c}i}$:
	$T_{\rm cn} = 5 \times 10^8~{\rm K}$,
	$T_{\rm cp} = 3 \times 10^9 ~{\rm K}$,
	$T_{\rm c\Xi^{-}} = T_{\rm c\Xi^{0}} = 5 \times 10^9 ~{\rm K}$.
These values do not contradict the results
	of microscopic calculations
	(see e.g., \citealt{1999SvPhU..42..737Y,
		2001LNP...578...30L,
		2013arXiv1302.6626P,
		2014arXiv1406.6109G}
		and references therein).
		
\begin{figure}
\centering
\begin{minipage}{.32\textwidth}
  \centering
  \includegraphics[width=1.\linewidth]{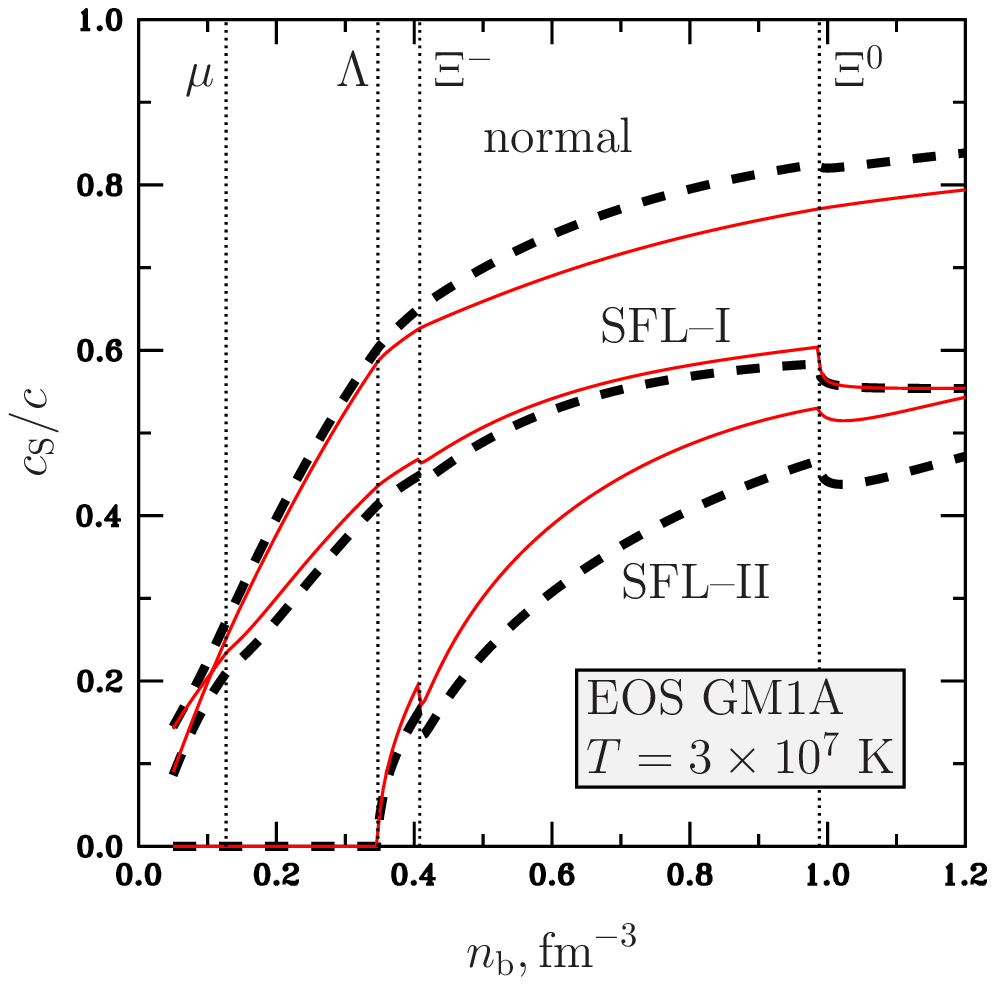}
  \label{fig:sound:Tcrit1:GM1A_logT=7.5}
\end{minipage}
\begin{minipage}{.32\textwidth}
  \centering
  \includegraphics[width=1.\linewidth]{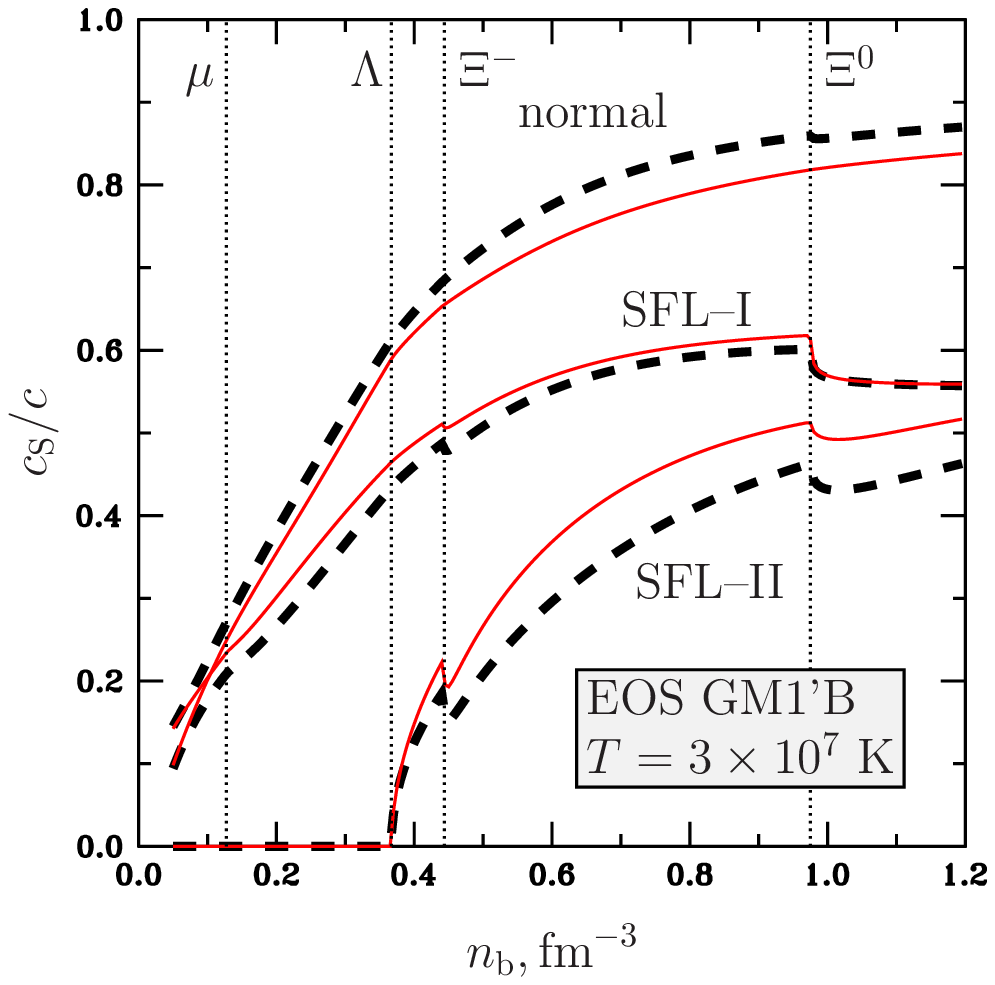}
	\label{fig:sound:Tcrit1:GM1B_logT=7.5}
\end{minipage}
\begin{minipage}{.32\textwidth}
  \centering
  \includegraphics[width=1.\linewidth]{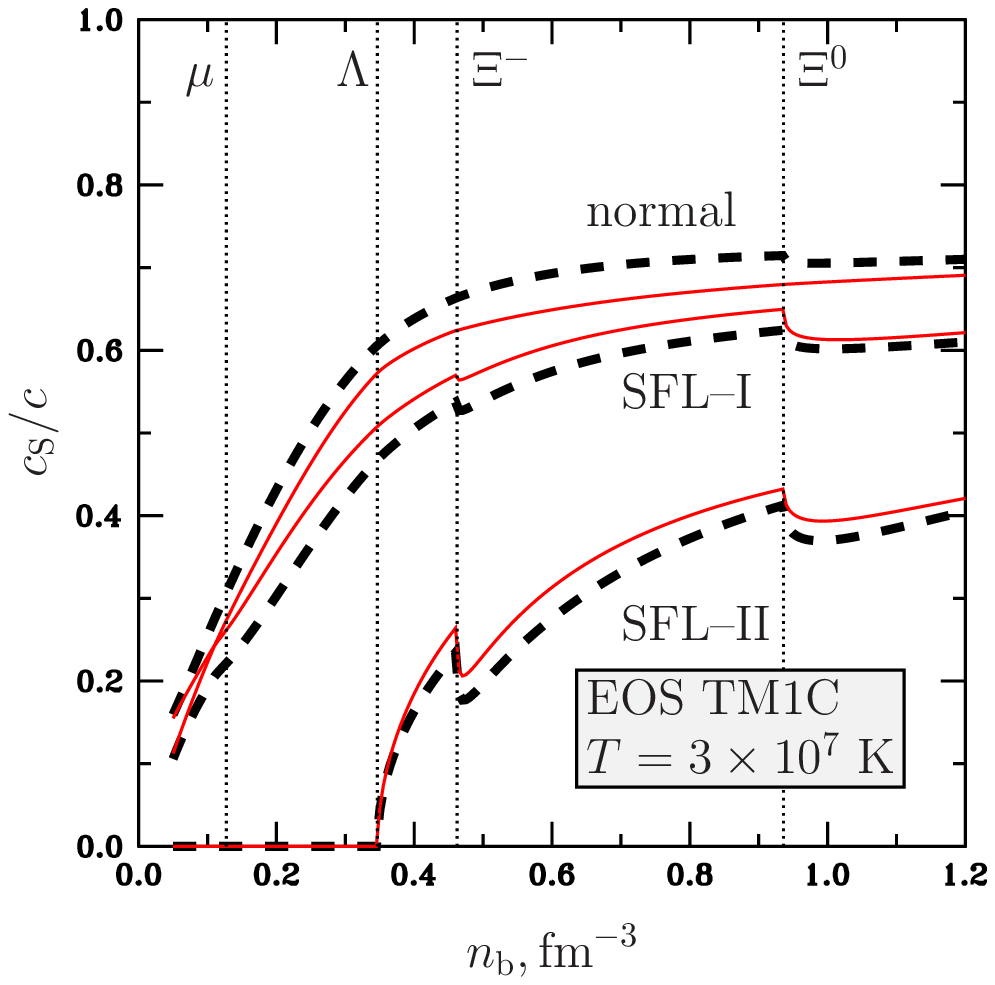}
	\label{fig:sound:Tcrit1:TM1C_logT=7.5}
\end{minipage}
\caption{
	Speed of sound $c_{\rm S}$ (in units of $c$)
		versus the baryon number density $n_{\rm b}$
		for the EOSs GM1A, GM1'B, TM1C
		at $T = 3 \times 10^7~{\rm K}$.
	Dashed lines: exact solution.
	Solid lines: decoupled solution.
	Vertical lines: thresholds for the appearance of muons
		and $\rm \Lambda$--,~${\rm \Xi}^{-}$--,~${\rm \Xi}^0$--hyperons.
	$\rm \Lambda$--hyperons are superfluid.
}
\label{fig:sound:Tcrit1:logT=7.5}
\end{figure}

\begin{figure}
\centering
\begin{minipage}{.32\textwidth}
  \centering
  \includegraphics[width=1.\linewidth]{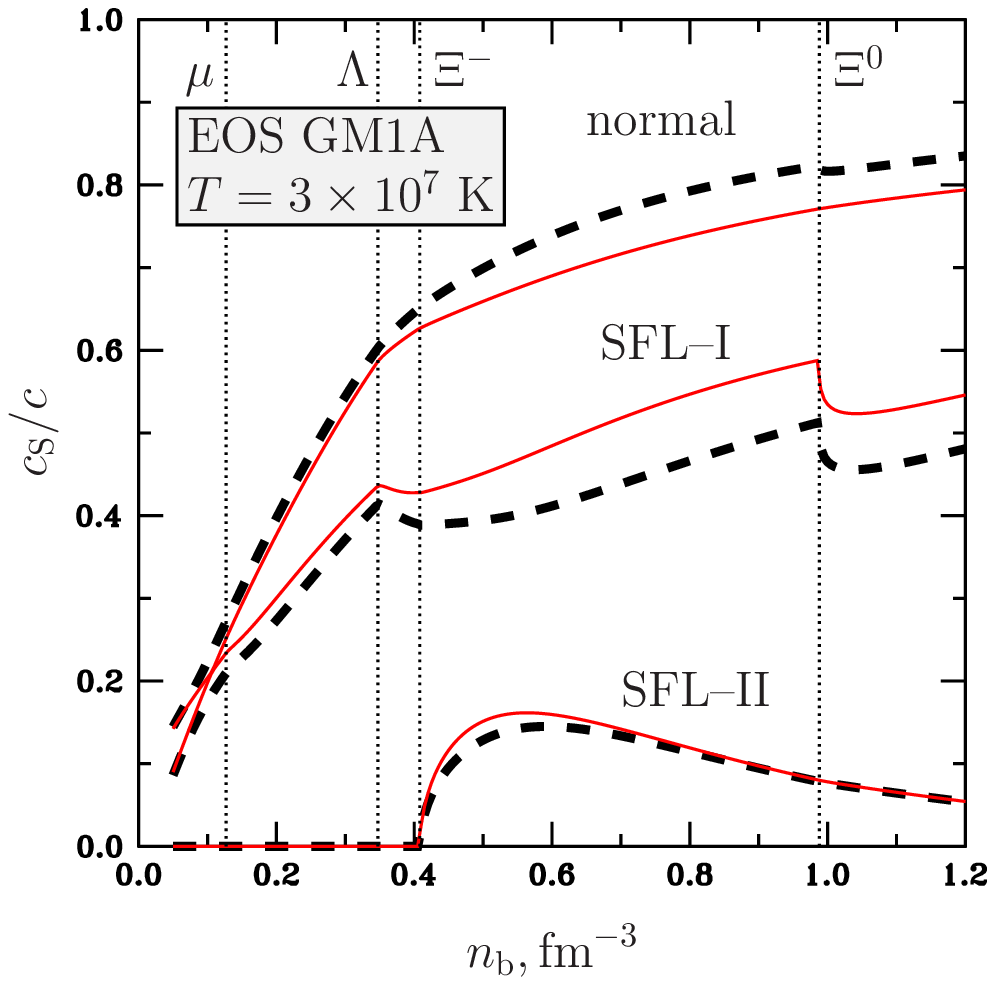}
  \label{fig:sound:Tcrit2:GM1A_logT=7.5}
\end{minipage}
\begin{minipage}{.32\textwidth}
  \centering
  \includegraphics[width=1.\linewidth]{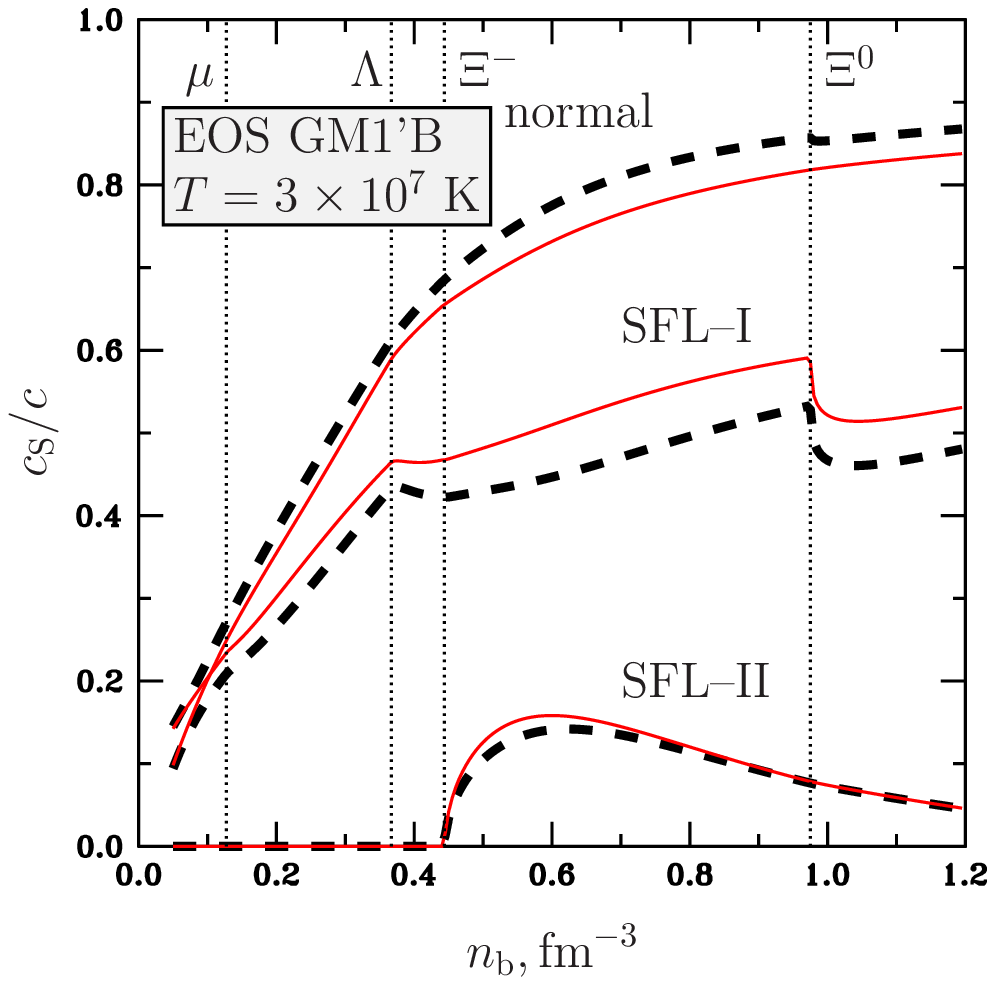}
	\label{fig:sound:Tcrit2:GM1B_logT=7.5}
\end{minipage}
\begin{minipage}{.32\textwidth}
  \centering
  \includegraphics[width=1.\linewidth]{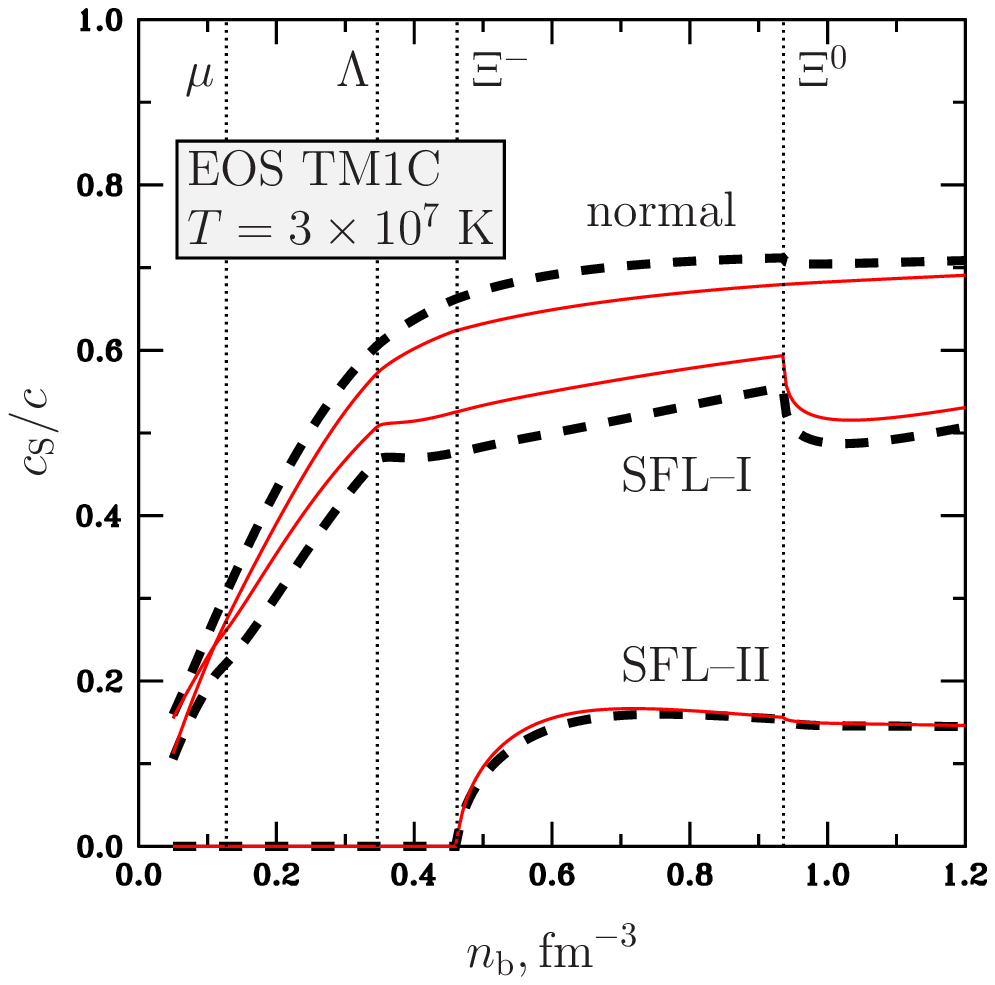}
	\label{fig:sound:Tcrit2:TM1C_logT=7.5}
\end{minipage}

\caption{
	Speed of sound $c_{\rm S}$ (in units of $c$)
		versus the baryon number density $n_{\rm b}$
		for the EOSs GM1A, GM1'B, TM1C
		at $T = 3 \times 10^7~{\rm K}$.
	Dashed lines: exact solution.
	Solid lines: decoupled solution.
	Vertical lines: thresholds for the appearance of muons
		and $\rm \Lambda$--,~${\rm \Xi}^{-}$--,~${\rm \Xi}^0$--hyperons.
	$\rm \Lambda$--hyperons are non-superfluid.
}
\label{fig:sound:Tcrit2:logT=7.5}
\end{figure}

\begin{figure}
\centering
\begin{minipage}{.32\textwidth}
  \centering
  \includegraphics[width=1.\linewidth]{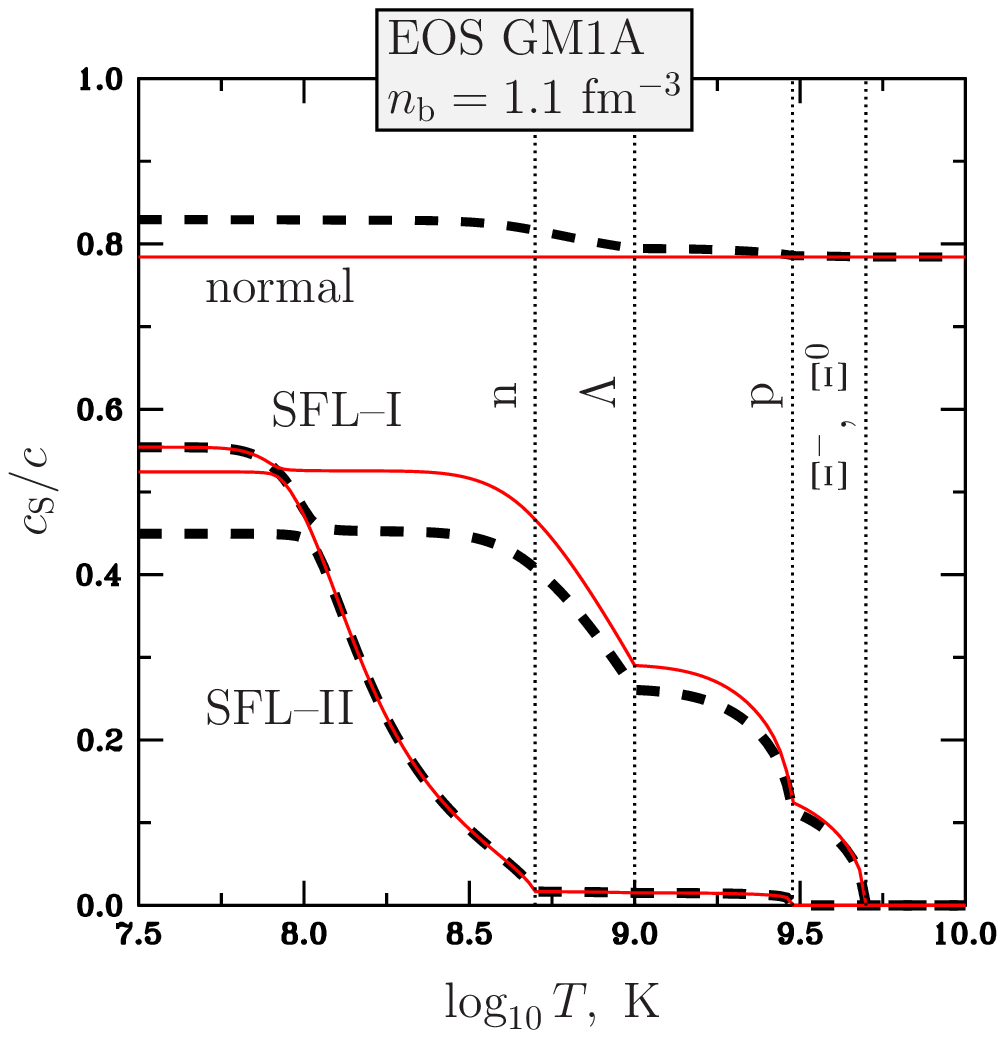}
  \label{fig:sound:Tcrit1:GM1A_nb=1.1}
\end{minipage}
\begin{minipage}{.32\textwidth}
  \centering
  \includegraphics[width=1.\linewidth]{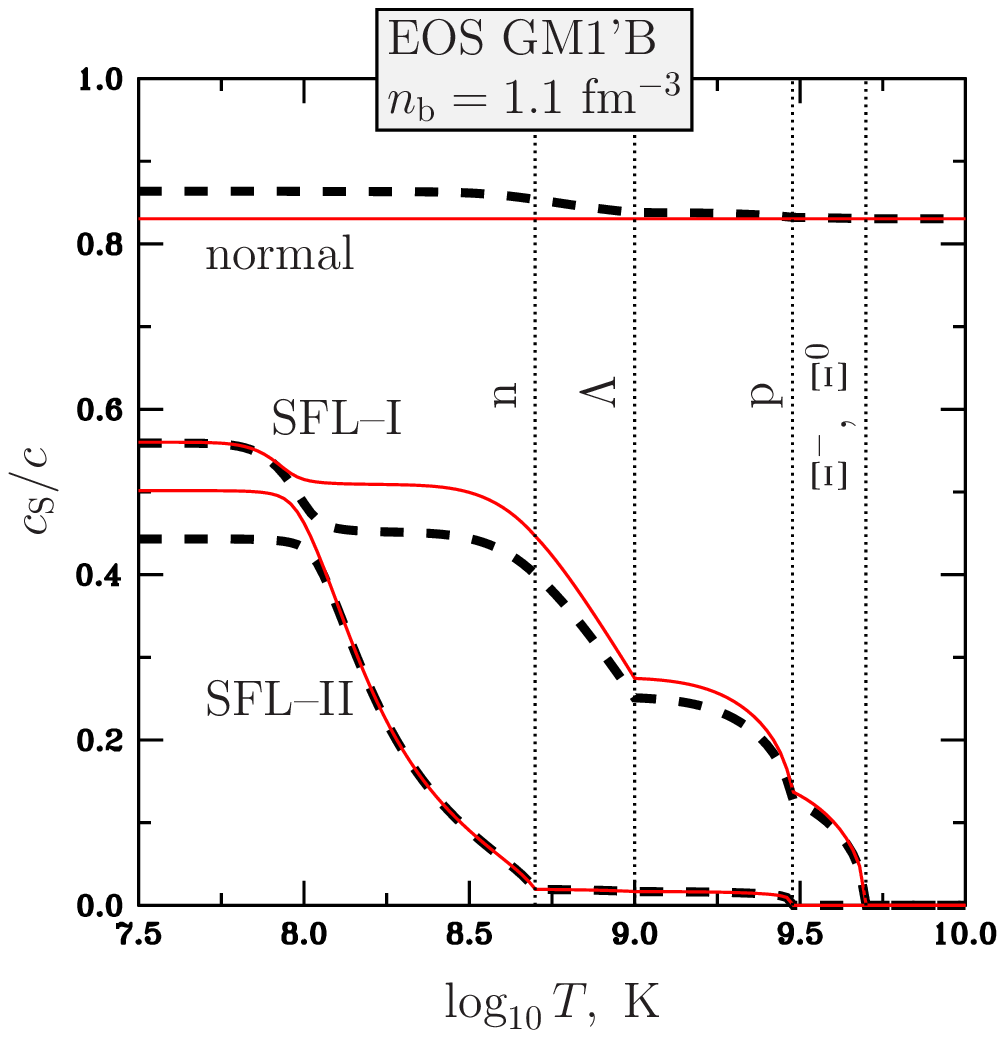}
	\label{fig:sound:Tcrit1:GM1B_nb=1.1}
\end{minipage}
\begin{minipage}{.32\textwidth}
  \centering
  \includegraphics[width=1.\linewidth]{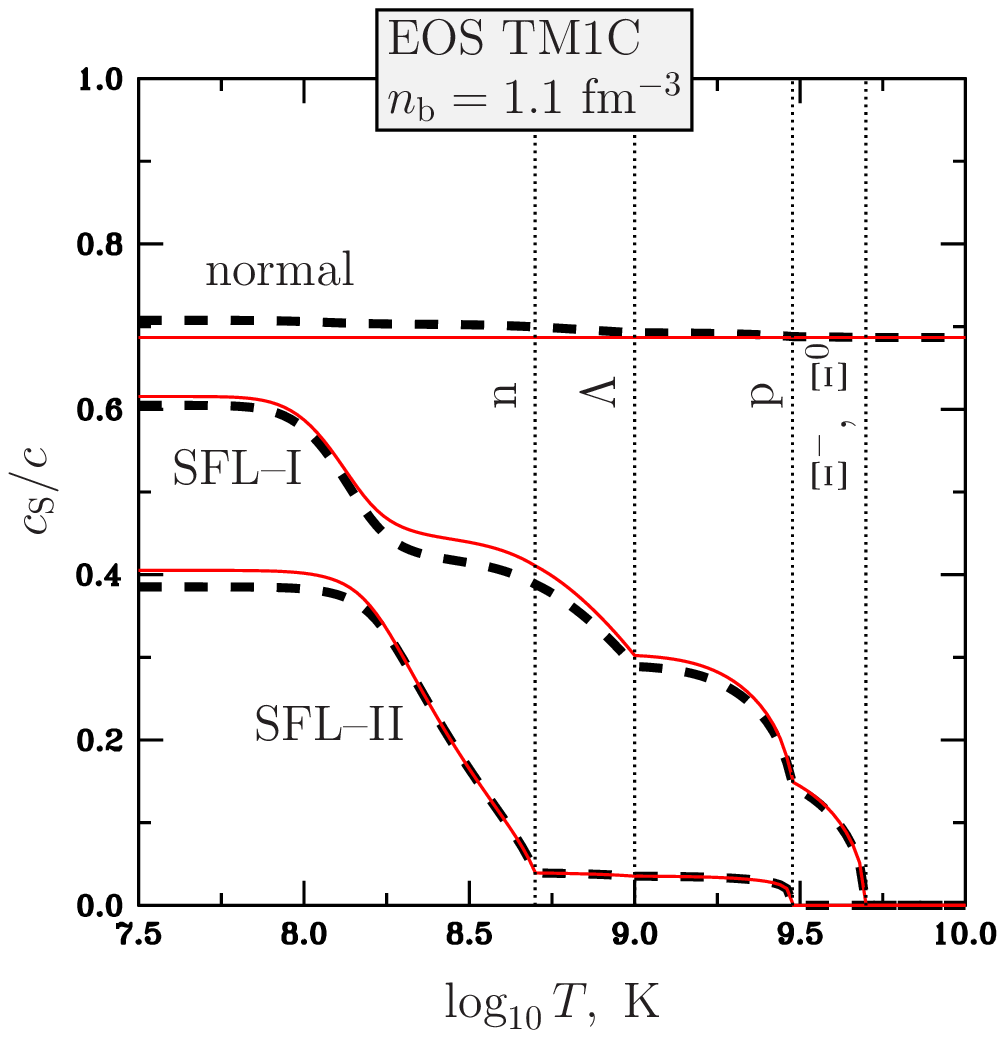}
	\label{fig:sound:Tcrit1:TM1C_nb=1.1}
\end{minipage}
\caption{
Speed of sound $c_{\rm S}$ (in units of $c$)
	versus temperature $\log_{10}T,~{\rm K}$
	for the EOSs GM1A, GM1'B, TM1C at $n_{\rm b} = 1.1 \;{\rm fm^{-3}}$.
Dashed lines: exact solution.
Solid lines: decoupled solution.
Vertical lines: critical temperatures for baryons.
$\rm \Lambda$--hyperons are superfluid.
}
\label{fig:sound:Tcrit1:nb=1.1}
\end{figure}

\begin{figure}
\centering
\begin{minipage}{.32\textwidth}
  \centering
  \includegraphics[width=1.\linewidth]{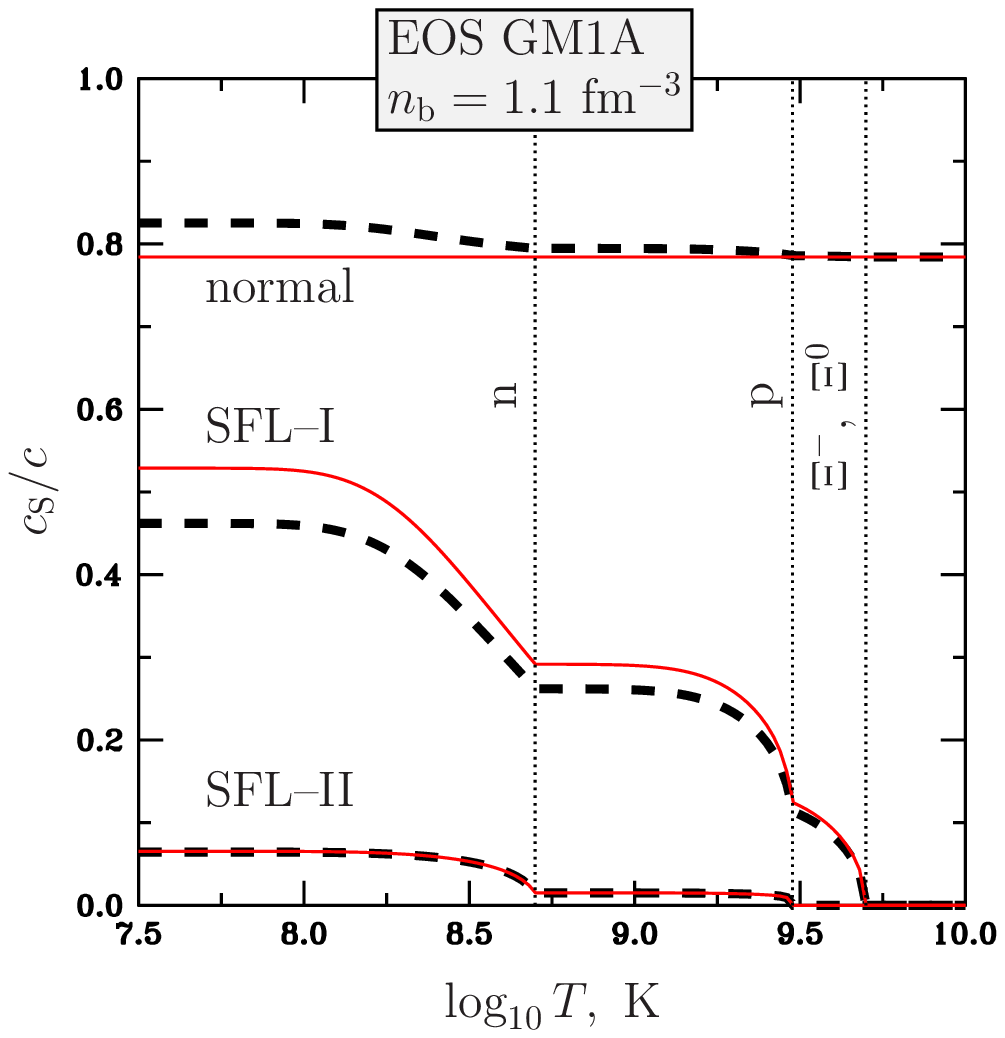}
  \label{fig:sound:Tcrit2:GM1A_nb=1.1}
\end{minipage}
\begin{minipage}{.32\textwidth}
  \centering
  \includegraphics[width=1.\linewidth]{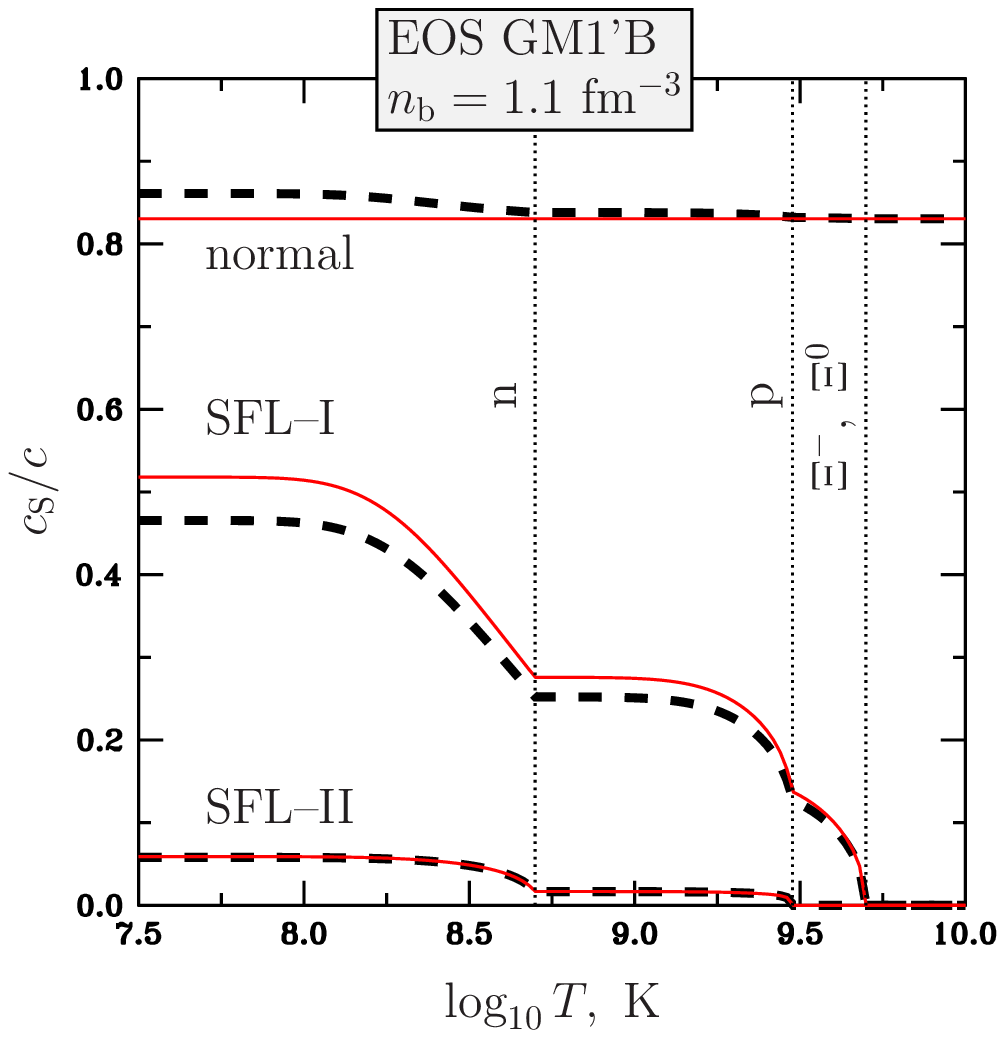}
	\label{fig:sound:Tcrit2:GM1B_nb=1.1}
\end{minipage}
\begin{minipage}{.32\textwidth}
  \centering
  \includegraphics[width=1.\linewidth]{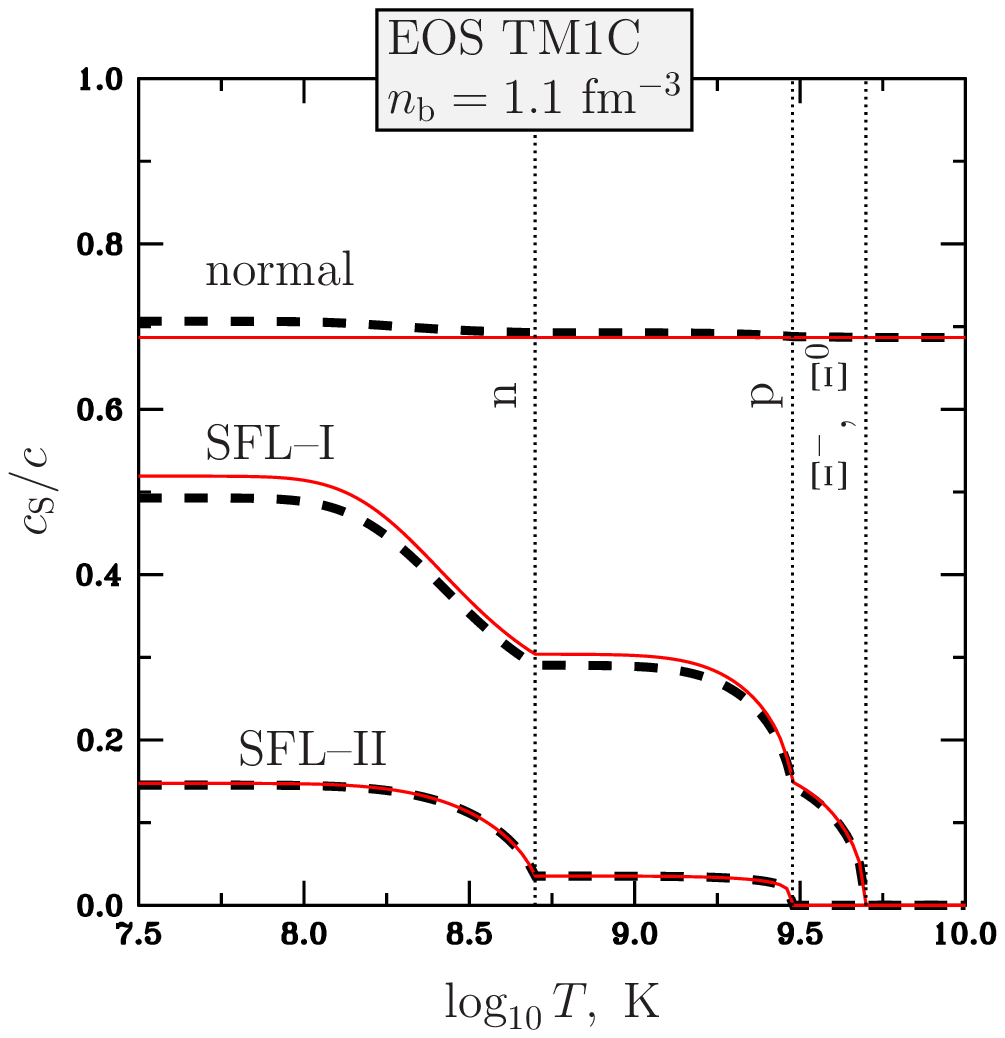}
	\label{fig:sound:Tcrit2:TM1C_nb=1.1}
\end{minipage}
\caption{
	Speed of sound $c_{\rm S}$ (in units of $c$)
		versus temperature $\log_{10} T,~{\rm K}$
		for the EOSs GM1A, GM1'B, TM1C at $n_{\rm b} = 1.1 \;{\rm fm^{-3}}$.
	Dashed lines: exact solution.
	Solid lines: decoupled solution.
	Vertical lines: critical temperatures for baryons.
	$\rm \Lambda$--hyperons are non-superfluid.
}
\label{fig:sound:Tcrit2:nb=1.1}
\end{figure}

As for $\rm \Lambda$--hyperons, we consider two different possibilities discussed in the literature 
(see e.g.,
	\citealt{2006PThPh.115..355T,
		2010PhRvC..81b5801W}):
\begin{enumerate}
	\item
		$\rm \Lambda$--hyperons are superfluid, $T_{\rm c\Lambda} = 10^9 \, {\rm K}$.
		The dependence of the sound speeds $c_{\rm S}$
			on the baryon number density $n_{\rm b}$
			and on the temperature $T$
			for this case is shown in 
			Figs.~\ref{fig:sound:Tcrit1:logT=7.5}
			and \ref{fig:sound:Tcrit1:nb=1.1}, respectively.
	\item
		$\rm \Lambda$--hyperons are normal at $T > 10^{7}\, {\rm K}$.
		The corresponding functions $c_{\rm S}(n_{\rm b})$ and $c_{\rm S}(T)$
		are demonstrated 
		in Figs.~\ref{fig:sound:Tcrit2:logT=7.5}
		and \ref{fig:sound:Tcrit2:nb=1.1}, respectively.
\end{enumerate}

%%%%%%%%%%%%%%%%%%
Let us discuss 
  Figs.~\ref{fig:sound:Tcrit1:logT=7.5}--\ref{fig:sound:Tcrit2:nb=1.1}
	in more detail.
Fig.~\ref{fig:sound:Tcrit1:logT=7.5} shows the dependence
	$c_{\rm S}(n_{\rm b})$ at fixed $T = 3 \times 10^7~{\rm K}$
	for the first case
	($T_{\rm c\Lambda} = 10^9 \, {\rm K}$).
The solid lines present sound speeds calculated in the decoupling approximation,
	the dashed lines show the exact results.
The vertical lines denote
	the thresholds for appearance of different particle species.
The highest sound speed on every plot
	is labelled `normal',
	because in the fully decoupled case
	it coincides with the sound speed in the non-superfluid matter.
Other modes appear only in superfluid matter
	and are therefore labelled `SFL'.
The number of superfluid sound modes
	is equal to the number of superfluid degrees of freedom,
	as discussed in Section~\ref{sec:sfl-degrees-of-freedom}.
The second superfluid mode arises after
	the appearance of $\rm \Lambda$--hyperons.
Note that the appearance of ${\rm \Xi}^-$ or ${\rm \Xi}^0$--hyperons 
	does not lead to any additional degrees of freedom
	(and, hence, to new sound modes)
	due to the constraints
	\eqref{eq:w-LLpXm} and \eqref{eq:w-LLnX0}.

Fig.~\ref{fig:sound:Tcrit2:logT=7.5}
	presents a similar plot 
	but for
	non-superfluid $\rm \Lambda$--hyperons (case ii).
Since $\rm \Lambda$--hyperons are normal,
	the second superfluid degree of freedom
	(associated with a `quasiparticle'
		$A = ({\rm p} + {\rm \Xi}^-) / 2$
		and its superfluid four-vector
		$w_{(A)\alpha} = (w_{({\rm p})\alpha} + w_{({\rm \Xi}^-)\alpha} ) / 2$;
		see Appendix~\ref{sec:appendix-generalized-sfl})
	exists only in the presence of ${\rm \Xi}^-$--hyperons.
Note that the second superfluid sound speed
	is much lower than in the case of superfluid $\rm \Lambda$--hyperons.
In Figs.~\ref{fig:sound:Tcrit1:logT=7.5} and \ref{fig:sound:Tcrit2:logT=7.5}
	(at low densities) one can see
	{\it crossing} of `normal' and `SFL-I' modes
	in the decoupling regime,
	while the exact solution
	shows the {\it avoided crossing}.
This feature, generic to superfluid stars,
	was also observed e.g., by \citet{2011PhRvD..83h1304G}
	and \citet{2011PhRvD..83j3008K}.

The dependence $c_{\rm S}(T)$
	at fixed $n_{\rm b} = 1.1~{\rm fm}^{-3}$
	is shown 
	for the cases of superfluid 
	and non-superfluid $\rm \Lambda$--hyperons
	in  Figs.~\ref{fig:sound:Tcrit1:nb=1.1}
	and~\ref{fig:sound:Tcrit2:nb=1.1},
	respectively.
The vertical lines in the figures denote the critical temperatures $T_{{\rm c}i}$
	for different baryon species.
At high temperatures, when all baryons become non-superfluid,
	the `decoupled' (normal) speed of sound
	is equal to the exact one (as it should be).
The sound modes depend on $T$ because of the temperature dependence of
	the entrainment matrix $Y_{ik}$.
The effect of finite temperatures on $Y_{ik}$
	was discussed by \citet{2009PhRvC..80a5803G}.
Since $Y_{ik} \rightarrow 0$ 
	as $T \rightarrow T_{{\rm c}i}$,
	superfluid speeds of sound
	also decrease with increasing temperature.
When protons become normal
	($T_{\rm cp} < T
				< T_{{\rm c} {\rm \Xi}^-},
				~T_{{\rm c} {\rm \Xi}^0}$),
	only one superfluid mode
	(associated with ${\rm \Xi}^0$--hyperons)
	survives.
One can see the avoided crossings of sound modes
	in Fig.~\ref{fig:sound:Tcrit1:nb=1.1}.

To sum up, our numerical results show that
	the decoupling scheme
	developed in Section~\ref{sec:decoupling}
	allows one to calculate the oscillation modes
	within reasonable accuracy,
	which is determined by the coupling parameters
	$s_{\rm e}$, $s_{\rm \mu}$, and $s_{\rm str}$.
At high densities the error
	is mainly due to the strange coupling parameter $s_{{\rm str}}$.
For the EOS TM1C, $s_{{\rm str}}$ is smaller
	than that for the EOSs GM1A and GM1'B (see Fig.\ \ref{fig:coupling}).
That is why the	difference
	between the exact and decoupled solution
	for the EOS TM1C is smaller.

%%%%%%%%%%%%%%%%%%%%%%%%%%%%%%%%%%%%%%%%%%%%%%%%%%%%%%%%%%%%%%%%%%%
%%%%%%%%%%%%%%%%%%%%%%%%%%%%%%%%%%%%%%%%%%%%%%%%%%%%%%%%%%%%%%%%%%%

%%%%%%%%%%%%%%%%%%%%%%%%%%%%%%%%%%%%%%%%%%%%%%%%%%%%%%%%%%%%%%%%%%%
%%%%%%%%%%%%%%%%%%%%%%%%%%%%%%%%%%%%%%%%%%%%%%%%%%%%%%%%%%%%%%%%%%%

%%%%%%%%%%%%%%%%%%%%%%%%%%%%%%%%%%%%%%%%%%%%%%%%%%%%%%%%%%%%%%%%%%%
%%%%%%%%%%%%%%%%%%%%%%%%%%%%%%%%%%%%%%%%%%%%%%%%%%%%%%%%%%%%%%%%%%%

%%%%%%%%%%%%%%%%%%%%%%%%%%%%%%%%%%%%%%%%%%%%%%%%%%%%%%%%%%%%%%%%%%%
%%%%%%%%%%%%%%%%%%%%%%%%%%%%%%%%%%%%%%%%%%%%%%%%%%%%%%%%%%%%%%%%%%%

%%%%%%%%%%%%%%%%%%%%%%%%%%%%%%%%%%%%%%%%%%%%%%%%%%%%%%%%%%%%%%%%%%%
\section{Composition g-modes in superfluid nucleon--hyperon matter}
\label{sec:g-modes}
%%%%%%%%%%%%%%%%%%%%%%%%%%%%%%%%%%%%%%%%%%%%%%%%%%%%%%%%%%%%%%%%%%%

The decoupling scheme developed and applied in the preceding sections
	can be used to calculate various oscillation modes of superfluid HSs, 
	e.g., f-, p-, and r-modes.
We postpone a detailed analysis
	of superfluid oscillation modes in the nucleon-hyperon matter
	for a future publication.

However, there is an important class of oscillations, namely, 
	the gravity modes (g-modes), that cannot be analysed 
	within the framework presented above.
It is easily verified
	that in the decoupling regime the g-modes exist and coincide
	with the g-modes of a non-superfluid HS.
This is so `by construction', because 
	the decoupled equations,
	which describe the normal modes, are exactly the same
	as those for a non-superfluid star. 
Unfortunately, this result is completely wrong:
	the local analysis of hydrodynamic equations 
	and numerical modelling
	show that the normal-like g-modes are artefacts of the adopted approximation.
Putting it differently, the decoupling approximation is too crude to find the real g-modes.
This conclusion is not surprising. 
For example, for a zero-temperature non-superfluid neutron star 
	with the $\rm npe$ composition of the core, the g-modes
	disappear from the oscillation spectrum
	if one neglects the dependence of the pressure $P$ on the electron number density $n_{\rm e}$
	(thus effectively treating a star as barotropic).
In the decoupling approximation we also 
	neglect the terms of this kind so that it is reasonable to expect that 
	this affects the g-modes somehow.	
The fact that the g-modes in superfluid stars will differ substantially
	from their normal counterparts also clearly follows from the thought experiment
	discussed in the section II in \citet{2014MNRAS.442L..90K}.

Meanwhile, the g-modes constitute a very interesting class of oscillations, 
	especially because it has been believed, until recently,
	that they do not exist in the zero-temperature superfluid neutron stars
	(see e.g.,
	\citealt{1995A&A...303..515L,
		2001MNRAS.328.1129A,
		2002A&A...393..949P}).
However, as demonstrated by Kantor \& Gusakov,
	this is generally not true
	(see also \citealt{2015arXiv150407470P}).
The g-modes, for example, can be excited in a superfluid $\rm npe\mu$ matter
	and their frequencies can be unusually large, up to $\sim 450$~Hz 
	(while the frequencies 
	of the ordinary composition g-modes in the non-superfluid 
	neutron stars do not exceed $50-150$~Hz;
	see e.g., \citealt{1992ApJ...395..240R}).
To the best of our knowledge, 
	these modes have never been studied
	for nucleon-hyperon matter,
	even for non-superfluid HSs.
This provides the motivation to study them here.

In this section, in all numerical calculations we employ
	the EOS GM1'B in the HS core
	and the EOS BSk21 (\citealt{2013A&A...560A..48P}) in the crust.
All numerical results are obtained for a neutron star with
	the mass $M = 1.634~\rm M_\odot$,
	the radius $R = 13.55 ~{\rm km}$,
	and the central density $\rho_{\rm c} = 8.1 \times 10^{14} \rm~g~cm^{-3}$.
The threshold for the $\rm \Lambda$--hyperon appearance 
	in such star lies at a distance
	$r \approx 5.29 \rm~km$ from the centre;
	other hyperons are absent.
We assume that
	$\rm \Lambda$--hyperons are normal (case ii in Section~\ref{sec:sound}),
	while the neutron and proton redshifted critical temperatures
	are constant throughout the core, 
	$T_{\rm cn}^\infty \equiv T_{\rm cn} {\rm e}^{\nu/2} = 5 \times 10^8 \rm K$ and
	$T_{\rm cp}^\infty \equiv T_{\rm cp} {\rm e}^{\nu/2} = 3 \times 10^9 \rm K$.
This simplifying assumption does not affect the g-mode spectrum 
	in the limit $T^\infty \ll T_{\rm cn}^\infty$ 
	($T^\infty$	is the redshifted internal stellar temperature),
	when the g-mode frequencies 
	reach a maximum value
	(\citealt{2014MNRAS.442L..90K}).
It is the limit we are mostly interested in here.

%%%%%%%%%%%%%%%%%%%%%%%%%%%%%%%%%%%%%%%%%%%%%%%%%%%%%%%%%%%%%%%%%%%%%%%%%%%%%
\subsection{Superfluid oscillation equations}
%%%%%%%%%%%%%%%%%%%%%%%%%%%%%%%%%%%%%%%%%%%%%%%%%%%%%%%%%%%%%%%%%%%%%%%%%%%%%

We examine the superfluid g-modes following an approach presented 
	recently by \citet{2014MNRAS.442L..90K}.
As mentioned above,
	we consider a model of a HS
	whose core consists of 
	neutrons, protons, electrons, muons, and $\rm \Lambda$--hyperons ($\rm npe\mu\Lambda$ matter),
	assuming that neutrons and protons are superfluid,
	while $\rm \Lambda$--hyperons are not.
We also assume that the metric is not perturbed during oscillations 
	-- this assumption, called the Cowling approximation (see \citealt{1941MNRAS.101..367C}), 
	works very well for the g-modes (see e.g., \citealt{2009PhRvD..80f4026G}).
We consider non-radial perturbations
	$\propto {\rm e}^{{\rm i}\omega t} Y_{lm}(\theta, \phi)$ 
	($Y_{lm}$ is a spherical harmonic)
	of a non-rotating spherically symmetric star
	with the Schwarzschild metric~\eqref{eq:Schwarzschild}.
Equations, governing such perturbations in the $\rm npe\mu$ matter, were derived
	in the paper by \citet{2014MNRAS.442L..90K}
	(see equations 7--10 there).
A straightforward generalization 
	of these equations to the case of $\rm npe\mu\Lambda$ matter 
	yields the following system of equations 
	(the terms arising due to the presence of $\rm \Lambda$--hyperons are underlined):
\begin{multline}
	\left(g \mu_{\rm n} n_{\rm b} \pd{n_{\rm b}}{P}
			+ g \mu_{\rm n} \pd{n_{\rm b}}{\mu_{\rm n}}
			-\pd{n_{\rm b}}{x_{\rm e\mu}} \nabla x_{\rm e\mu}
			-\underline{\pd{n_{\rm b}}{x_{\rm e\Lambda}} \nabla x_{\rm e\Lambda}}
		\right)\,\xi^r_{(\rm b)}
		-\frac{n_{\rm b}}{{\rm e}^{\lambda/2}r^2}\pd{}{r}\left({\rm e}^{\lambda/2} r^2 \xi^r_{(\rm b)}\right)
		+\frac{n_{\rm b} l(l+1){\rm e}^{\nu}}{r^2 \omega^2(P+\varepsilon)} \delta P
	\\=
	\pd{n_{\rm b}}{P} \delta P + \pd{n_{\rm b}}{\mu_{\rm n}}\delta \mu_{\rm n}
		- \frac{\partial n_{\rm b}}{\partial x_{\rm e\mu}} \nabla x_{\rm e\mu}\, \xi^r
		- \underline{\frac{\partial n_{\rm b}}{\partial x_{\rm e\Lambda}} \nabla x_{\rm e\Lambda}\, \xi^r},
\label{eq:g-modes:1}
\end{multline}
\begin{gather}
	-\omega^2 \mu_{\rm n} n_{\rm b} {\rm e}^{\lambda-\nu}\xi_{(\rm b)}^r + \frac{\partial \delta P}{\partial r}
	+g\left(
			\pd{w}{P} \delta P
			+\pd{w}{\mu_{\rm n}}\delta \mu_{\rm n}
			-\frac{\partial w}{\partial x_{\rm e\mu}} \nabla x_{\rm e\mu} \xi^r
			-\underline{\frac{\partial w}{\partial x_{\rm e\Lambda}} \nabla x_{\rm e\Lambda} \xi^r}
		\right)
	=0,
\label{eq:g-modes:2}
\\
	{\rm e}^{\nu/2}\pd{}{r}
		\left(
			\delta \mu_{\rm n} {\rm e}^{\nu/2}
		\right)
	-\omega^2 {\rm e}^{\lambda} \mu_{\rm n}
		\left[
			(y+1) \xi_{(\rm b)}^r-y \xi^r
		\right]
	= 0,
\label{eq:g-modes:3}
\\
		\left(
			g \mu_{\rm n} n_{\rm b} \pd{n_{\rm e}}{P} + g \mu_{\rm n} \pd{n_{\rm e}}{\mu_{\rm n}}
		\right)\,\xi^r
		-\frac{n_{\rm e}}{{\rm e}^{\lambda/2}r^2}
			\pd{}{r}\left( {\rm e}^{\lambda/2} r^2 \xi^r \right)
		+\frac{n_{\rm e} l(l+1){\rm e}^{\nu}}{r^2 \omega^2 y(P+\varepsilon)} \left[(y+1)\delta P-n_{\rm b} \delta \mu_{\rm n}\right]
	=
		\pd{n_{\rm e}}{P} \delta P+\pd{n_{\rm e}}{\mu_{\rm n}}\delta \mu_{\rm n}
.
\label{eq:g-modes:4}
\end{gather}
Here all the quantities except for
	$\xi^r$, $\xi_{({\rm b})}^r$, $\delta P$, and $\delta\mu_{\rm n}$
	are taken in equilibrium.
$\delta P$ and $\delta\mu_{\rm n}$
		are the Eulerian perturbations of the pressure and the neutron chemical potential, respectively;
	$\nabla \equiv {\rm d}/{\rm d}r$;
	$g = \nabla \nu/2$;
	$w = P + \varepsilon$;
	$x_{\rm e\mu} = n_{\rm \mu}/n_{\rm e}$;
	$x_{\rm e\Lambda} = n_{\rm \Lambda}/n_{\rm e}$;
	and the parameter $y$ is expressed through the entrainment matrix $Y_{ik}$ as
\begin{gather}
	y = \frac{n_{\rm b} Y_{\rm pp}}{\mu_{\rm n} \left(Y_{\rm nn} Y_{\rm pp} - Y_{\rm np}^2 \right)} - 1
.
\label{eq:g-modes:y}
\end{gather}
Finally, $\xi^r$ and $\xi_{({\rm b})}^r$ are the radial components of the Lagrangian displacements
	for the normal liquid component and baryons, respectively. 
	They are defined by
\begin{gather}
	u^r = {\rm i} \omega {\rm e}^{-\nu/2} \xi^r
	,\quad
	U_{({\rm b})}^r = {\rm i} \omega {\rm e}^{-\nu/2} \xi_{({\rm b})}^r
.
\label{eq:g-modes:xi}
\end{gather}
%

%%%%%%%%%%%%%%%%%%%%%%%%%%%%%%%%%%%%%%%%%%%%%%%%%%%%%%%%%%%%%%%%%%%%%
\subsection{Non-superfluid equations and boundary conditions}
%%%%%%%%%%%%%%%%%%%%%%%%%%%%%%%%%%%%%%%%%%%%%%%%%%%%%%%%%%%%%%%%%%%%%

Equations \eqref{eq:g-modes:1}--\eqref{eq:g-modes:4}
	describe the oscillations in the internal superfluid region of the star.
To calculate the eigenfrequencies of global oscillations
	(or to calculate the g-mode spectrum of a non-superfluid star)
	one should also consider the equations governing the oscillations of the non-superfluid matter
	(see e.g., \citealt{1983ApJ...268..837M,1992ApJ...395..240R}):
\begin{gather}
	-\frac{1}{{\rm e}^{\lambda/2}r^2} \pd{}{r}\left[{\rm e}^{\lambda/2}r^2 \xi_{(\rm b)}^r \right]
	+\frac{l(l+1){\rm e}^{\nu}}{r^2 \omega^2} \frac{\delta P}{P+\varepsilon} -\frac{\delta P+\nabla P\, \xi_{(\rm b)}^r}{\gamma P}
	= 0,
\label{eq:g-modes:1nsf}
	\\
	\pd{\delta P}{r}
		+ g\left(
			1 + \frac{1}{c_{\rm s}^2
		}\right)\delta P
		+ {\rm e}^{\lambda-\nu}(P+\varepsilon)(\mathcal{N}_{\rm nsf}^2-\omega^2) \xi_{(\rm b)}^r
	= 0.
\label{eq:g-modes:2nsf}
\end{gather}

Here $\mathcal{N}_{\rm nsf}$ is the 
	Brunt-V$\ddot{\rm a}$is$\ddot{\rm a}$l$\ddot{\rm a}$ frequency for the non-superfluid matter.
	In the (normal) core it is given by
\begin{gather}
	\mathcal{N}_{\rm nsf}^2
	= g^2\,\left(\frac{1}{c_{\rm eq}^2}-\frac{1}{c_s^2}\right)\, {\rm e}^{\nu-\lambda},
\label{eq:g-modes:Nnsf}
\end{gather}
where
	$c_{\rm eq}^2=\nabla P /(\mu_{\rm n} \nabla n_{\rm b})$;
	$c_{\rm s}^2\equiv \gamma P/(\mu_{\rm n} n_{\rm b})$;
	and
	$\gamma= (n_{\rm b}/P) \,\partial P(n_{\rm b}, n_{\rm e}/n_{\rm b}, n_{\rm \mu}/n_{\rm b}, n_{\rm str}/n_{\rm b})/\partial n_{\rm b}$ 
	is the (frozen) adiabatic index.
In the crust we set
	$\mathcal{N}_{\rm nsf} = 0$,
	thus ignoring possible surface g-modes
	localized at
	the interfaces between phases
	with different chemical composition
	%in the crust
	%there
	(\citealt{1987MNRAS.227..265F}).

Equations \eqref{eq:g-modes:1}--\eqref{eq:g-modes:4} and \eqref{eq:g-modes:1nsf}--\eqref{eq:g-modes:2nsf}
	should be supplied by the following boundary conditions.
\begin{enumerate}
\item 
	The existence of the finite solution to equations \eqref{eq:g-modes:1}--\eqref{eq:g-modes:4}
	implies that at the stellar centre
	\begin{gather}
		\xi^r \propto r^{l-1}
		,\quad
		\xi_{(\rm b)}^r \propto r^{l-1}
		,\quad
		\delta P \propto r^l
		,\quad
		\delta \mu_{\rm n} \propto r^l.
	\label{eq:g-modes:boundary-centre}
	\end{gather}
\item 
	The continuity of the electron (or muon) current as well as 
		the continuity of the energy and momentum currents
		through the superfluid/non-superfluid interface 
		result in
	\begin{gather}
		\xi_{(\rm b)}^r(r_0 - 0) = \xi_{(\rm b)}^r(r_0 + 0)
	\label{eq:g-modes:boundary-rcc-xi_b}
	,\\
		\delta P(r_0 - 0) = \delta P(r_0 + 0)
	\label{eq:g-modes:boundary-rcc-dP}
	,\\
		\xi^r_{(\rm b)}(r_0 - 0) = \xi^r(r_0 - 0),
	\label{eq:g-modes:boundary-rcc-xi}
	\end{gather}
	where $r_0$ is the radial coordinate of the interface. 
\item 
	Vanishing of the pressure $P$ at the stellar surface means 
	\begin{gather}
		\left. \delta P \right|_{r=R}
		+ \left. \xi^r \nabla P \right|_{r=R} = 0.
	\label{eq:g-modes:boundary-surface}
	\end{gather}
\end{enumerate}
A solution to the oscillation equations with these boundary conditions 
allows one to determine stellar eigenfrequencies and eigenfunctions in the Cowling approximation.

%%%%%%%%%%%%%%%%%%%%%%%%%%%%%%%%%%%%%%%%%%%%%%%%%%%%%%%%%%%%%%%%%%%%%%%%%%%%%%%%%%%%%%%%%%%%%%%%%%%%
\subsection{Local analysis and the Brunt-V$\ddot{\rm a}$is$\ddot{\rm a}$l$\ddot{\rm a}$ frequency}
%%%%%%%%%%%%%%%%%%%%%%%%%%%%%%%%%%%%%%%%%%%%%%%%%%%%%%%%%%%%%%%%%%%%%%%%%%%%%%%%%%%%%%%%%%%%%%%%%%%%

\begin{figure}
  \centering
  \includegraphics[width=.5\linewidth]{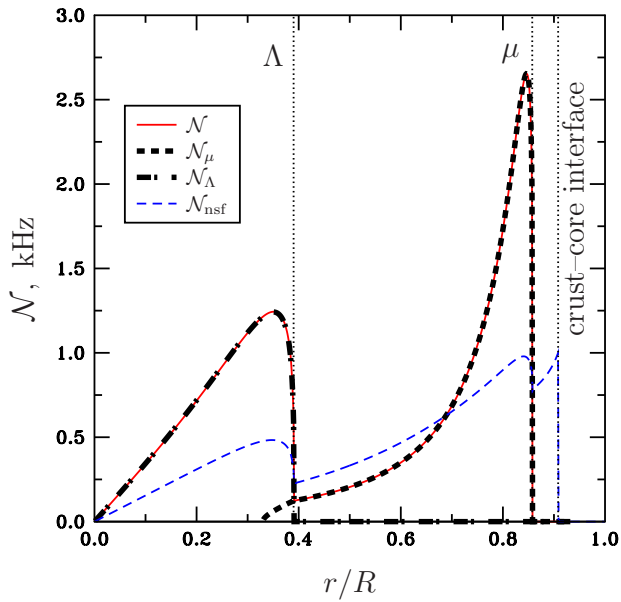}
\caption{
	Brunt-V$\ddot{\rm a}$is$\ddot{\rm a}$l$\ddot{\rm a}$ frequencies
		versus $r$ (in units of $R$).
	Red-shifted internal temperature is chosen to be
		$T^\infty = 10^7 ~\rm K$.
	Solid line:
		the Brunt-V$\ddot{\rm a}$is$\ddot{\rm a}$l$\ddot{\rm a}$
		frequency $\mathcal N$ (in kHz)
		for a superfluid HS.
	Dashed line:
		the Brunt-V$\ddot{\rm a}$is$\ddot{\rm a}$l$\ddot{\rm a}$
		frequency ${\mathcal N}_{\rm nsf}$ (in kHz)
		for a non-superfluid HS.
	Thick dashed line: the `muonic'
		Brunt-V$\ddot{\rm a}$is$\ddot{\rm a}$l$\ddot{\rm a}$
		frequency ${\mathcal N}_{\rm \mu}$.
	Thick dot--dashed line: the `$\rm \Lambda$--hyperonic'
		Brunt-V$\ddot{\rm a}$is$\ddot{\rm a}$l$\ddot{\rm a}$
		frequency ${\mathcal N}_{\rm \Lambda}$.
	Vertical dotted lines: the thresholds for the appearance of muons and $\rm \Lambda$--hyperons.
}
\label{fig:g-modes:N_BV}
\end{figure}

Examining short-wave perturbations of the system \eqref{eq:g-modes:1}--\eqref{eq:g-modes:4}, 
	proportional to ${\exp}[{\rm i} \int^r {\rm d}r' k(r')]$
	(WKB approximation, $k\gg |{\rm d} \, \ln k/{\rm d}r|$),
	one can find the standard 
	(see e.g., \citealt{1983ApJ...268..837M}) short-wave g-mode dispersion relation,
\begin{gather}
	\omega^2 = {\mathcal N}^2 \,
				\frac{l(l+1){\rm e}^{\lambda}}
					 {l (l+1) {\rm e}^{\lambda} + k^2 r^2},
	\label{eq:g-modes:dispersion}
\end{gather}
where
\begin{gather}
\label{eq:g-modes:N2}
	{\mathcal N}^2 = 
	-\frac{g}{\mu_{\rm n} n_{\rm b}} {\rm e}^{\nu-\lambda}\, \frac{(1+y)}{y} \,
	\left[
		\pd{w(P,\, \mu_{\rm n}, \, x_{\rm e\mu}, \, x_{\rm e\Lambda})}{x_{\rm e\mu}}  \,\, \nabla x_{\rm e\mu}
		+ \pd{w(P,\, \mu_{\rm n}, \, x_{\rm e\mu}, \, x_{\rm e\Lambda})}{x_{\rm e\Lambda}}  \,\, \nabla x_{\rm e\Lambda}
	\right]
\end{gather}
is the corresponding
	Brunt-V$\ddot{\rm a}$is$\ddot{\rm a}$l$\ddot{\rm a}$ 
	frequency squared.
It can be written as a sum of two terms,
	$\mathcal{N}^2 = \mathcal{N}_{\rm\mu}^2 + \mathcal{N}_{\rm\Lambda}^2$,
	where $\mathcal{N}_{\rm \mu}^2$
	and $\mathcal{N}_{\rm \Lambda}^2$
	correspond, respectively, 
	to the first and second terms
	in the square brackets
	in equation~\eqref{eq:g-modes:N2}.
The frequencies
	${\mathcal N} (r)$ (solid line),
	${\mathcal N}_{\rm\mu} (r)$ (thick dashed line)
	and ${\mathcal N}_{\rm\Lambda} (r)$  (thick dot--dashed line)
	are plotted in Fig.~\ref{fig:g-modes:N_BV}.
Since ${\mathcal N}_{\rm\mu}^2 (r) < 0$
	at $r/R < 0.33$,
	${\mathcal N}_{\rm\mu}$ becomes imaginary 
	and we do not plot it in this region.
However, the fact that ${\mathcal N}_{\rm\mu}^2 < 0$
	does not lead to convective instability,
	because ${\mathcal N}^2$
	and, therefore,
	$\omega^2$
	are still positive.
Note also that in the inner core
	${\mathcal N}_{\rm\mu}^2$
	is much smaller than
	${\mathcal N}_{\rm\Lambda}^2$,
	hence
	 ${\mathcal N}$
	is approximately equal to ${\mathcal N}_{\rm\Lambda}$
	in that region.

Since ${\mathcal N} (r)$
	has two peaks, associated with
	${\mathcal N}_{\rm \mu}$
	and ${\mathcal N}_{\rm \Lambda}$,
	we can expect the existence of
	two types of modes,
	which it is convenient to call `muonic' and `$\rm \Lambda$-hyperonic' g-modes.
The main difference between them is in their localization.
The muonic g-modes
	should be
	localized in the region where muons exist
	($r/R < 0.857$ for the considered HS model),
	whereas the $\rm \Lambda$-hyperonic modes
	should be
	localized only in the inner core, where $\rm \Lambda$--hyperons are present
		($r/R < 0.39$).
As we show below,
	numerical calculations 
	confirm this hypothesis.

If a star is non-superfluid,
	the local analysis of equations
	\eqref{eq:g-modes:1nsf} and \eqref{eq:g-modes:2nsf}
	leads to a similar dispersion relation~\eqref{eq:g-modes:dispersion}
	with ${\mathcal N}^2 = {\mathcal N}_{\rm nsf}^2$,
	where
	the Brunt-V$\ddot{\rm a}$is$\ddot{\rm a}$l$\ddot{\rm a}$
	frequency for non-superfluid matter,
	${\mathcal N}_{\rm nsf}$, is defined
	by equation~\eqref{eq:g-modes:Nnsf}.
${\mathcal N}_{\rm nsf} (r)$ (dashed line)
	is also shown in Fig.~\ref{fig:g-modes:N_BV}.

%%%%%%%%%%%%%%%%%%%%%%%%%%%%%%%%%%%%%%%%%%%%%%%%%%%%%%%%%%%%%%%%%%%
%%%%%%%%%%%%%%%%%%%%%%%%%%%%%%%%%%%%%%%%%%%%%%%%%%%%%%%%%%%%%%%%%%%

%%%%%%%%%%%%%%%%%%%%%%%%%%%%%%%%%%%%%%%%%%%%%%%%%%%%%%%%%%%%%
\subsection{Numerical results}
%%%%%%%%%%%%%%%%%%%%%%%%%%%%%%%%%%%%%%%%%%%%%%%%%%%%%%%%%%%%%

\begin{figure}
  \centering
  \includegraphics[width=.5\linewidth]{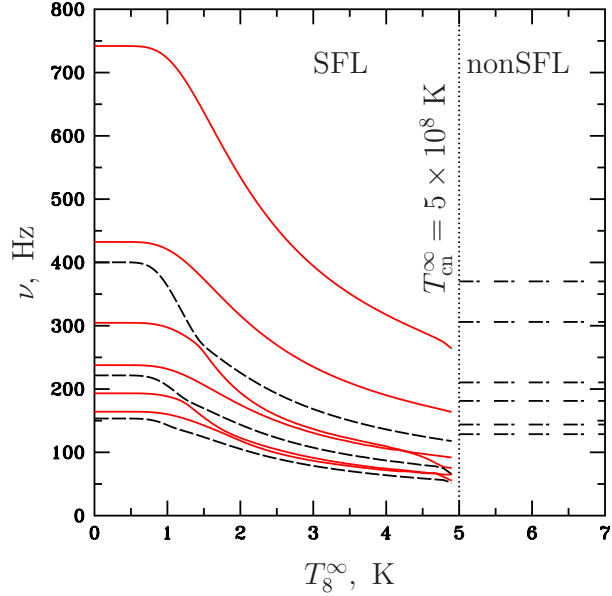}
\caption{
	Spectrum of quadrupolar ($l = 2$) g-modes 
		versus $T_8^\infty \equiv T^\infty/(10^8 \,\rm K)$.
	Critical temperatures:
		$T_{\rm cn}^\infty = 5 \times 10^8 ~\rm K$,
		$T_{\rm cp}^\infty = 3 \times 10^9 ~\rm K$.
		$\rm \Lambda$--hyperons are assumed to be non-superfluid.
	Solid lines: eigenfrequencies (in Hz) for the g-modes which are `$\rm \Lambda$--hyperonic' at $T^\infty=0$.
	Dashed lines: eigenfrequencies for the g-modes which are `muonic' at $T^\infty=0$.
	Dot--dashed lines: eigenfrequencies for
		the g-modes in a non-superfluid HS.
	Vertical dotted line: the redshifted critical temperature for neutrons.
}
\label{fig:g-modes:spectrum}
\end{figure}

The spectrum
	of the first nine quadrupolar ($l = 2$) g-modes
	for a chosen HS model
	is shown in 
	Fig.~\ref{fig:g-modes:spectrum} as a function of $T^\infty$.
The solid lines present the eigenfrequencies $\nu = \omega/(2\pi)$ 
	for the g-modes which are `$\rm \Lambda$--hyperonic' at $T^\infty=0$,
	the dashed lines present eigenfrequencies 
	for the g-modes which are `muonic' at $T^\infty=0$ 
	[because of numerous avoided crossings of the modes 
		(see Fig.~\ref{fig:g-modes:spectrum} and a discussion below) 
		any muonic g-mode may turn into a $\rm \Lambda$-hyperonic g-mode
		with growing $T^\infty$ (and vice versa)].
The dot--dashed lines show
	the g-mode eigenfrequencies
	%which are plotted
	for	a non-superfluid HS of the same mass.
As one could expect, they do not depend on $T^\infty$.

\begin{figure}
  \centering
  \includegraphics[width=.5\linewidth]{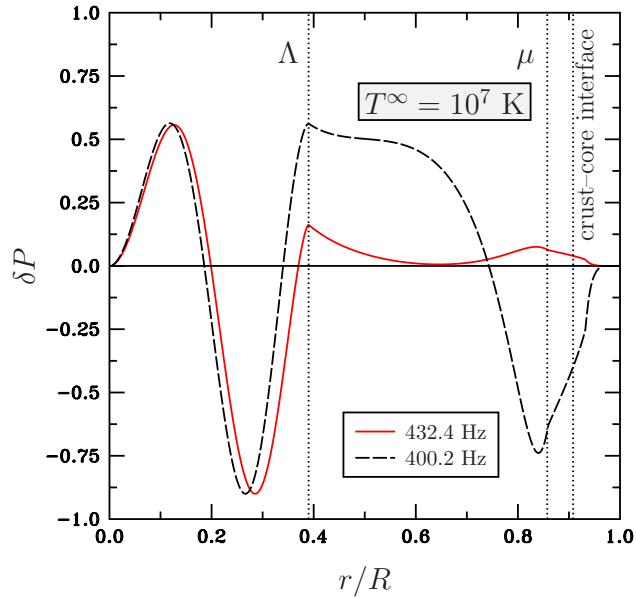}
\caption{
	Perturbation of the pressure $\delta P$ (dimensionless)
	versus $r$ (in units of $R$)
		for the second `$\rm \Lambda$--hyperonic' g-mode ($\nu \approx 432~\rm Hz$, solid line)
		and the first `muonic' g-mode ($\nu \approx 400~\rm Hz$, dashed line).
	%EOS GM1'B~\citet{2014MNRAS.439..318G}.
	The red-shifted internal temperature is chosen to be
	$T^\infty = 10^7 ~\rm K$.
	%$T_{\rm cn}^\infty = 5 \times 10^8 ~\rm K$,
	%	$T_{\rm cp}^\infty = 5 \times 10^9 ~\rm K$,
%		$\rm \Lambda$--hyperons are non-superfluid.
	Vertical dotted lines: the threshold for the appearance of muons and $\rm \Lambda$--hyperons.
}
\label{fig:g-modes:eigenfuncs}
\end{figure}

The difference between the muonic
	and $\rm \Lambda$--hyperonic
	superfluid g-modes
	is illustrated in Fig.~\ref{fig:g-modes:eigenfuncs},
	where the eigenfunctions $\delta P (r)$ (dimensionless) are 
	plotted for the two modes with close frequencies:
		$\nu \approx 432~\rm Hz$ (solid line) and $\nu \approx 400~\rm Hz$ (dashed line).
The red-shifted internal temperature is chosen to be
	$T^\infty = 10^7 ~\rm K$.
One can see that the mode with 
	$\nu \approx 432~\rm Hz$ 
	is localized only in the inner core,
	where $\rm \Lambda$--hyperons exist,
	and hence it could be called $\rm \Lambda$--hyperonic.
In contrast,
	an area of localization of
	the $\nu \approx 400~\rm Hz$ mode
	coincides with the region where muons are present,  
	hence we call it muonic g-mode.

When the frequencies of two different modes come close to each other, 
	they demonstrate, as in the case of sound modes, an avoided crossing.
Since the eigenfrequencies of two neighbouring modes
	near an avoided crossing
	may differ by just a few Hz
	(as e.g., in the case of the fourth and fifth modes in Fig.~\ref{fig:g-modes:spectrum}
	at $T^\infty \sim 4 \times 10^8~{\rm K}$),
	it is sometimes hard to distinguish the avoided crossing
	from the ordinary crossing of modes in the plot.

Although the superfluid HS matter is strongly degenerate,
	the g-mode frequencies
	(as well as	the Brunt-V$\ddot{\rm a}$is$\ddot{\rm a}$l$\ddot{\rm a}$ frequency)
	strongly depend on $T^\infty$
	through the parameter $y$,
	which, in turn, can be expressed through
	the temperature-dependent
	entrainment matrix $Y_{ik}$.
Temperature dependence of g-mode frequencies 
in Fig.~\ref{fig:g-modes:spectrum}
	is very similar to a dependence shown in fig.~4
	in~\citet{2014MNRAS.442L..90K}.
That plot, as well as ours,
	was obtained under the assumption that
	$T_{{\rm c} i}^\infty$
	are constant throughout the core.
However, one should keep in mind
	that, adopting a more realistic superfluidity model,
	in which $T_{\rm cn}^\infty$ and $T_{\rm cp}^\infty$ depend on density,
	will lead to a different behaviour of the spectrum
	at $T^{\infty}$ close to $T_{\rm cn}^\infty$.
Namely, the superfluid g-modes do not vanish
	at $T^\infty \rightarrow T_{\rm cn}^\infty$
	but turn into ordinary g-modes of a non-superfluid star
	(see fig.~5 and its discussion
	in \citealt{2014MNRAS.442L..90K}
	for details).

Note that the eigenfrequency of the fundamental quadrupolar ($l=2$) g-mode 
	turns out to be exceptionally large ($\nu \approx 742 ~{\rm Hz}$
	in the low-temperature limit,
	$T^\infty \ll T_{\rm cn}^\infty,~T_{\rm cp}^\infty$).
For comparison, the eigenfrequency of the corresponding mode
	in a neutron star with the $\rm npe\mu$ core composition,
	calculated by \citet{2014MNRAS.442L..90K},
	equals $\nu \approx 462 \rm~Hz$.
The g-mode frequencies for non-superfluid HSs
	are also quite large
	(up to $\sim 370 ~{\rm Hz}$)
	in comparison to those for non-superfluid neutron stars with the $\rm npe$ core composition
	($\sim 50-150$~Hz; see e.g., \citealt{1992ApJ...395..240R}).
Such high frequencies arise in HSs because of the strong stratification,
	which leads to a large value of 
	the Brunt-V$\ddot{\rm a}$is$\ddot{\rm a}$l$\ddot{\rm a}$ frequency
	and, hence, to large oscillation frequencies.

%%%%%%%%%%%%%%%%%%%%%%%%%%%%%%%%%%%%%%%%%%%%%%%%%%%%%%%%%%%%%%%%%%%
%%%%%%%%%%%%%%%%%%%%%%%%%%%%%%%%%%%%%%%%%%%%%%%%%%%%%%%%%%%%%%%%%%%
%%%%%%%%%%%%%%%%%%%%%%%%%%%%%%%%%%%%%%%%%%%%%%%%%%%%%%%%%%%%%%%%%%%
%%%%%%%%%%%%%%%%%%%%%%%%%%%%%%%%%%%%%%%%%%%%%%%%%%%%%%%%%%%%%%%%%%%

%%%%%%%%%%%%%%%%%%%%%%%%%%%%%%%%%%%%%%%%%%%%%%%%%%%%%%%%%%%%%%%%%%%
\section{Summary}
\label{sec:summary}
%%%%%%%%%%%%%%%%%%%%%%%%%%%%%%%%%%%%%%%%%%%%%%%%%%%%%%%%%%%%%%%%%%%

In this paper we generalized to the case of
	the nucleon-hyperon matter an approximate method of decoupling
	of superfluid and normal degrees of freedom, 
	suggested by \citet{2011PhRvD..83h1304G}
	and \citet{2013MNRAS.428.1518G}.
We showed that the equations
	governing the oscillations
	of superfluid hyperon stars (HSs)
	can be split into two weakly coupled
	systems of equations
	with the coupling parameters
	$s_{\rm e}$, $s_{\rm \mu}$, and $s_{\rm str}$,
	given by Eq.~\eqref{eq:ss}.
These two systems describe the
	`normal' and `superfluid'
	oscillation modes.
Neglecting the rather small coupling terms
	(i.e. putting $s_{\rm e} = s_{\rm \mu} = s_{\rm str} = 0$; 
	the so called `decoupling approximation')
	allows one to drastically simplify
	the calculations of the oscillation spectra.
Namely, we have shown that in the decoupling regime
	the normal modes coincide with
	the ordinary modes of a non-superfluid HS
	and can be calculated
	within the non-superfluid hydrodynamics.
As for the superfluid modes,
	in this approximation
	they can be calculated
	by using only two `superfluid' equations
	\eqref{eq:sfl-osc-n} and \eqref{eq:sfl-osc-L}
	(along with the continuity equations~\eqref{eq:continuity}
		and the conditions
		$\delta U_{({\rm b})}^\alpha = \delta g^{\alpha\beta} = 0$).
These modes do not perturb metric, pressure,
	baryon current density,
	and are localized in the superfluid region of a star.
	It is shown how the proposed approach can be modified 
	to study the oscillations in {\it rotating} HSs, 
	containing arrays
	of Feynman-Onsager vortices.

An efficiency of the presented decoupling scheme
	is illustrated in Section~\ref{sec:sound}
	by the calculation, using modern hyperonic EOSs, 
	of the sound speeds in the superfluid nucleon-hyperon matter
	at arbitrary temperature.
It is shown that the approximate approach qualitatively well 
	reproduces the results of the accurate calculation.
Summarizing, the decoupling scheme presented here  
	can be used to study various 
	oscillation modes 
	in rotating superfluid HSs 
	(e.g., p-, f-, and r-modes).
Such a detailed analysis 
	is beyond the scope of the present paper.
	
Unfortunately, there exist a class of oscillations, namely the gravity modes (g-modes), 
	that cannot be treated
	within the proposed simple scheme
	and should be considered separately.
We have performed such an analysis in Section~\ref{sec:g-modes},
	where we, for the first time,  
	discussed the composition g-modes 
	in a star with a superfluid $\rm npe\mu\Lambda$ core.
Our consideration complements the results of \citet{2014MNRAS.442L..90K} 
	who analysed the g-modes in 
	superfluid neutron stars with an $\rm npe\mu$ core.
We showed that such a HS harbours two types of superfluid g-modes,
	which we call `muonic' and `$\rm \Lambda$--hyperonic'.
The eigenfrequencies of g-modes in superfluid HSs
	turn out to be exceptionally large
	(up to $\nu \approx 742~{\rm Hz}$ for the considered HS model).
	This may have a strong impact on the 
	properties of inertial-gravity modes
	in rotating stars, and, as a consequence, 
	on damping and saturation of r-modes with which they can interact.
Also, the g-modes analysed in this paper
	may substantially modify gravitational-wave signal
	from coalescing HS--compact star
	(or HS--black hole) binaries
	(see
		\citealt{1999MNRAS.307.1001L,
			1999MNRAS.308..153H}).
More details on these issues as well as 
	other possible implications of our result
	are discussed by
	\citet{2014MNRAS.442L..90K}.
We hope to address some of these problems in the near future.

%%%%%%%%%%%%%%%%%%%%%%%%%%%%%%%%%%%%%%%%%%%%%%%%%%%%%%%%%%%%%%%%%%%
\section{Acknowledgements}
%%%%%%%%%%%%%%%%%%%%%%%%%%%%%%%%%%%%%%%%%%%%%%%%%%%%%%%%%%%%%%%%%%%
We are very grateful to D.P.~Barsukov, E.M.~Kantor, and D.G.~Yakovlev
	for reading the draft of the paper and many valuable suggestions,
	and to A.I.~Chugunov for discussions.
This work was partially supported by
 RFBR (grants 14-02-00868-a and 14-02-31616-mol-a),
 by RF president programme
(grants MK-506.2014.2 and NSh-294.2014.2),
 and by the Dynasty Foundation.

\bibliography{hyperons}

\begin{thebibliography}{}
\makeatletter
\relax
\def\mn@urlcharsother{\let\do\@makeother \do\$\do\&\do\#\do\^\do\_\do\%\do\~}
\def\mn@doi{\begingroup\mn@urlcharsother \@ifnextchar [ {\mn@doi@}
  {\mn@doi@[]}}
\def\mn@doi@[#1]#2{\def\@tempa{#1}\ifx\@tempa\@empty \href
  {http://dx.doi.org/#2} {doi:#2}\else \href {http://dx.doi.org/#2} {#1}\fi
  \endgroup}
\def\mn@eprint#1#2{\mn@eprint@#1:#2::\@nil}
\def\mn@eprint@arXiv#1{\href {http://arxiv.org/abs/#1} {{\tt arXiv:#1}}}
\def\mn@eprint@dblp#1{\href {http://dblp.uni-trier.de/rec/bibtex/#1.xml}
  {dblp:#1}}
\def\mn@eprint@#1:#2:#3:#4\@nil{\def\@tempa {#1}\def\@tempb {#2}\def\@tempc
  {#3}\ifx \@tempc \@empty \let \@tempc \@tempb \let \@tempb \@tempa \fi \ifx
  \@tempb \@empty \def\@tempb {arXiv}\fi \@ifundefined
  {mn@eprint@\@tempb}{\@tempb:\@tempc}{\expandafter \expandafter \csname
  mn@eprint@\@tempb\endcsname \expandafter{\@tempc}}}

\bibitem[\protect\citeauthoryear{{Andersson}}{{Andersson}}{2003}]{2003CQGra..20R.105A}
{Andersson} N.,  2003, Classical and Quantum Gravity, \href
  {http://adsabs.harvard.edu/abs/2003CQGra..20R.105A} {20, 105}

\bibitem[\protect\citeauthoryear{{Andersson} \& {Comer}}{{Andersson} \&
  {Comer}}{2001}]{2001MNRAS.328.1129A}
{Andersson} N.,  {Comer} G.~L.,  2001, \mn@doi [\mnras]
  {10.1046/j.1365-8711.2001.04923.x}, \href
  {http://adsabs.harvard.edu/abs/2001MNRAS.328.1129A} {328, 1129}

\bibitem[\protect\citeauthoryear{{Andersson}, {Ferrari}, {Jones}, {Kokkotas},
  {Krishnan}, {Read}, {Rezzolla}  \& {Zink}}{{Andersson}
  et~al.}{2011}]{2011GReGr..43..409A}
{Andersson} N.,  {Ferrari} V.,  {Jones} D.~I.,  {Kokkotas} K.~D.,  {Krishnan}
  B.,  {Read} J.~S.,  {Rezzolla} L.,   {Zink} B.,  2011, \mn@doi [General
  Relativity and Gravitation] {10.1007/s10714-010-1059-4}, \href
  {http://adsabs.harvard.edu/abs/2011GReGr..43..409A} {43, 409}

\bibitem[\protect\citeauthoryear{{Andersson} et~al.,}{{Andersson}
  et~al.}{2013}]{2013CQGra..30s3002A}
{Andersson} N.,  et~al., 2013, \mn@doi [Classical and Quantum Gravity]
  {10.1088/0264-9381/30/19/193002}, \href
  {http://adsabs.harvard.edu/abs/2013CQGra..30s3002A} {30, 193002}

\bibitem[\protect\citeauthoryear{{Andersson}, {Jones}  \& {Ho}}{{Andersson}
  et~al.}{2014}]{2014MNRAS.442.1786A}
{Andersson} N.,  {Jones} D.~I.,   {Ho} W.~C.~G.,  2014, \mn@doi [\mnras]
  {10.1093/mnras/stu870}, \href
  {http://adsabs.harvard.edu/abs/2014MNRAS.442.1786A} {442, 1786}

\bibitem[\protect\citeauthoryear{{Bednarek}, {Haensel}, {Zdunik}, {Bejger}  \&
  {Ma{\'n}ka}}{{Bednarek} et~al.}{2012}]{2012A&A...543A.157B}
{Bednarek} I.,  {Haensel} P.,  {Zdunik} J.~L.,  {Bejger} M.,   {Ma{\'n}ka} R.,
  2012, \mn@doi [\aap] {10.1051/0004-6361/201118560}, \href
  {http://adsabs.harvard.edu/abs/2012A%26A...543A.157B} {543, A157}

\bibitem[\protect\citeauthoryear{{Benhar}, {Ferrari}  \& {Gualtieri}}{{Benhar}
  et~al.}{2004}]{2004PhRvD..70l4015B}
{Benhar} O.,  {Ferrari} V.,   {Gualtieri} L.,  2004, \mn@doi [\prd]
  {10.1103/PhysRevD.70.124015}, \href
  {http://adsabs.harvard.edu/abs/2004PhRvD..70l4015B} {70, 124015}

\bibitem[\protect\citeauthoryear{{Bl{\'a}zquez-Salcedo}, {Gonz{\'a}lez-Romero}
  \& {Navarro-L{\'e}rida}}{{Bl{\'a}zquez-Salcedo}
  et~al.}{2014}]{2014PhRvD..89d4006B}
{Bl{\'a}zquez-Salcedo} J.~L.,  {Gonz{\'a}lez-Romero} L.~M.,
  {Navarro-L{\'e}rida} F.,  2014, \mn@doi [\prd] {10.1103/PhysRevD.89.044006},
  \href {http://adsabs.harvard.edu/abs/2014PhRvD..89d4006B} {89, 044006}

\bibitem[\protect\citeauthoryear{{Bondarescu}, {Teukolsky}  \&
  {Wasserman}}{{Bondarescu} et~al.}{2007}]{2007PhRvD..76f4019B}
{Bondarescu} R.,  {Teukolsky} S.~A.,   {Wasserman} I.,  2007, \mn@doi [\prd]
  {10.1103/PhysRevD.76.064019}, \href
  {http://adsabs.harvard.edu/abs/2007PhRvD..76f4019B} {76, 064019}

\bibitem[\protect\citeauthoryear{{Chirenti}, {de Souza}  \&
  {Kastaun}}{{Chirenti} et~al.}{2015}]{2015PhRvD..91d4034C}
{Chirenti} C.,  {de Souza} G.~H.,   {Kastaun} W.,  2015, \mn@doi [\prd]
  {10.1103/PhysRevD.91.044034}, \href
  {http://adsabs.harvard.edu/abs/2015PhRvD..91d4034C} {91, 044034}

\bibitem[\protect\citeauthoryear{{Chugunov} \& {Gusakov}}{{Chugunov} \&
  {Gusakov}}{2011}]{2011MNRAS.418L..54C}
{Chugunov} A.~I.,  {Gusakov} M.~E.,  2011, \mn@doi [\mnras]
  {10.1111/j.1745-3933.2011.01142.x}, \href
  {http://adsabs.harvard.edu/abs/2011MNRAS.418L..54C} {418, L54}

\bibitem[\protect\citeauthoryear{{Cowling}}{{Cowling}}{1941}]{1941MNRAS.101..367C}
{Cowling} T.~G.,  1941, \mnras, \href
  {http://adsabs.harvard.edu/abs/1941MNRAS.101..367C} {101, 367}

\bibitem[\protect\citeauthoryear{{Finn}}{{Finn}}{1987}]{1987MNRAS.227..265F}
{Finn} L.~S.,  1987, \mnras, \href
  {http://adsabs.harvard.edu/abs/1987MNRAS.227..265F} {227, 265}

\bibitem[\protect\citeauthoryear{{Gaertig} \& {Kokkotas}}{{Gaertig} \&
  {Kokkotas}}{2009}]{2009PhRvD..80f4026G}
{Gaertig} E.,  {Kokkotas} K.~D.,  2009, \mn@doi [\prd]
  {10.1103/PhysRevD.80.064026}, \href
  {http://adsabs.harvard.edu/abs/2009PhRvD..80f4026G} {80, 064026}

\bibitem[\protect\citeauthoryear{{Gezerlis}, {Pethick}  \&
  {Schwenk}}{{Gezerlis} et~al.}{2014}]{2014arXiv1406.6109G}
{Gezerlis} A.,  {Pethick} C.~J.,   {Schwenk} A.,  2014, preprint, \href
  {http://adsabs.harvard.edu/abs/2014arXiv1406.6109G} {} (\mn@eprint {arXiv}
  {1406.6109})

\bibitem[\protect\citeauthoryear{{Gualtieri}, {Kantor}, {Gusakov}  \&
  {Chugunov}}{{Gualtieri} et~al.}{2014}]{2014PhRvD..90b4010G}
{Gualtieri} L.,  {Kantor} E.~M.,  {Gusakov} M.~E.,   {Chugunov} A.~I.,  2014,
  \mn@doi [\prd] {10.1103/PhysRevD.90.024010}, \href
  {http://adsabs.harvard.edu/abs/2014PhRvD..90b4010G} {90, 024010}

\bibitem[\protect\citeauthoryear{{Gusakov}}{{Gusakov}}{2007}]{2007PhRvD..76h3001G}
{Gusakov} M.~E.,  2007, \mn@doi [\prd] {10.1103/PhysRevD.76.083001}, \href
  {http://adsabs.harvard.edu/abs/2007PhRvD..76h3001G} {76, 083001}

\bibitem[\protect\citeauthoryear{{Gusakov} \& {Andersson}}{{Gusakov} \&
  {Andersson}}{2006}]{2006MNRAS.372.1776G}
{Gusakov} M.~E.,  {Andersson} N.,  2006, \mn@doi [\mnras]
  {10.1111/j.1365-2966.2006.10982.x}, \href
  {http://adsabs.harvard.edu/abs/2006MNRAS.372.1776G} {372, 1776}

\bibitem[\protect\citeauthoryear{{Gusakov} \& {Kantor}}{{Gusakov} \&
  {Kantor}}{2008}]{2008PhRvD..78h3006G}
{Gusakov} M.~E.,  {Kantor} E.~M.,  2008, \mn@doi [\prd]
  {10.1103/PhysRevD.78.083006}, \href
  {http://adsabs.harvard.edu/abs/2008PhRvD..78h3006G} {78, 083006}

\bibitem[\protect\citeauthoryear{{Gusakov} \& {Kantor}}{{Gusakov} \&
  {Kantor}}{2011}]{2011PhRvD..83h1304G}
{Gusakov} M.~E.,  {Kantor} E.~M.,  2011, \mn@doi [\prd]
  {10.1103/PhysRevD.83.081304}, \href
  {http://adsabs.harvard.edu/abs/2011PhRvD..83h1304G} {83, 081304}

\bibitem[\protect\citeauthoryear{{Gusakov}, {Kantor}  \& {Haensel}}{{Gusakov}
  et~al.}{2009a}]{2009PhRvC..79e5806G}
{Gusakov} M.~E.,  {Kantor} E.~M.,   {Haensel} P.,  2009a, \mn@doi [\prc]
  {10.1103/PhysRevC.79.055806}, \href
  {http://adsabs.harvard.edu/abs/2009PhRvC..79e5806G} {79, 055806}

\bibitem[\protect\citeauthoryear{{Gusakov}, {Kantor}  \& {Haensel}}{{Gusakov}
  et~al.}{2009b}]{2009PhRvC..80a5803G}
{Gusakov} M.~E.,  {Kantor} E.~M.,   {Haensel} P.,  2009b, \mn@doi [\prc]
  {10.1103/PhysRevC.80.015803}, \href
  {http://adsabs.harvard.edu/abs/2009PhRvC..80a5803G} {80, 015803}

\bibitem[\protect\citeauthoryear{{Gusakov}, {Kantor}, {Chugunov}  \&
  {Gualtieri}}{{Gusakov} et~al.}{2013}]{2013MNRAS.428.1518G}
{Gusakov} M.~E.,  {Kantor} E.~M.,  {Chugunov} A.~I.,   {Gualtieri} L.,  2013,
  \mn@doi [\mnras] {10.1093/mnras/sts129}, \href
  {http://adsabs.harvard.edu/abs/2013MNRAS.428.1518G} {428, 1518}

\bibitem[\protect\citeauthoryear{{Gusakov}, {Haensel}  \& {Kantor}}{{Gusakov}
  et~al.}{2014}]{2014MNRAS.439..318G}
{Gusakov} M.~E.,  {Haensel} P.,   {Kantor} E.~M.,  2014, \mn@doi [\mnras]
  {10.1093/mnras/stt2438}, \href
  {http://adsabs.harvard.edu/abs/2014MNRAS.439..318G} {439, 318}

\bibitem[\protect\citeauthoryear{{Haensel}, {Levenfish}  \&
  {Yakovlev}}{{Haensel} et~al.}{2002}]{2002A&A...394..213H}
{Haensel} P.,  {Levenfish} K.~P.,   {Yakovlev} D.~G.,  2002, \mn@doi [\aap]
  {10.1051/0004-6361:20021112}, \href
  {http://adsabs.harvard.edu/abs/2002A%26A...394..213H} {394, 213}

\bibitem[\protect\citeauthoryear{{Haskell} \& {Andersson}}{{Haskell} \&
  {Andersson}}{2010}]{2010MNRAS.408.1897H}
{Haskell} B.,  {Andersson} N.,  2010, \mn@doi [\mnras]
  {10.1111/j.1365-2966.2010.17255.x}, \href
  {http://adsabs.harvard.edu/abs/2010MNRAS.408.1897H} {408, 1897}

\bibitem[\protect\citeauthoryear{{Haskell}, {Andersson}  \& {Comer}}{{Haskell}
  et~al.}{2012}]{2012PhRvD..86f3002H}
{Haskell} B.,  {Andersson} N.,   {Comer} G.~L.,  2012, \mn@doi [\prd]
  {10.1103/PhysRevD.86.063002}, \href
  {http://adsabs.harvard.edu/abs/2012PhRvD..86f3002H} {86, 063002}

\bibitem[\protect\citeauthoryear{{Ho} \& {Lai}}{{Ho} \&
  {Lai}}{1999}]{1999MNRAS.308..153H}
{Ho} W.~C.~G.,  {Lai} D.,  1999, \mn@doi [\mnras]
  {10.1046/j.1365-8711.1999.02703.x}, \href
  {http://adsabs.harvard.edu/abs/1999MNRAS.308..153H} {308, 153}

\bibitem[\protect\citeauthoryear{{Israel} et~al.,}{{Israel}
  et~al.}{2005}]{2005ApJ...628L..53I}
{Israel} G.~L.,  et~al., 2005, \mn@doi [\apjl] {10.1086/432615}, \href
  {http://adsabs.harvard.edu/abs/2005ApJ...628L..53I} {628, L53}

\bibitem[\protect\citeauthoryear{{Kantor} \& {Gusakov}}{{Kantor} \&
  {Gusakov}}{2009}]{2009PhRvD..79d3004K}
{Kantor} E.~M.,  {Gusakov} M.~E.,  2009, \mn@doi [\prd]
  {10.1103/PhysRevD.79.043004}, \href
  {http://adsabs.harvard.edu/abs/2009PhRvD..79d3004K} {79, 043004}

\bibitem[\protect\citeauthoryear{{Kantor} \& {Gusakov}}{{Kantor} \&
  {Gusakov}}{2011}]{2011PhRvD..83j3008K}
{Kantor} E.~M.,  {Gusakov} M.~E.,  2011, \mn@doi [\prd]
  {10.1103/PhysRevD.83.103008}, \href
  {http://adsabs.harvard.edu/abs/2011PhRvD..83j3008K} {83, 103008}

\bibitem[\protect\citeauthoryear{{Kantor} \& {Gusakov}}{{Kantor} \&
  {Gusakov}}{2012}]{2012ASPC..466..211K}
{Kantor} E.~M.,  {Gusakov} M.~E.,  2012, in {Lewandowski} W.,  {Maron} O.,
  {Kijak} J.,  eds,  Astronomical Society of the Pacific Conference Series Vol.
  466, Electromagnetic Radiation from Pulsars and Magnetars. p.~211

\bibitem[\protect\citeauthoryear{{Kantor} \& {Gusakov}}{{Kantor} \&
  {Gusakov}}{2014}]{2014MNRAS.442L..90K}
{Kantor} E.~M.,  {Gusakov} M.~E.,  2014, \mn@doi [\mnras]
  {10.1093/mnrasl/slu061}, \href
  {http://adsabs.harvard.edu/abs/2014MNRAS.442L..90K} {442, L90}

\bibitem[\protect\citeauthoryear{{Khalatnikov} \& {Bekarevich}}{{Khalatnikov}
  \& {Bekarevich}}{1961}]{1961JETP..40..920K}
{Khalatnikov} I.~M.,  {Bekarevich} I.~L.,  1961, JETP, 40, 920

\bibitem[\protect\citeauthoryear{{Lai}}{{Lai}}{1999}]{1999MNRAS.307.1001L}
{Lai} D.,  1999, \mn@doi [\mnras] {10.1046/j.1365-8711.1999.02723.x}, \href
  {http://adsabs.harvard.edu/abs/1999MNRAS.307.1001L} {307, 1001}

\bibitem[\protect\citeauthoryear{{Landau} \& {Lifshitz}}{{Landau} \&
  {Lifshitz}}{1980}]{1980stph.book.....L}
{Landau} L.~D.,  {Lifshitz} E.~M.,  1980, {Statistical physics. Pt.1, Pt.2}

\bibitem[\protect\citeauthoryear{{Lee}}{{Lee}}{1995}]{1995A&A...303..515L}
{Lee} U.,  1995, \aap, \href
  {http://adsabs.harvard.edu/abs/1995A%26A...303..515L} {303, 515}

\bibitem[\protect\citeauthoryear{{Lee}}{{Lee}}{2014}]{2014MNRAS.442.3037L}
{Lee} U.,  2014, \mn@doi [\mnras] {10.1093/mnras/stu1077}, \href
  {http://adsabs.harvard.edu/abs/2014MNRAS.442.3037L} {442, 3037}

\bibitem[\protect\citeauthoryear{{Lindblom} \& {Owen}}{{Lindblom} \&
  {Owen}}{2002}]{2002PhRvD..65f3006L}
{Lindblom} L.,  {Owen} B.~J.,  2002, \mn@doi [\prd]
  {10.1103/PhysRevD.65.063006}, \href
  {http://adsabs.harvard.edu/abs/2002PhRvD..65f3006L} {65, 063006}

\bibitem[\protect\citeauthoryear{{Lindblom} \& {Splinter}}{{Lindblom} \&
  {Splinter}}{1990}]{1990ApJ...348..198L}
{Lindblom} L.,  {Splinter} R.~J.,  1990, \mn@doi [\apj] {10.1086/168227}, \href
  {http://adsabs.harvard.edu/abs/1990ApJ...348..198L} {348, 198}

\bibitem[\protect\citeauthoryear{{Lombardo} \& {Schulze}}{{Lombardo} \&
  {Schulze}}{2001}]{2001LNP...578...30L}
{Lombardo} U.,  {Schulze} H.-J.,  2001, in {Blaschke} D.,  {Glendenning} N.~K.,
    {Sedrakian} A.,  eds,  Lecture Notes in Physics, Berlin Springer Verlag
  Vol. 578, Physics of Neutron Star Interiors. p.~30 (\mn@eprint {}
  {astro-ph/0012209})

\bibitem[\protect\citeauthoryear{{McDermott}, {van Horn}  \&
  {Scholl}}{{McDermott} et~al.}{1983}]{1983ApJ...268..837M}
{McDermott} P.~N.,  {van Horn} H.~M.,   {Scholl} J.~F.,  1983, \mn@doi [\apj]
  {10.1086/161006}, \href {http://adsabs.harvard.edu/abs/1983ApJ...268..837M}
  {268, 837}

\bibitem[\protect\citeauthoryear{{Mendell} \& {Lindblom}}{{Mendell} \&
  {Lindblom}}{1991}]{1991AnPhy.205..110M}
{Mendell} G.,  {Lindblom} L.,  1991, \mn@doi [Annals of Physics]
  {10.1016/0003-4916(91)90239-5}, \href
  {http://adsabs.harvard.edu/abs/1991AnPhy.205..110M} {205, 110}

\bibitem[\protect\citeauthoryear{{Nayyar} \& {Owen}}{{Nayyar} \&
  {Owen}}{2006}]{2006PhRvD..73h4001N}
{Nayyar} M.,  {Owen} B.~J.,  2006, \mn@doi [\prd] {10.1103/PhysRevD.73.084001},
  \href {http://adsabs.harvard.edu/abs/2006PhRvD..73h4001N} {73, 084001}

\bibitem[\protect\citeauthoryear{{Page}, {Lattimer}, {Prakash}  \&
  {Steiner}}{{Page} et~al.}{2013}]{2013arXiv1302.6626P}
{Page} D.,  {Lattimer} J.~M.,  {Prakash} M.,   {Steiner} A.~W.,  2013,
  preprint, \href {http://adsabs.harvard.edu/abs/2013arXiv1302.6626P} {}
  (\mn@eprint {arXiv} {1302.6626})

\bibitem[\protect\citeauthoryear{{Passamonti}, {Andersson}  \&
  {Ho}}{{Passamonti} et~al.}{2015}]{2015arXiv150407470P}
{Passamonti} A.,  {Andersson} N.,   {Ho} W.~C.~G.,  2015, preprint, \href
  {http://adsabs.harvard.edu/abs/2015arXiv150407470P} {} (\mn@eprint {arXiv}
  {1504.07470})

\bibitem[\protect\citeauthoryear{{Potekhin}, {Fantina}, {Chamel}, {Pearson}  \&
  {Goriely}}{{Potekhin} et~al.}{2013}]{2013A&A...560A..48P}
{Potekhin} A.~Y.,  {Fantina} A.~F.,  {Chamel} N.,  {Pearson} J.~M.,   {Goriely}
  S.,  2013, \mn@doi [\aap] {10.1051/0004-6361/201321697}, \href
  {http://adsabs.harvard.edu/abs/2013A%26A...560A..48P} {560, A48}

\bibitem[\protect\citeauthoryear{{Prix} \& {Rieutord}}{{Prix} \&
  {Rieutord}}{2002}]{2002A&A...393..949P}
{Prix} R.,  {Rieutord} M.,  2002, \mn@doi [\aap] {10.1051/0004-6361:20021049},
  \href {http://adsabs.harvard.edu/abs/2002A%26A...393..949P} {393, 949}

\bibitem[\protect\citeauthoryear{{Reisenegger} \& {Goldreich}}{{Reisenegger} \&
  {Goldreich}}{1992}]{1992ApJ...395..240R}
{Reisenegger} A.,  {Goldreich} P.,  1992, \mn@doi [\apj] {10.1086/171645},
  \href {http://adsabs.harvard.edu/abs/1992ApJ...395..240R} {395, 240}

\bibitem[\protect\citeauthoryear{{Sathyaprakash} et~al.,}{{Sathyaprakash}
  et~al.}{2012}]{2012CQGra..29l4013S}
{Sathyaprakash} B.,  et~al., 2012, \mn@doi [Classical and Quantum Gravity]
  {10.1088/0264-9381/29/12/124013}, \href
  {http://adsabs.harvard.edu/abs/2012CQGra..29l4013S} {29, 124013}

\bibitem[\protect\citeauthoryear{{Strohmayer} \& {Mahmoodifar}}{{Strohmayer} \&
  {Mahmoodifar}}{2014a}]{2014ApJ...784...72S}
{Strohmayer} T.,  {Mahmoodifar} S.,  2014a, \mn@doi [\apj]
  {10.1088/0004-637X/784/1/72}, \href
  {http://adsabs.harvard.edu/abs/2014ApJ...784...72S} {784, 72}

\bibitem[\protect\citeauthoryear{{Strohmayer} \& {Mahmoodifar}}{{Strohmayer} \&
  {Mahmoodifar}}{2014b}]{2014ApJ...793L..38S}
{Strohmayer} T.,  {Mahmoodifar} S.,  2014b, \mn@doi [\apjl]
  {10.1088/2041-8205/793/2/L38}, \href
  {http://adsabs.harvard.edu/abs/2014ApJ...793L..38S} {793, L38}

\bibitem[\protect\citeauthoryear{{Strohmayer} \& {Watts}}{{Strohmayer} \&
  {Watts}}{2006}]{2006ApJ...653..593S}
{Strohmayer} T.~E.,  {Watts} A.~L.,  2006, \mn@doi [\apj] {10.1086/508703},
  \href {http://adsabs.harvard.edu/abs/2006ApJ...653..593S} {653, 593}

\bibitem[\protect\citeauthoryear{{Takatsuka}, {Nishizaki}, {Yamamoto}  \&
  {Tamagaki}}{{Takatsuka} et~al.}{2006}]{2006PThPh.115..355T}
{Takatsuka} T.,  {Nishizaki} S.,  {Yamamoto} Y.,   {Tamagaki} R.,  2006,
  \mn@doi [Progress of Theoretical Physics] {10.1143/PTP.115.355}, \href
  {http://adsabs.harvard.edu/abs/2006PThPh.115..355T} {115, 355}

\bibitem[\protect\citeauthoryear{{Wang} \& {Shen}}{{Wang} \&
  {Shen}}{2010}]{2010PhRvC..81b5801W}
{Wang} Y.~N.,  {Shen} H.,  2010, \mn@doi [\prc] {10.1103/PhysRevC.81.025801},
  \href {http://adsabs.harvard.edu/abs/2010PhRvC..81b5801W} {81, 025801}

\bibitem[\protect\citeauthoryear{{Watts} \& {Strohmayer}}{{Watts} \&
  {Strohmayer}}{2007a}]{2007AdSpR..40.1446W}
{Watts} A.~L.,  {Strohmayer} T.~E.,  2007a, \mn@doi [Advances in Space
  Research] {10.1016/j.asr.2006.12.021}, \href
  {http://adsabs.harvard.edu/abs/2007AdSpR..40.1446W} {40, 1446}

\bibitem[\protect\citeauthoryear{{Watts} \& {Strohmayer}}{{Watts} \&
  {Strohmayer}}{2007b}]{2007Ap&SS.308..625W}
{Watts} A.~L.,  {Strohmayer} T.~E.,  2007b, \mn@doi [\apss]
  {10.1007/s10509-007-9296-z}, \href
  {http://adsabs.harvard.edu/abs/2007Ap%26SS.308..625W} {308, 625}

\bibitem[\protect\citeauthoryear{{Weissenborn}, {Chatterjee}  \&
  {Schaffner-Bielich}}{{Weissenborn} et~al.}{2012a}]{2012PhRvC..85f5802W}
{Weissenborn} S.,  {Chatterjee} D.,   {Schaffner-Bielich} J.,  2012a, \mn@doi
  [\prc] {10.1103/PhysRevC.85.065802}, \href
  {http://adsabs.harvard.edu/abs/2012PhRvC..85f5802W} {85, 065802}

\bibitem[\protect\citeauthoryear{{Weissenborn}, {Chatterjee}  \&
  {Schaffner-Bielich}}{{Weissenborn} et~al.}{2012b}]{2012NuPhA.881...62W}
{Weissenborn} S.,  {Chatterjee} D.,   {Schaffner-Bielich} J.,  2012b, \mn@doi
  [Nuclear Physics A] {10.1016/j.nuclphysa.2012.02.012}, \href
  {http://adsabs.harvard.edu/abs/2012NuPhA.881...62W} {881, 62}

\bibitem[\protect\citeauthoryear{{Yakovlev}, {Levenfish}  \&
  {Shibanov}}{{Yakovlev} et~al.}{1999}]{1999SvPhU..42..737Y}
{Yakovlev} D.~G.,  {Levenfish} K.~P.,   {Shibanov} Y.~A.,  1999, Soviet Physics
  Uspekhi, \href {http://adsabs.harvard.edu/abs/1999SvPhU..42..737Y} {42, 737}

\makeatother
\end{thebibliography}
\bibliographystyle{mnras}

\appendix

%%%%%%%%%%%%%%%%%%%%%%%%%%%%%%%%%%%%%%%%%%%%%%%%%%%%%%%%%%%%%%%%%%%%%%%%%%%%%%%%%%%%%%%%%%%%%%%
\section{Error estimates for superfluid modes in the decoupling approximation}
\label{sec:appendix-decoupling}
%%%%%%%%%%%%%%%%%%%%%%%%%%%%%%%%%%%%%%%%%%%%%%%%%%%%%%%%%%%%%%%%%%%%%%%%%%%%%%%%%%%%%%%%%%%%%%%

As mentioned in Section~\ref{sec:decoupling-sfl},
	if all the coupling parameters 
	are strictly zero
	($s_{\rm e} = s_{\rm \mu} = s_{\rm str} = 0$),
	then the superfluid oscillation modes can be studied
	by making use of the potentiality conditions 
	for motion of superfluid components \eqref{eq:potentiality}
	together with the continuity equations \eqref{eq:continuity}
	and the conditions $\delta U_{({\rm b})}^\alpha = \delta g^{\alpha\beta} = 0$.

However, if the coupling parameters 
	are finite (but small), 
	which is the case for realistic EOSs,
	then the application of the decoupling approximation scheme
	directly to equation \eqref{eq:potentiality}
	will lead to significant errors and hence is not appropriate.
In this section we briefly explain this fact
	and demonstrate	that use of `superfluid' equations
	\eqref{eq:sfl-osc-n} and \eqref{eq:sfl-osc-L} instead of \eqref{eq:potentiality}
	substantially reduce errors and thus
	is more suitable for calculations of superfluid modes in the decoupling approximation.

Suppose that we have calculated some superfluid oscillation modes in the decoupling regime, assuming
	$\delta U_{({\rm b})}^\alpha = \delta g^{\alpha\beta} = 0$.
How good is this approximation
	if the coupling parameters are small but finite?
To estimate an error one has to compare the various terms depending on
	the baryon four-velocity perturbation $\delta U_{({\rm b})}^\alpha$
	and on the superfluid vectors $w_{(i)}^\alpha$\footnote{
		To simplify our consideration,
			we ignore in what follows
			a metric perturbation $\delta g^{\alpha\beta}$,
			which typically has a smaller effect 
			on oscillations than
			$\delta U_{({\rm b})}^\alpha$
			(see e.g., \citealt{1990ApJ...348..198L}).
	}.
Since in the fully decoupled case
	$\delta U_{({\rm b})}^\alpha$ vanishes for the superfluid modes,
	it should be small,
	$\sim O(|s_{\rm e}|+|s_{\mu}|+|s_{\rm str}|)$, 
	in the exact calculation.
In other words, 
	for the superfluid modes one can make the following estimate,
	$\delta U_{({\rm b})} \sim s W$,
	where
	$s = |s_{\rm e}|+|s_{\rm \mu}|+|s_{\rm str}|$,
	and $\delta U_{({\rm b})}$ and $W$ are the absolute values of the perturbations
	of the baryon four-velocity $\delta U_{({\rm b})}^\alpha$
	and the superfluid four-vector $W^\alpha = \sum_i Y_{ik} w_{(k)}^\alpha$,
	respectively:
	$\delta U_{({\rm b})} \equiv
		\sqrt{| \delta U_{({\rm b})}^\alpha \delta U_{({\rm b})\alpha} |}$,
	$W \equiv 1 / n_{\rm b}  \sqrt{| W^\alpha W_\alpha |}$.

First, we show 
	that the use of the potentiality conditions
	\eqref{eq:potentiality} leads to large errors
	if calculations are made in the fully decoupled case ($s=0$).
Let us consider a harmonic perturbation
	($\propto {\rm e}^{{\rm i} \omega t}$)
	of a non-rotating star,
	assuming, for simplicity, that only $\rm \Lambda$--hyperons are superfluid.
Equation \eqref{eq:potentiality} for $\rm \Lambda$--hyperons 
	in the linear approximation reads
	(we take $\beta = 0$ and $\alpha = 1,2,3$)
\begin{gather}
\label{eq:app:potentiality}
	{\rm i} \omega \left(
			w_{({\rm \Lambda})\alpha}
			+ \mu_{\rm \Lambda} \delta u_\alpha
		\right)
	-\pd{}{x^\alpha}\left(
			u_0 \delta \mu_\Lambda
		\right)
	= 0
.
\end{gather}
Using the definitions for $U_{({\rm b})\alpha}$ \eqref{eq:continuity-b}
	and $W_\alpha$,
	one can rewrite equation~\eqref{eq:app:potentiality}
	as
\begin{gather}
\label{eq:app:sfl-osc}
	{\rm i} \omega \left[
			\frac{n_{\rm b}}{Y_{\Lambda\Lambda}} W_\alpha
			+ \mu_\Lambda \left( \delta U_{({\rm b})\alpha} - W_\alpha \right)
		\right]
	= \pd{}{x^\alpha}\left(
			u_0 \delta \mu_\Lambda
		\right)
.
\end{gather}
In the left-hand side of equation~\eqref{eq:app:sfl-osc}
	the `superfluid' terms, depending on $W_\alpha$,
	are much greater than the
	`normal' terms, depending on $\delta U_{({\rm b})\alpha}$,
	because $\delta U_{({\rm b})} \sim sW$.
The approximation $\delta U_{({\rm b})\alpha} = 0$
	is valid only if the same is true
	also for the terms in the right-hand side
	of this equation, namely
	for the quantity $\delta\mu_{\rm \Lambda}$.
Generally, $\delta \mu_\Lambda$ depends on both
	the baryon velocity $\delta U_{({\rm b})\alpha}$
	and superfluid four-vector $W_\alpha$.
Let us express $\delta\mu_{\rm \Lambda}$ through the number density perturbations
	$\delta n_{\rm b}$, $\delta n_{\rm e}$, $\delta n_{\rm \mu}$, and $\delta n_{\rm str}$:
\begin{gather}
\label{eq:app:dmu_L}
	\delta\mu_{\rm \Lambda}
	= \frac{\delta n_{\rm b}}{n_{\rm b}} \pd{\mu_\Lambda}{\ln n_{\rm b}}
		+ \frac{\delta n_{\rm e}}{n_{\rm e}} \pd{\mu_\Lambda}{\ln n_{\rm e}}
		+ \frac{\delta n_{\rm \mu}}{n_{\rm \mu}} \pd{\mu_\Lambda}{\ln n_{\rm \mu}}
		+ \frac{\delta n_{\rm str}}{n_{\rm str}} \pd{\mu_\Lambda}{\ln n_{\rm str}}
.
\end{gather}
These perturbations can in turn be expressed through
	$\delta U_{({\rm b})}^\alpha$ and $W^\alpha$
	using the continuity equations
	\eqref{eq:continuity-b} and \eqref{eq:continuity}:
\begin{gather}
\label{eq:app:dn_b}
	\frac{\delta n_{\rm b}}{n_{\rm b}}
	= - \frac{1}{{\rm i} \omega} {\rm e}^{\nu/2}
		\left(
			\covd{\delta U_{({\rm b})}^\alpha}{\alpha} + \d{\ln n_{\rm b}}{r} \delta U_{({\rm b})}^r
		\right)
,\\
\label{eq:app:dn_e}
	\frac{\delta n_{\rm e}}{n_{\rm e}}
	= - \frac{1}{{\rm i} \omega} {\rm e}^{\nu/2}
		\left[
			\covd{\left(\delta U_{({\rm b})}^\alpha - W^\alpha\right)}{\alpha}
			+ \d{\ln n_{\rm e}}{r} \left(\delta U_{({\rm b})}^r - W^r\right)
		\right]
,\\
\label{eq:app:dn_mu}
	\frac{\delta n_{\rm \mu}}{n_{\rm \mu}}
	= - \frac{1}{{\rm i} \omega} {\rm e}^{\nu/2}
		\left[
			\covd{\left(\delta U_{({\rm b})}^\alpha - W^\alpha \right)}{\alpha}
			+ \d{\ln n_{\rm \mu}}{r} \left(\delta U_{({\rm b})}^r - W^r\right)
		\right]
,\\
\label{eq:app:dn_str}
	\frac{\delta n_{{\rm str}}}{n_{{\rm str}}}
	= - \frac{1}{{\rm i} \omega} {\rm e}^{\nu/2}
		\left[
			\covd{\left(
				\delta U_{({\rm b})}^\alpha + \frac{n_{\rm b} - n_{{\rm str}}}{n_{{\rm str}}} W^\alpha
			\right)}{\alpha}
			+ \d{\ln n_{{\rm str}}}{r} \left(
					\delta U_{({\rm b})}^r + \frac{n_{\rm b} - n_{{\rm str}}}{n_{{\rm str}}}W^r
				\right)
		\right]
.
\end{gather}
After substituting \eqref{eq:app:dn_b}--\eqref{eq:app:dn_str}
	into \eqref{eq:app:dmu_L},
	one can roughly estimate
	the ratio of normal to superfluid terms
	in $\delta\mu_{\rm \Lambda}$ as
\begin{gather}
\label{eq:app:dmu_L-norm-to-SFL}
	\frac{\text{normal~terms~in~} \delta\mu_{\rm \Lambda} }{\text{SFL~terms~in~} \delta\mu_{\rm \Lambda}}
		\sim z \frac{\delta U_{({\rm b})}}{W} \sim z s
\end{gather}
(remember that $\delta U_{({\rm b})} \sim s W$), where
\begin{gather}
\label{eq:app:dmu_L-z}
	z =- \left(
			\pd{\mu_{\rm \Lambda}}{\ln n_{\rm b}}
			+\pd{\mu_{\rm \Lambda}}{\ln n_{\rm e}}
			+\pd{\mu_{\rm \Lambda}}{\ln n_{\rm \mu}}
			+\pd{\mu_{\rm \Lambda}}{\ln n_{\rm str}}
		\right)
		/
		\left(
			\pd{\mu_{\rm \Lambda}}{\ln n_{\rm e}}
			+\pd{\mu_{\rm \Lambda}}{\ln n_{\rm \mu}}
			+\pd{\mu_{\rm \Lambda}}{\ln n_{\rm str}} \frac{n_{\rm str} - n_{\rm b}}{n_{\rm str}}
		\right)
.
\end{gather}
For the EOSs GM1A,~GM1'B, and TM1C 
$zs$ can be larger than unity even when $s$ is small.
For example, for the EOS TM1C
	$z \approx 14.6$ and
	$|z| |s| \approx 1.66$
	at $n_{\rm b} = 0.5~{\rm fm}^{-3}$ .
Thus, for superfluid modes 
	the terms depending on $\delta U_{({\rm b})}^\alpha$ 
	can be even greater than the terms depending on $W^\alpha$.
This means that the approximation
	$\delta U_{({\rm b})}^\alpha = 0$
	leads to completely wrong results 
	if we use it together with the potentiality conditions \eqref{eq:potentiality}.

Now let us check whether the approximation
	$\delta U_{({\rm b})}^\alpha = 0$
	is suitable for calculating the superfluid modes 
	within the approach presented in Section~\ref{sec:decoupling},
	when we use equation~\eqref{eq:sfl-osc-L}
	instead of equation~\eqref{eq:app:potentiality}.
We have to compare the `normal' and `superfluid' terms
	entering the expressions for $\Delta\mu_j$,
	where $j = {\rm e}$, ${\rm \mu}$, ${\rm \Lambda}$.
One can write out an expansion
	for $\Delta\mu_j$
	similar to equation~\eqref{eq:app:dmu_L},
\begin{gather}
\label{eq:app:dmu_j}
	\Delta\mu_j
	= \frac{\delta n_{\rm b}}{n_{\rm b}} \pd{\Delta\mu_j}{\ln n_{\rm b}}
		+ \frac{\delta n_{\rm e}}{n_{\rm e}} \pd{\Delta\mu_j}{\ln n_{\rm e}}
		+ \frac{\delta n_{\rm \mu}}{n_{\rm \mu}} \pd{\Delta\mu_j}{\ln n_{\rm \mu}}
		+ \frac{\delta n_{\rm str}}{n_{\rm str}} \pd{\Delta\mu_j}{\ln n_{\rm str}}
.
\end{gather}
Using then equations \eqref{eq:app:dn_b}--\eqref{eq:app:dn_str},
	one can estimate the ratio of the `normal' to `superfluid' terms
	in $\Delta\mu_j$
	as
\begin{gather}
\label{eq:app:dmu_e-norm-to-SFL}
	\frac{\text{normal~terms~in~} \Delta\mu_j }{\text{SFL~terms~in~} \Delta\mu_j}
		\sim z_j \frac{\delta U_{({\rm b})}}{W} \sim z_j s
,
\end{gather}
where 
\begin{gather}
\label{eq:app:zj}
	z_j =- \left(
			\pd{\Delta\mu_j}{\ln n_{\rm b}}
			+\pd{\Delta\mu_j}{\ln n_{\rm e}}
			+\pd{\Delta\mu_j}{\ln n_{\rm \mu}}
			+\pd{\Delta\mu_j}{\ln n_{\rm str}}
		\right)
		/
		\left(
			\pd{\Delta\mu_j}{\ln n_{\rm e}}
			+\pd{\Delta\mu_j}{\ln n_{\rm \mu}}
			+\pd{\Delta\mu_j}{\ln n_{\rm str}} \frac{n_{\rm str} - n_{\rm b}}{n_{\rm str}}
		\right)
.
\end{gather}
As a result, the total error of the approximation 
$\delta U_{({\rm b})}^\alpha = 0$
	in equation~\eqref{eq:sfl-osc-L}
	can be estimated as
	$(|z_{\rm e}|+|z_{\rm \mu}|+|z_{\rm \Lambda}|) |s|$,
	which is the sum of errors arising from the three terms
	in the right-hand side of that equation.
For the hyperonic EOSs GM1A, GM1'B, and TM1C
	$|z_{\rm e}|, |z_{\rm \mu}|, |z_{\rm \Lambda}| \sim 1$,
	whereas the coupling parameters
	$|s_{\rm e}|, |s_{\rm \mu}|, |s_{\rm str}| \ll 1$, 
	so our perturbative scheme is valid.
For example, for the EOS TM1C at $n_{\rm b} = 0.5~{\rm fm}^{-3}$ 
	$(|z_{\rm e}|+|z_{\rm \mu}|+|z_{\rm \Lambda}|) |s|
		\approx
		0.23$, 
	hence our decoupling scheme developed in Section~\ref{sec:decoupling}
	calculates the superfluid modes within the accuracy of $\sim 20\%$.
Estimates presented here are supported by calculations 
	of sound speeds in Section~\ref{sec:sound}.

%%%%%%%%%%%%%%%%%%%%%%%%%%%%%%%%%%%%%%%%%%%%%%%%%%%%%%%%%%%%%%%%%%
\section{Superfluid oscillation equations for different sets of superfluid particle species}
\label{sec:appendix-generalized-sfl}
%%%%%%%%%%%%%%%%%%%%%%%%%%%%%%%%%%%%%%%%%%%%%%%%%%%%%%%%%%%%%%%%%%

In Section~\ref{sec:decoupling-sfl} 
	we derived the superfluid equation~\eqref{eq:sfl-osc-general2}
	using the potentiality condition~\eqref{eq:potentiality} 
	for the neutron superfluid four-vector
	$w_{({\rm n})}^\alpha$
	as well as the energy-momentum conservation law~\eqref{eq:Tabb}.
In addition, we used the fact that
	$P + \varepsilon - \mu_{\rm n} n_{\rm b}$
	and $\partial_\beta P - n_{\rm b} \partial_\beta \mu_{\rm n}$
	are small quantities, vanishing in equilibrium.
The same derivation can be performed for any baryon species $i$, 
	if the following two conditions are fulfilled:
\begin{enumerate}
	\item the superfluid four-vector $w_{(i)\alpha}$ satisfies the potentiality equation
			$
				\covd{\left( w_{(i)\alpha} + \mu_i u_\alpha \right)}{\beta}
				-\covd{\left( w_{(i)\beta} + \mu_i u_\beta \right)}{\alpha} = 0
			$;
			
	\item the difference of chemical potentials $\mu_i - \mu_{\rm n}$ is a small quantity, 
	vanishing in equilibrium.
\end{enumerate}
These conditions are fulfilled e.g., for $\rm \Lambda$--~or~${\rm \Xi}^0$--hyperons.

Furthermore, one can take not only a single
	superfluid four-vector $w_{(i)\alpha}$
	and chemical potential $\mu_i$,
	but also an appropriate 
	linear combination of $w_{(i)\alpha}$ 
	and the corresponding linear combination of $\mu_i$.
For example, if we introduce
	a `quasiparticle' $({\rm p} + {\rm \Sigma}^-)/2$,
	with 
	$w_{\alpha} = (w_{({\rm p})\alpha} + w_{({\rm \Sigma}^{-})\alpha})/2$
	and the chemical potential
	$\mu = (\mu_{\rm p} + \mu_{{\rm \Sigma}^-})/2$,
	then it meets the conditions (i) and (ii).
Therefore, even in this case we can derive a superfluid equation 
	for this `quasiparticle'.

Let us demonstrate this for an arbitrary 
	particle (or `quasiparticle') $A$, which meets the conditions (i) and (ii) 
	[in particular, $\mu_A-\mu_{\rm n}$ vanishes in equilibrium].
The derivation is the same as that for equations 
	\eqref{eq:sfl-osc-general2}--\eqref{eq:sfl-osc-L}.
Now we shall outline it,
	underlining, for clarity, the additional terms 
	that have not appeared in the derivation of 
	\eqref{eq:sfl-osc-general2}--\eqref{eq:sfl-osc-L}.
Using the energy-momentum conservation law \eqref{eq:Tabb} 
together with the potentiality condition \eqref{eq:potentiality} 
	for (quasi)particle $A$, one obtains an equation similar to \eqref{eq:sfl-osc-general},
\begin{multline}
\label{eq:sfl-osc-general-A}
	\covd{{T_\alpha}^\beta}{\beta} + u_\alpha u_\gamma  \covd{T^{\gamma\beta}}{\beta}
	- n_{\rm b} u^\beta
		\left[
			\covd{\left( w_{(A)\alpha} + \mu_A u_\alpha \right)}{\beta}
			-\covd{\left( w_{(A)\beta} + \mu_A u_\beta \right)}{\alpha}
		\right]
	\\=
	(P+\varepsilon - \mu_A n_{\rm b}) u^\beta \covd{u_\alpha}{\beta}
		+ \left(
				\covd{P}{\beta} - n_{\rm b} \covd{\mu_A }{\beta}
			\right) u_\alpha u^\beta
		+ \left(
				\covd{P}{\alpha} - n_{\rm b} \covd{\mu_A }{\alpha}
			\right)
		\\
		+ (g_{\alpha\gamma} + u_\alpha u_\gamma ) u^\beta \covd{( \mu_{\rm n} n_{\rm b} W^\gamma )}{\beta}
		+\mu_{\rm n} n_{\rm b}
		\left(
			\covd{{u^\beta}}{\beta}  W_\alpha
			+ \covd{u_{\alpha}}{\beta} W^\beta
		\right)
	- n_{\rm b} u^\beta
	\left[
		\covd{w_{(A)\alpha}}{\beta}
		-\covd{w_{(A)\beta}}{\alpha}
	\right]
	= 0
.
\end{multline}
It follows from equations \eqref{eq:thermodyn1} and \eqref{eq:thermodyn3} that
\begin{gather}
\label{eq:sfl-gen:thermodyn1}
	P + \varepsilon - \mu_A n_{\rm b} 
		%= P + \varepsilon - \mu_{\rm n} n_{\rm b} + \underline{\Delta\mu_A n_{\rm b}}
		= - \Delta\mu_{\rm e} n_{\rm e} 
		  - \Delta\mu_{\rm \mu} n_{\rm \mu}
		  - \Delta\mu_{\rm \Lambda} n_{\rm str} 
		  + \underline{\Delta\mu_A n_{\rm b}}
	,\quad
		\Delta\mu_A \equiv \mu_{\rm n} - \mu_A
	,\\
\label{eq:sfl-gen:thermodyn2}
	{\rm d}P - n_{\rm b} {\rm d}\mu_A
		= - n_{\rm e} {\rm d}\Delta\mu_{\rm e}
		  - n_{\rm \mu} {\rm d}\Delta\mu_{\rm \mu}
		  - n_{\rm str} {\rm d}\Delta\mu_{\rm \Lambda}
		  + \underline{n_{\rm b} {\rm d}\Delta\mu_A}
.
\end{gather}
Using these relations one can rewrite equation \eqref{eq:sfl-osc-general-A} as
\begin{multline}
\label{eq:sfl-osc-general-A2}
		\left(
			- \Delta\mu_{\rm e} n_{\rm e}  - \Delta\mu_{\rm \mu} n_{\rm \mu} - \Delta\mu_{\rm \Lambda} n_{\rm str}
				+ \underline{\Delta\mu_A n_{\rm b}} 
			\right) u^\beta \covd{u_\alpha}{\beta}
		+ \left(
			- n_{\rm e} \partial_\beta{\Delta\mu_{\rm e}} - n_{\rm \mu} \partial_\beta{\Delta\mu_{\rm \mu}} - n_{\rm str} \partial_\beta{\Delta\mu_{\rm \Lambda}}
			+ \underline{n_{\rm b} \partial_\beta \Delta\mu_A} 
			\right) u_\alpha u^\beta
		\\+ \left(
				- n_{\rm e} \partial_\alpha{\Delta\mu_{\rm e}}- n_{\rm \mu} \partial_\alpha{\Delta\mu_{\rm \mu}} - n_{\rm str} \partial_\alpha{\Delta\mu_{\rm \Lambda}}
				+ \underline{n_{\rm b} \partial_\alpha \Delta\mu_A} 
			\right)
		+ (g_{\alpha\gamma} + u_\alpha u_\gamma ) u^\beta \covd{( \mu_{\rm n} n_{\rm b} W^\gamma )}{\beta}
		+\mu_{\rm n} n_{\rm b}
		\left(
			\covd{{u^\beta}}{\beta}  W_\alpha
			+ \covd{u_{\alpha}}{\beta} W^\beta
		\right)
	\\- n_{\rm b} u^\beta
	\left[
		\covd{w_{(A)\alpha}}{\beta}
		-\covd{w_{(A)\beta}}{\alpha}
	\right]
	= 0
.
\end{multline}
In the case of
	a non-rotating star with the Schwarzschild metric,
	when all perturbations depend on time as ${\rm e}^{{\rm i}\omega t}$,
	the spatial components ($\alpha = 1,2,3$)
	of this equation
	take the following final form:
\begin{gather}
\label{eq:sfl-osc-A}
	{\rm i} \omega n_{\rm b} ( \mu_{\rm n} W_\alpha - w_{(A)\alpha})
	=
		n_{\rm e}     \pd{}{x^\alpha}\left( \Delta\mu_{\rm e} {\rm e}^{\nu/2} \right)
		+ n_{\rm \mu} \pd{}{x^\alpha}\left( \Delta\mu_{\rm \mu} {\rm e}^{\nu/2} \right)
		+ n_{\rm str} \pd{}{x^\alpha}\left( \Delta\mu_{\rm \Lambda} {\rm e}^{\nu/2} \right)
		-\underline{ n_{\rm b} \pd{}{x^\alpha}\left( \Delta\mu_A {\rm e}^{\nu/2} \right) }
	,\quad \alpha = 1,2,3
.
\end{gather}

Now let us focus on the following question.
In a real neutron star, 
	depending on a density and temperature,
	some particle species are superfluid,
	some are present but non-superfluid,
	while others are absent.
How many different equations do we need to cover all the cases?
It turns out that, if the thresholds for the appearance of hyperons $n_{({\rm b})}^{(i)}$ 
	satisfy the inequality
	$n_{({\rm b})}^{({\rm \Lambda})}
		< n_{({\rm b})}^{({\rm \Xi}^{-})}
		< n_{({\rm b})}^{({\rm \Xi}^{0})}
		< n_{({\rm b})}^{({\rm \Sigma}^{-})}
	$
	(which is true for many modern equations of state, 
		including GM1A, GM1`B, and TM1C),
	then in all the situations there are no more than two superfluid degrees of freedom.
Interestingly, all the cases except one (see below)
	can be covered with the only four choices
	of (quasi)particle $A$ in equation
	\eqref{eq:sfl-osc-general-A2} (or \ref{eq:sfl-osc-A}),
\begin{enumerate}
	\item $A = {\rm n}$
	\item $A = {\rm \Lambda}$
	\item $A = {\rm \Xi}^0$%:
	\item $A = ({\rm p} + {\rm \Sigma}^-)/2$.
\end{enumerate}

The special case is when
	${\rm \Xi}^-$-- and ${\rm \Sigma}^-$--hyperons
	are the \textit{only} superfluid species
	in the system.
Then
	we can construct superfluid equation by subtracting
	the potentiality condition \eqref{eq:potentiality} for ${\rm \Sigma}^-$--hyperons
	from the potentiality condition for ${\rm \Xi}^-$-- hyperons.
Using then the fact that
	$\mu_{{\rm \Sigma}^-} - \mu_{{\rm \Xi}^-} = \mu_{\rm n} - \mu_{\rm \Lambda} = \Delta\mu_{\rm \Lambda}$
	(see equations \ref{eq:dmu-LLnX0}--\ref{eq:dmu-nLpSm}), one gets
\begin{gather}
\label{eq:sfl-gen:osc-XS}	
	{\rm i} \omega (w_{({\rm \Xi}^-)\alpha} - w_{({\rm \Sigma}^-)\alpha}) = \pd{}{x^\alpha}\left( \Delta\mu_{\rm \Lambda} {\rm e}^{\nu/2} \right)
	,\quad \alpha = 1,2,3.
\end{gather}

Let us illustrate the above statements by considering
a few possible situations.

(1) Assume that the protons	as well as
	the ${\rm \Xi}^0$-- and ${\rm \Xi}^{-}$--hyperons
	are superfluid, while other particles are not.
Then
	there are two superfluid degrees of freedom;
	the use of the superfluid four-vector $w_{({\rm n})\alpha}$ and, 
	as a consequence, of the superfluid equation for neutrons, seems to be incorrect (neutrons are normal!).
However, one can {\it formally} introduce a variable 
	$w_{({\rm n})\alpha} \equiv w_{({\rm p})\alpha} + w_{({\rm \Xi}^-)\alpha} - w_{({\rm \Xi}^0)\alpha}$,
	such that the superfluid equation for `neutrons' remains valid
	(remember that $\mu_{\rm n} = \mu_{\rm p} + \mu_{{\rm \Xi}^-} - \mu_{{\rm \Xi}^0}$).
Moreover, proceeding in the same way, one can  introduce a variable
	$w_{({\rm \Lambda})\alpha} \equiv (w_{({\rm p})\alpha} + w_{({\rm \Xi}^-)\alpha})/2$, 
	and use the standard superfluid equation for `$\rm \Lambda$--hyperons' \eqref{eq:sfl-osc-L}.
As a result,
	we cover this case
	by formally introducing superfluid `quasiparticles'
	${\rm n} = {\rm p} + {\rm \Xi}^- - {\rm \Xi}^0$ (case i)
	and ${\rm \Lambda} = ({\rm p} + {\rm \Xi}^-) / 2$ (case ii).

(2) Assume now that only the neutrons, protons, and ${\rm \Sigma}^-$--hyperons are superfluid. 
In this situation one also has two superfluid degrees of freedom 
	and, therefore, two superfluid equations.
The first is the equation for neutrons (case i), 
	while the second is the superfluid equation
	for a quasiparticle
	$A = ({\rm p} + {\rm \Sigma}^-)/2$ (case iv).

\end{document}